# Supercurrent Time Division Multiplexing with Solid-State Integrated Hybrid Superconducting Electronics

Alessandro Paghi[1*], Laura Borgongino[1], Simone Tortorella[1,2], Giorgio De Simoni[1], Elia Strambini[1], Lucia Sorba[1], and Francesco Giazotto[1*]

[1]Istituto Nanoscienze-CNR and Scuola Normale Superiore, Piazza San Silvestro 12, 56127 Pisa, Italy.

[2]Dipartimento di Ingegneria Civile e Industriale, Università di Pisa, Largo Lucio Lazzarino, 56122 Pisa, Italy

[*]Corresponding authors: alessandro.paghi@nano.cnr.it, francesco.giazotto@sns.it

## Abstract

Time-division multiplexing of cryogenic signals is a promising approach to reduce space requirements, shorten cooldown times, and increase the number of quantum devices measured per cooldown. We demonstrate time-division multiplexing of non-dissipative supercurrents using voltage-controlled hybrid superconducting demultiplexers. These chips integrate superconducting Josephson Field Effect Transistors including Al superconducting electrodes, proximitized semiconducting InAs channels, and hafnium oxide gate insulators. Each transistor fully suppresses the switching current and increases the resistance 20 times under a gate voltage of -4.5 V. A demultiplexer with one input and eight outputs showed a non-dissipative input range of ±2 μA, operates up to 100 MHz in signal frequency, and 100 kHz in switching frequency at 50 mK. It achieved near-zero insertion loss in the superconducting state and an $\frac{ON}{OFF}$ ratio of 17.5 dB. By optimizing the signal layout, the operation was extended up to 4 GHz using a demultiplexer with two outputs.

## Introduction

In the past decades, there has been tremendous progress in the theoretical and experimental development of a wide range of quantum devices operating at cryogenic temperatures, where phase transitions, reduced thermal noise, and minimized interactions with the surrounding environment can be achieved. Superconducting [1–3] and spin qubits [2,4,5], parametric amplifiers [6–9], diodes [10–12], and single photon detectors [13–15], are only few examples of the devices proposed and manufactured. Along the same side, cryogenic temperatures are mandatory to study hot topics in quantum physics, such as spin-dependent supercurrent [16], topological phase transitions [17–19], superconductivity-quantum Hall effect interaction [20–22], and many-body electron systems [23–25].

In a conventional cryostat, N signal lines come from (to) room temperature electronics and travel to (from) the low-temperature plates connecting to the N ports of the quantum device under test (QDUT) (Fig. 1a). Each signal line significantly impacts the cryostat's costs, space occupied, and temperature stability. Thus, the number of cables is a significant constraint in testing and using of very-large-scale-integrated (VLSI) quantum devices. For instance, one of the most advanced quantum machines, the quantum computer Google Sycamore, uses more than 200 RF coaxial cables to control only 54 qubits [26,27]. To reduce the required space, minimize the cooldown time, and increase the number of measurable devices per cooldown, time division multiplexing (TDM) of signal lines is needed to route signals to the QDUT. In a TDM-cryostat, only one signal line reaches the lowest temperature plate to connect $2^N$ ports of the QPU using N control lines (Fig. 1b).

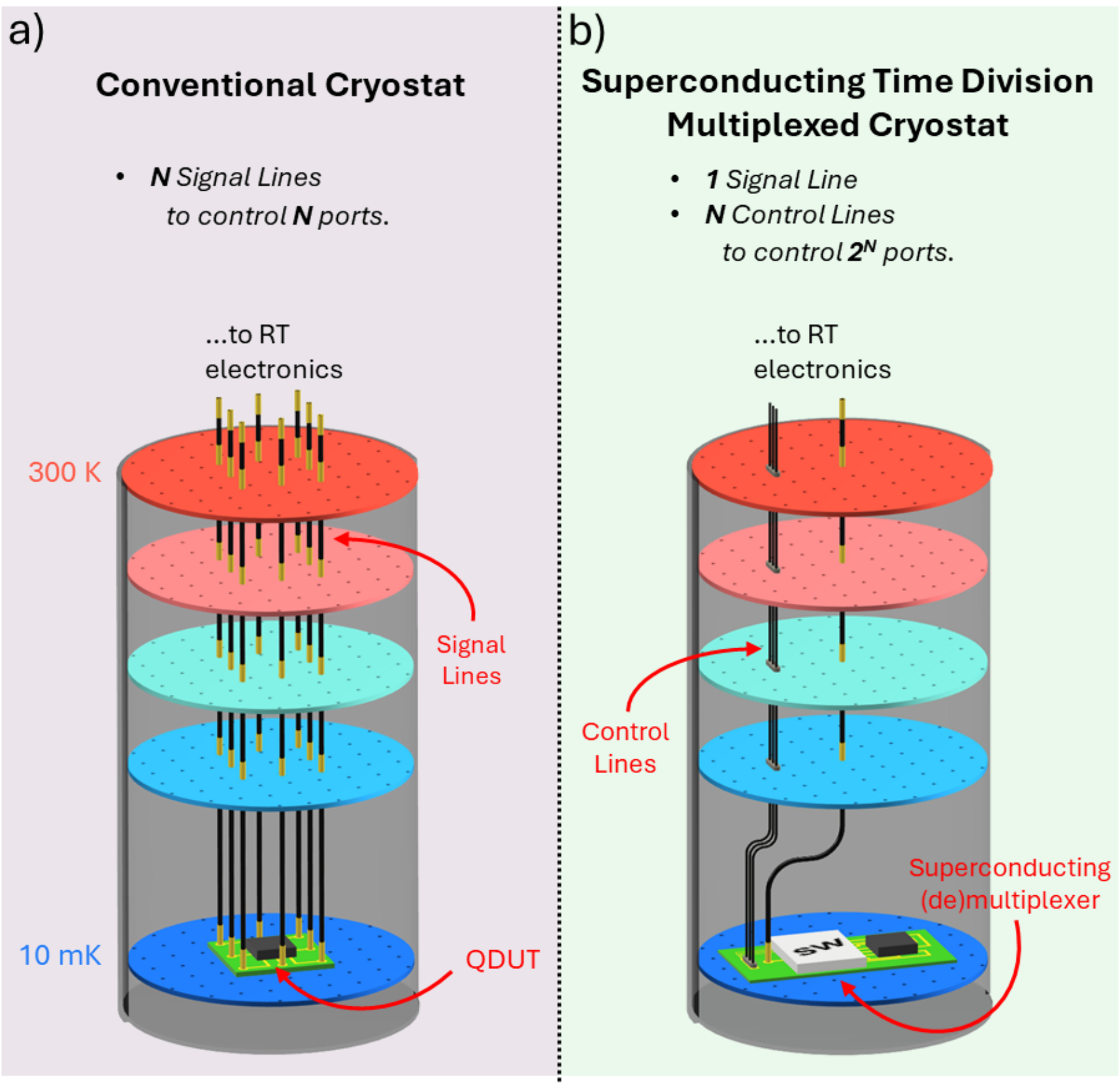

**Fig. 1: Comparison between conventional and superconducting TDM cryostats.** a) A conventional cryostat needs N signal lines to control N ports of the quantum device under test (QDUT). b) A superconducting TDM-cryostat features N control lines and one signal line to control $2^N$ ports of the QDUT. The schematic representations are intended to represent the input or the output lines of the QDUT.

Nowadays, electromechanical switches are usually involved in TDM of signals at cryogenic temperatures down to a few mK, with the drawbacks of a large amount of volume occupied (~100 $cm^{-3}$), reduced I/O lines, and a high switching time (~10 ms) [28,29,30]. Cryo-CMOS solid-state solutions offer the advantage of a reduced volume together with an increased number of I/O lines and faster-switching events (~1 ns), allowing at the same time to achieve line-to-line isolations of 30 ÷ 40 dBs [31,32,33,34]. On the other hand, static power dissipation accounts for the increase of the temperature of the cold plate of tens of mKs [31,33], while an insertion loss of 1 ÷ 3 dBs is usual in the ON-state due to the Ohmic behavior of the chip [33,34]. In quantum computing, it is noteworthy that the deployment of TDM using cryo-CMOS platforms remains incompatible with scalable, high-fidelity control of superconducting transmon qubits. This incompatibility arises from significant thermal radiation emitted by the multiplexer, necessitating the thermalization of signals between the demultiplexer's output and the quantum chip via a well-thermalized attenuator[33,34]. Along with the previous solutions, more exotic TDM demonstrations are proposed using platforms based on SiGe-2DEG [35], GaAs-HEMTs [36,37], and InAsNWs [38]. In this context, TDM with a superconducting solid-state multiplexer promises short downtimes between control switching events, near-to-0W static power dissipation, negligible heat injection, and a minimized insertion loss. In recent years, few demonstrations of proof-of-concept superconducting switches have been documented [39–41]. On the other hand, considerable insertion loss and static power required in the ON-state [39], reduced operating band [40,41], and a limited number of output ports (2 at maximum) [39–41], are the major drawbacks highlighted.

Hybrid superconductor-semiconductor architectures enable ideal platforms to develop Josephson Field Effect Transistors (JoFETs) for supercurrent multiplexing [42,43,44,45,46]. The hybrid super-semi JoFET is the key to implementing a two-state ON/OFF superconducting/normal building block [47]. The gate tunability of the critical current, as well as the gate tunability of the normal-state resistance, are the fundamental aspects involved in the supercurrent multiplexing operation. In the III-V compound group, InAs is the most common semiconductor employed in the manufacturing of JoFETs thanks to the optimal superconducting proximity effect [42,43,44,48,49,50,51], which provides the electron gas hosted by the InAs layer the ability to support a supercurrent. Among InAs-based platforms, we recently proposed InAs on Insulator (InAsOI) as a new convenient platform to develop planar hybrid superconducting electronics [52–54]. Unlike other InAs architectures, the InAsOI features a InAs epilayer grown onto a cryogenic electrical insulating InAlAs metamorphic buffer. The InAsOI allows the electrical decoupling of surface-exposed adjacent devices together with the capability of achieving the superconducting proximity effect with high critical current density integration. Irrespective of the fabrication methodology customary to GaAs-based heterostructures, the InAsOI epitomizes the counterpart of the silicon-on-insulator (SOI) architecture apt for superconducting technologies, wherein proximitized InAs and the cryogenic electrical insulating InAlAs substitute silicon and silicon oxide layers of SOI, respectively.

Here, we report on the TDM of non-dissipative supercurrents with voltage-actuated hybrid superconducting demultiplexers able to reduce the number of signal lines of a conventional cryostat. As ON-OFF building blocks, we developed InAsOI-based JoFETs featuring Al as superconductor and $HfO_2$ as gate insulator able to suppress the switching current entirely and to increase the normal-state resistance by 20 times with a gate voltage of -4.5 V. We proposed a superconducting demultiplexer with 1 input and 8 outputs operating at 50 mK up to 100 MHz in signal frequency and 100 kHz in switching frequency, featuring an insertion loss of ~ 0 dB in the superconducting state, and an $\frac{ON}{OFF}$ ratio of ~ 17.5 dB. The operational range of signal frequency was effectively broadened by designing a superconducting demultiplexer with a single input and dual outputs, incorporating an optimized microwave signal transport layout that concurrently mitigates the influence of parasitic electrical elements. The constructed device underwent validation at frequencies reaching up to 4 GHz at 50 mK, demonstrating an insertion loss of approximately 0 dB within the superconducting state and exhibiting an $\frac{ON}{OFF}$ ratio of ~ 16 dB. The reciprocity of the circuits allows, in principle, reversed operations of the proposed devices as a multiplexer by inverting the I/O ports.

## Results and Discussion

### *InAsOI-Based Josephson Field Effect Transistor Fabrication*

The cross-section of a Josephson Field Effect Transistor (JoFET) fabricated on the InAs on Insulator (InAsOI) platform is shown in Fig. 2a.

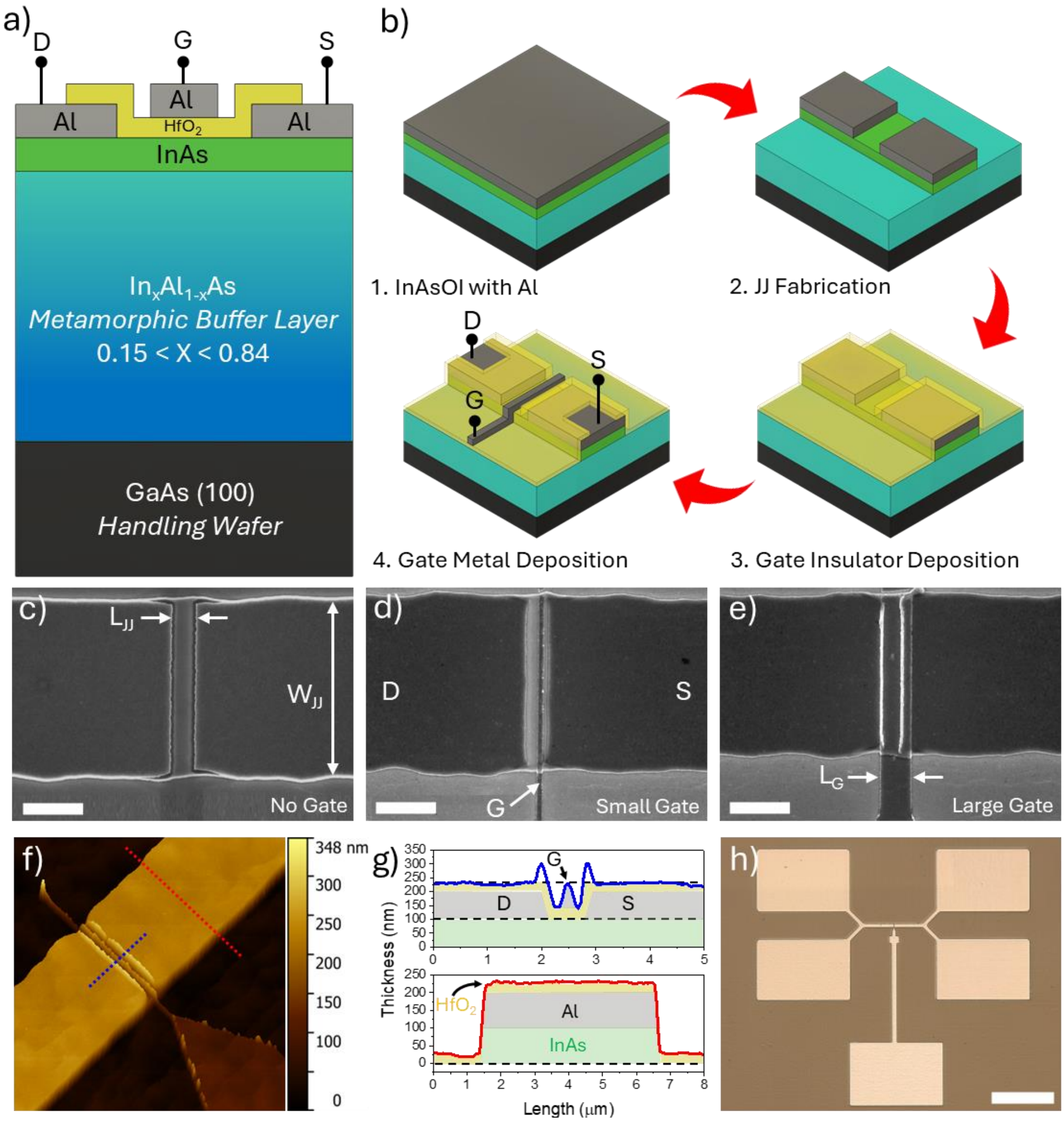

**Fig. 2: Josephson Field Effect Transistors with InAsOI: concept, manufacturing process, and morphologic characterization.** a) Cross-section structure of a JoFET with InAsOI. b) Fabrication process of a JoFET with InAsOI: 1. Superconductive Al deposition onto InAsOI; 2. JJ fabrication via

MESA and Al wet etching; 3. Gate insulator deposition via ALD; 4. Gate metal deposition. c,d,e) Top-view SEM images (10k ×) of JoFETs with $W_{JJ}$ = 6 µm and $L_{JJ}$ = 800 nm before gate fabrication (c), and after small ($L_G$ = 200 nm) (d) and large ($L_G$ = 1000 nm) (d) gate manufacturing; the scalebar is 2 µm. f) xy-plane tilted AFM z-profile of a JoFET featuring $W_{JJ}$ = 6 µm, $L_{JJ}$ = 1200 nm, and $L_G$ = 100 nm. g) AFM thickness profiles extrapolated from the red and blue dashed line in (f). h) Optical microscope image of the 4-terminal JoFET architecture. The scalebar is 100 µm.

The JoFET is built upon a Molecular Beam Epitaxy (MBE)-grown InAsOI stack consisting, from bottom to top, of a 500-µm-thick semi-insulating GaAs (001) substrate, a ~1.5-µm-thick step-graded $In_xAl_{1-x}As$ metamorphic buffer with X increasing from 0.15 to 0.84, and a 100-nm-thick InAs semiconductive epilayer [52]. The InAs layer is *n*-type doped (extrinsically undoped) due to the InAs natural surface electron accumulation layer [55,56] and deep energy donor levels in the InAlAs band gap [57]. For the InAs epilayer, a sheet electron density ($n_{2D}$) of $1.35\times10^{12}$ $cm^{-2}$ and a mobility ($\mu_n$) of $6.7\times10^3$ $cm^2/Vs$ were measured at 3K from Hall measurements. The InAlAs metamorphic buffer behaves as an insulator at cryogenic temperatures [52]. Fig. 2b depicts the fabrication process of the JoFET. A 100-nm-thick Al layer was deposited onto InAsOI, which acts as the superconductor of the JoFET (Fig. 2b.1). JoFETs were fabricated via aligned lithographic steps. First, Al and InAs MESA were defined by UV-lithography and manufactured by successive Al and InAs wet etching, setting the Josephson Junction (JJ) width ($W_{JJ}$) and exposing the InAlAs layer. Then, the JJ length ($L_{JJ}$) was defined by electron-beam-lithography and Al wet-etching, leaving the underneath InAs unaffected (Fig. 2b.2). A ~30-nm-thick $HfO_2$ dielectric layer was conformably deposited via Atomic Layer Deposition (ALD) and used as the gate insulator (Fig. 2b.3) [58]. The ALD was performed at an optimized temperature of 130° C to preserve the Al/InAs superconducting and transport properties [59]. Eventually, the metal gate was fabricated by electron-beam-lithography and tilted Ti/Al deposition (Fig. 2b.4), which allowed to conformably cover acute angles left by the InAs wet etching (Fig. S1). We realized superconductor-semiconductor-superconductor Al-InAs-Al JoFETs with $W_{JJ}$ of 6 µm, $L_{JJ}$ from 500 to 1250 nm, and gate lengths ($L_G$) from 100 to 1000 nm. Fig. 2c-e show Scanning Electron Microscopy (SEM) images of JoFETs before and after gate fabrication, from which one can appreciate that the metal gate is aligned and deposited in the middle of the JJ with a precision of ±150 nm, regardless of the gate length. SEM analysis of JoFETs with different values of $L_{JJ}$ and $L_G$ is reported in Fig. S2. Fig. 2f,g report the thickness profile of the JoFET surface obtained through Atomic Force Microscopy (AFM) analysis. Finally, an optical microscope photograph of the entire device is shown in Fig. 2h, from which the 4-terminals architecture of the JoFET, meant for the electrical characterization, can be appreciated.

*InAsOI-Based Josephson Field Effect Transistor Electrical Characterization*

JoFETs were measured in a dilution fridge equipped with a z-axis superconducting magnet and a DC measurement setup (Fig. S3); the electrical characterization was performed at 50 mK. Fig. 3a shows gate-dependent V-I characteristics of a JoFET featuring $W_{JJ} = 6$ µm, $L_{JJ} = 500$ nm, and $L_G = 1000$ nm. The switching current ($I_S$) was extracted from the V-I curves as the largest current at 0-voltage drop, while the normal state resistance ($R_N$) is estimated as the V-I curve slope above $I_S$.

By decreasing the gate voltage, a reduction of $I_S$ and an increase of $R_N$ is observed. This behavior is confirmed for all the JoFETs tested, regardless of their morphological properties (Fig. S4 left column).

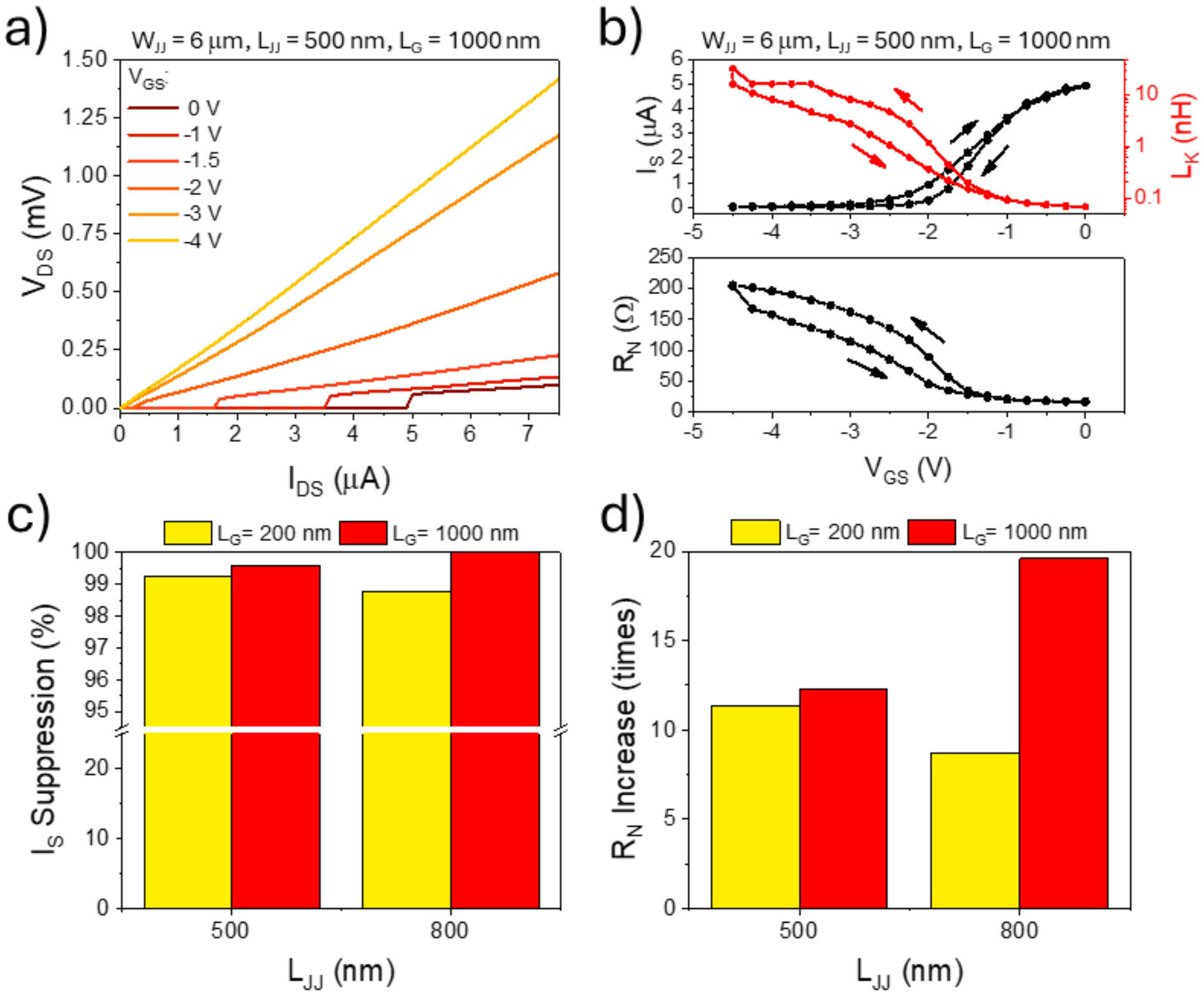

**Fig. 3: InAsOI-based Josephson Field Effect Transistors electrical characterization.** a) Gate-dependent voltage vs. current characteristics of a JoFET featuring $W_{JJ} = 6$ µm, $L_{JJ} = 500$ nm, and $L_G = 1000$ nm. b) Upward and downward gate-dependent switching current, kinetic inductance, and normal-state resistance of a JoFET featuring $W_{JJ} = 6$ µm, $L_{JJ} = 500$ nm, and $L_G = 1000$ nm. c,d) Switching current suppression (c) and normal-state resistance increase (d) of JoFET featuring different $L_{JJ}$ and $L_G$ values.

Fig. 3b shows the $I_S$ and $R_N$ gate dependence extracted from Fig. 3a. Fig. 3b top also reports the resulting kinetic inductance ($L_K = \frac{\hbar}{2e\times I_S}$ , where $\hbar$ is the reduced Plank's constant and $e$ is the elementary charge). For all the curves, a sigmoidal behavior was observed with the gate voltage, with an evident hysteresis between the upward and downward scans presumably related to electrical traps in the insulator or at the InAs/insulator interface [60,61]. In summary, for all the JoFETs fabricated, the higher the $V_{GS}$ value, the higher the switching current (the lower the normal state resistance), and vice versa (Fig. S4 middle column). The zero-voltage and gate-dependent electrical properties of JoFETs rely on the specific transistor morphology. With the increase of $L_{JJ}$, a reduction of $I_S$ and an increase of $R_N$ is obtained at $V_{GS}$ = 0 V, as expected (Fig. S5). Moreover, at $V_{GS}$ = 0 V, the larger the gate length, the lower the switching current (the higher the normal state resistance), regardless of the interelectrode separation (Fig. S5). To quantify the ability of the gate to tune the superconducting and transport properties of the JoFETs, we defined the $I_S$ suppression factor, namely $I_{Sup} = 1 - \frac{I_{S@V_{GS}=-4.5V}}{I_{S@V_{GS}=0V}}$ (Fig. 3c), and the $R_N$ increase factor, namely $R_{Inc} = \frac{R_{N@V_{GS}=-4.5V}}{R_{N@V_{GS}=0V}}$ (Fig. 3d). For both the interelectrode separations tested ($L_{JJ}$ = 500 or 800 nm), higher $I_{Sup}$ and $R_{Inc}$ factors were obtained in the case of wider gates compared to the thinner ones, namely $L_G$=1000 nm instead of $L_G$=200 nm. In addition, longer JJs and more extended gates together gave the best $I_{Sup}$ and $R_{Inc}$. In the case of the JoFET with $L_G$=1000 nm and $L_{JJ}$=850 nm, a 100 % switching current suppression together with ~20 times increase of the normal state resistance was obtained, changing the gate voltage in the range from 0 to -4.5 V. The higher $R_{Inc}$ factor obtained with wider gates is directly related to larger part of InAs tuned in electron density with the applied gate electric field. At the same time, the larger gate decreases uniformly the Fermi velocity in the semiconductor channel resulting in a larger effective barrier between the semiconductor and the superconductor. This leads to a less efficient Andreev conversion at the super/semi interface with a stronger suppression of the supercurrent compared to the narrower gate configuration. Remarkably, a negligible leakage current was collected in the gate voltage operation range (Fig. S4 right column). Additional information about the temperature and magnetic field-dependent electrical behavior of the fabricate JoFETs can be found here [54]. $I_{Sup}$ and $R_{Inc}$ achieved for InAsOI-based JoFETs are on par with the top values obtained in other gate-tunable-critical-current platforms, such as those based on InAs[59,62,63] and Ge[45,64,65] quantum wells (QWs), InAs[44] and metallic[66] nanowires (NWs), InAs nanosheets (NSs)[67], and graphene[46,68] (for a comparison see Fig. S6 and Table S1). This makes InAsOI a valuable candidate for implementing cryogenic hybrid superconducting electronics, and the InAsOI-based JoFET a promising circuital element acting as a two-state ON/OFF superconducting/normal building blocks.

We now focus on developing a superconducting 1-input-to-8-outputs (1I8O) analog demultiplexer operating at 50 mK. The roles of the I/O ports can be, in principle, reversed to operate the proposed device as a multiplexer. Fig. 4a depicts the schematic diagram of the hierarchical architecture of the demultiplexer, where three signal splitting levels are obtained involving 14 JoFETs ($JoFET_0$ to $JoFET_{13}$), one input line ($I_0$), eight output lines ($O_0$ to $O_7$), and six control lines ($G_{00}$ to $G_{21}$). Two indexes characterize control lines; the first one, ranging from 0 to 2, refers to the three splitting levels (ranging from left to right in the scheme of Fig. 4a), the second identifying the bottom (value of 0) or top (value of 1) output branch of each splitting node. Signal and control lines in the schematic representation are highlighted with red and blue colors, respectively, to improve the readability of the diagram.

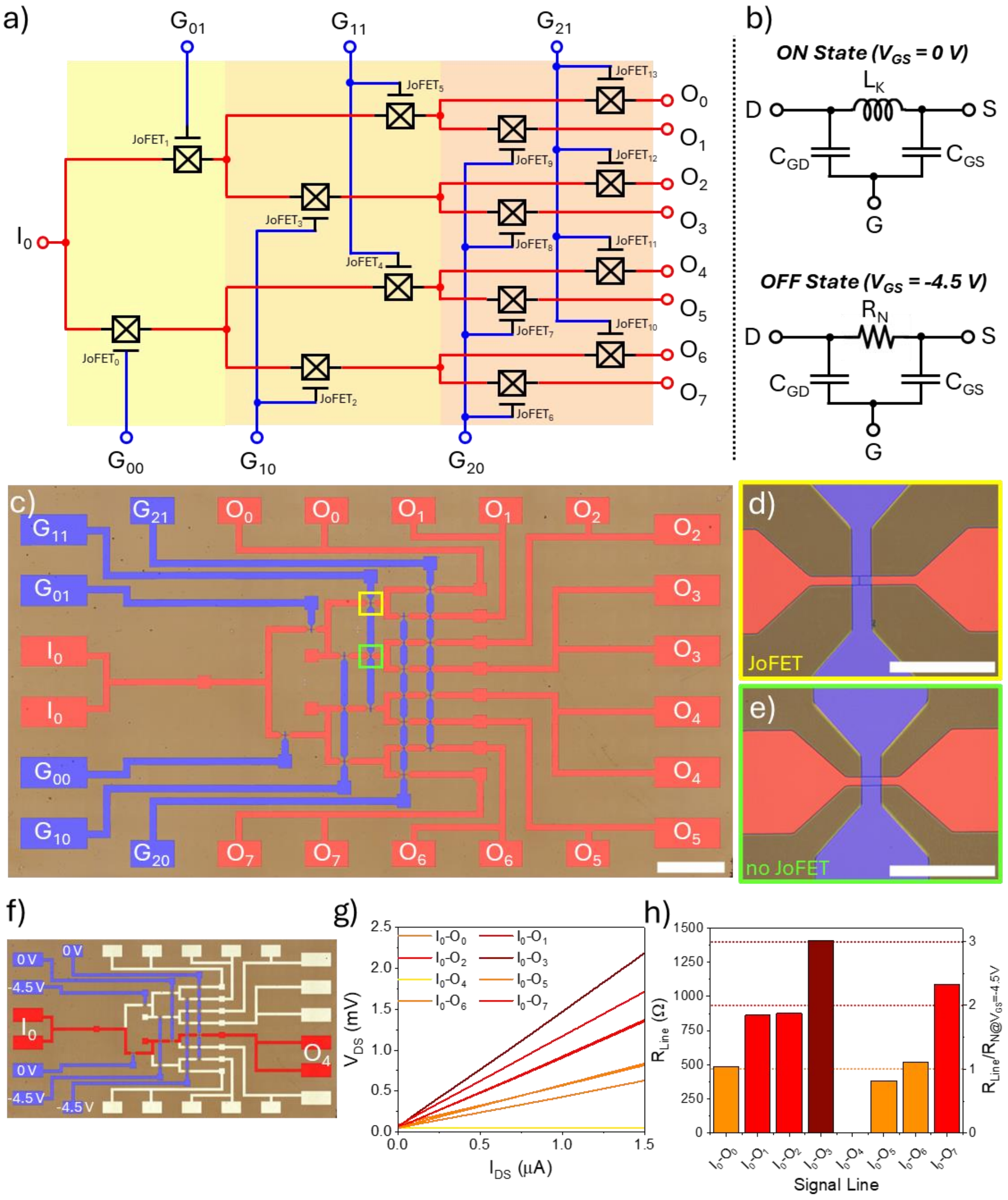

**Fig. 4: Concept, morphological, and DC electrical characterization of the superconducting 1-input-8-outputs analog demultiplexer.** a) Schematic diagram of the superconducting 1I8O demultiplexer. b) Electrical lumped model of the InAsOI-based JoFET. c) False-color optical microscope image of superconducting 1I8O demultiplexer; the scalebar is 500 μm. d,e) False-colors optical microscope image of an overlap between a signal and a gate control line with (d) or without (e) a JoFET; the scalebar is 50 μm. f) False-color optical microscope image of the superconducting 1I8O demultiplexer highlighting the chosen superconducting path and the gate control configuration. g) V-I characteristics of all the 1I8O demultiplexer signal lines with the gate configuration shown in (f). h) Line resistances (left) and normalized line resistances (right) of all the 1I8O demultiplexer signal lines with the gate configuration shown in (f).

The superconducting signal can be hierarchically divided from the input to one of the two following outputs depending on the gate control settings of each signal-splitting node. The number of required JoFETs (U) increases with the number of splitting levels (M) as $U = \sum_{i=1}^{M} 2^i = \frac{1-2^{M+1}}{-1} - 1$, while the number of output signals (O) increases with the number of control lines (N) as $O = 2^{\frac{N}{2}}$. Remarkably, the number of outputs can be improved to $2^N$ by using cold NOT logic gates to invert one of the controls in each signal-splitting level. To effectively accomplish a reduction in control lines, it is imperative that the NOT logic gates, which function at cryogenic temperatures, be situated at the same cold stage as the superconducting demultiplexer.

Fig. 4b depicts the schematic representation of the JoFET in the non-gated superconducting state, i.e., the ON state, and in the gated normal state, i.e., the OFF state. In the ON state ($V_{GS}$ = 0 V), the JoFET acts as an inductor with value $L_K$; conversely, in the OFF state ($V_{GS}$ = -4.5 V), the JoFET works as a resistor with value $R_N$. The two capacitors between the gate and drain ($C_{GD}$) and gate and source ($C_{GS}$) terminals represent the major parasitic elements for both gate settings. The contribution of the JJ capacitance is considered negligible [69].

In the DC ON state, the current (and the power) through the JoFET totally flows into the inductive branch (Fig. 4b top). This trend is suppressed with the increase of the operating frequency by the gate capacitors, providing a path to the AC-grounded gate terminal. We defined the JoFET ON-state maximum operational frequency ($f_{MAX_{ON}}$) as the frequency at which the power provided by the drain to the source decreases by 10% (-0.46 dB) compared to its DC value. Based on the values reported in Table S2, $f_{MAX_{ON}} = 6\ GHz$ was retrieved from the simulations. Similarly, in the OFF state the parasitic gate capacitors limit the maximum operation frequency of the JoFET ($f_{MAX_{OFF}} = 17\ GHz$) shunting $R_N$. The overall maximum operating frequency ($f_{MAX}$) of the JoFET can be expressed as min ($f_{MAX_{ON}}, f_{MAX_{OFF}}$), resulting in $f_{MAX} = f_{MAX_{ON}}$. A detailed discussion is provided in Section 5 of the Supporting Information. $f_{MAX}$ can be increased to higher frequencies designing JoFETs with a reduced gate capacitance by downsizing the gate capacitor area. Operating JoFETs at higher frequencies also requires considering the wavelength period in GaAs ($\lambda_{GaAs}$), namely 25 ÷ 2.5 mm for 1 ÷ 10 GHz frequencies. If $\lambda_{GaAs}$ is comparable to the dimension of the fabricated chip, a specific layout with electromagnetic wave transport through waveguides is mandatory to achieve negligible insertion loss.

Fig. 4c shows a false-color optical microscope image of the 1I8O demultiplexer fabricated on InAsOI. The chip is 5.5×3 $mm^2$ and features two routing levels of 50-µm-width Al superconducting traces: the underneath for the signal lines (red traces) and the overlying for the gate control lines (blue traces).

A $HfO_2$ layer with 30 nm thickness was used as a gate and metal trace insulator. Each signal line is provided with two pads to perform 4-terminal electrical measurements (Fig. S7). JoFETs with $W_{JJ}$ = 4 μm, $L_{JJ}$ = 500 nm, and $L_G$ = 9.5 μm were fabricated as ON/OFF building blocks. Fig. 4d shows an overlap between a signal and a gate control line where a JJ is present, i.e., the JoFET. On the other hand, Fig. 4e exhibits an overlap without active elements, which is used to reach positions on the chip with control lines that cannot be reached without two routing levels.

Table S3 reports the 1I8O demultiplexer truth table. For each gate control configuration, only one input-to-output path is fully superconductive, while all the remaining have at least one JoFET in the dissipative state. The signal line resistance is 0 Ω in the case of the non-dissipative path or equal to $1\times R_N$, $2\times R_N$, or $3\times R_N$ according to the number of JoFETs in series in the OFF state along the specific input-to-output line. As a case of study, we focused on having the $I_0$-$O_4$ line superconductive (Fig. 4f), with gate control lines set to $V_{G_{00}} = V_{G_{11}} = V_{G_{21}} = 0\ V$ and $V_{G_{01}} = V_{G_{10}} = V_{G_{20}} = -4.5\ V$. Fig. S8 depicts the schematic lumped element diagram of the demultiplexer with the applied gate settings to clarify the electrical configuration obtained. Fig. 4g shows V-I characteristics of all the 1I8O demultiplexer signal lines. As desired, only $I_0$-$O_4$ exhibits a 0 Ω resistance path up to the switching current, while all the other lines present a dissipative path. The non-dissipative dynamic input range of the superconducting demultiplexer consequently goes from $-I_S$ to $I_S$. Line resistances ($R_{Line}$) are summarized in Fig. 4h, from which we can observe that the $R_{Line}/R_{N@V_{GS}=-4.5V}$ ratio agrees with what is expected (Table S3). By changing the gate control lines setting, the non-dissipative path can be transferred to one of the other signal lines, as reported in Fig. S9 where $I_0$-$O_6$ was selected as fully superconductive. This results from the good reproducibility of the electrical properties of both JoFETs and signal lines (Fig. S10). Over the 14 JoFETs integrated, a switching current of 1.98 ± 0.52 μA (average ± standard deviation) is calculated, leading to a percentage coefficient of variation (CV) of 26.2 % (Fig. S10a). Only the $JoFET_7$ exhibited a switching current strongly far from the average; by removing this contribution, a switching current of 2.09 ± 0.31 μA (CV = 14.8 %) is obtained. Differently, the signal line resistance for input currents higher than the switching current, which is related to the reliability of the transport properties of the JoFETs, was calculated as 84.2 ± 3.4 Ω (CV = 4.0 %) (Fig. S10b). The achieved results indicate that by introducing technological steps typical of those of the large-scale integration, a superconducting demultiplexer with a higher number of reliable output lines can be fabricated.

We then evaluated the time-resolved electrical behavior of the superconducting 1I8O demultiplexer. We simultaneously measured signal lines $I_0$-$O_4$ and $I_0$-$O_6$ while switching between two gating configurations to have only one line in the non-dissipative state at a time. The gate voltages were switched at a switching frequency ($f_{SW}$) of 100 Hz from $V_{G_{10}} = -4.5\,V$ and $V_{G_{11}} = 0\,V$ to cancel the resistance of line $I_0$-$O_4$, to $V_{G_{10}} = 0\,V$ and $V_{G_{11}} = -4.5\,V$, to make the path $I_0$-$O_6$ superconductive, while leaving all the other gate control lines to $V_{G_{00}} = V_{G_{21}} = 0\,V$ and $V_{G_{01}} = V_{G_{20}} = -4.5\,V$ (see Table S3 and Fig. 5a top). The electrical characterization was performed in a dilution fridge equipped with a very-large-frequency (VLF) measurement setup (cut-off frequency ~ 45 kHz) with load resistances ($R_L$) of ~200 Ω due to the cryostat wire impedance (Fig. S11). The signal-line input current ($I_{IN}$) is ~1.3 μA. The central panel of Fig. 5a shows signal line voltages: only one line per time exhibits a zero voltage drop corresponding to the line with all the JoFETs in the ON state. Voltage drops measured for both the lines in the dissipative state are not identical due to the small differences in the gated line resistances (Fig. 4h, S10c). Correspondingly, the input current is divided into the two signal lines in agreement with the total line resistance (see Fig. 5a bottom). Specifically, the signal line current does not reach 0 A when the line is in the dissipative state due to the cryostat wiring impedance in series to the switch ($R_L$ ~ 200 Ω). This is a remarkable point to consider when selecting the impedance loads that the superconducting switch can drive. We defined the rise ($t_r$) and fall ($t_f$) time as the time the signal line voltage takes to go from 10% to 90% of its peak value, and vice versa, for the rise and fall time, respectively. We evaluated $t_r$=$t_f$~50-75 μs. Since for a square signal, the contribution of the frequency harmonics higher than the carrier frequency are relevant only in the transition between the peak-to-peak value, the maximum switching frequency ($f_{SW_{MAX}}$) achievable for the TDM can be calculated as $f_{SW_{MAX}} = \frac{1}{t_r} = 15 \div 20\,kHz$. The switching speed is indeed limited by the cut-off frequency of the setup used to characterize the fabricated die (Fig. S11).

To complete the electrical characterization of the superconducting 1I8O demultiplexer, we also evaluated the AC electrical performance. We injected into the signal line $I_0$-$O_4$ an AC input current with 1 kHz frequency, collecting the voltage drop across the line. Fig. 5b shows the time-resolved signal line voltages measured by changing the peak-to-peak value of the input current from 2 μA to 16 μA. When the input current instantaneously overcomes the minimum switching current of the JoFETs in the $I_0$-$O_4$ path (i.e., $I_S$ = 1.9 μA of $JoFET_4$), a finite voltage drop is detected. The higher the peak-to-peak value of the input current, the higher the signal line voltage drop obtained.

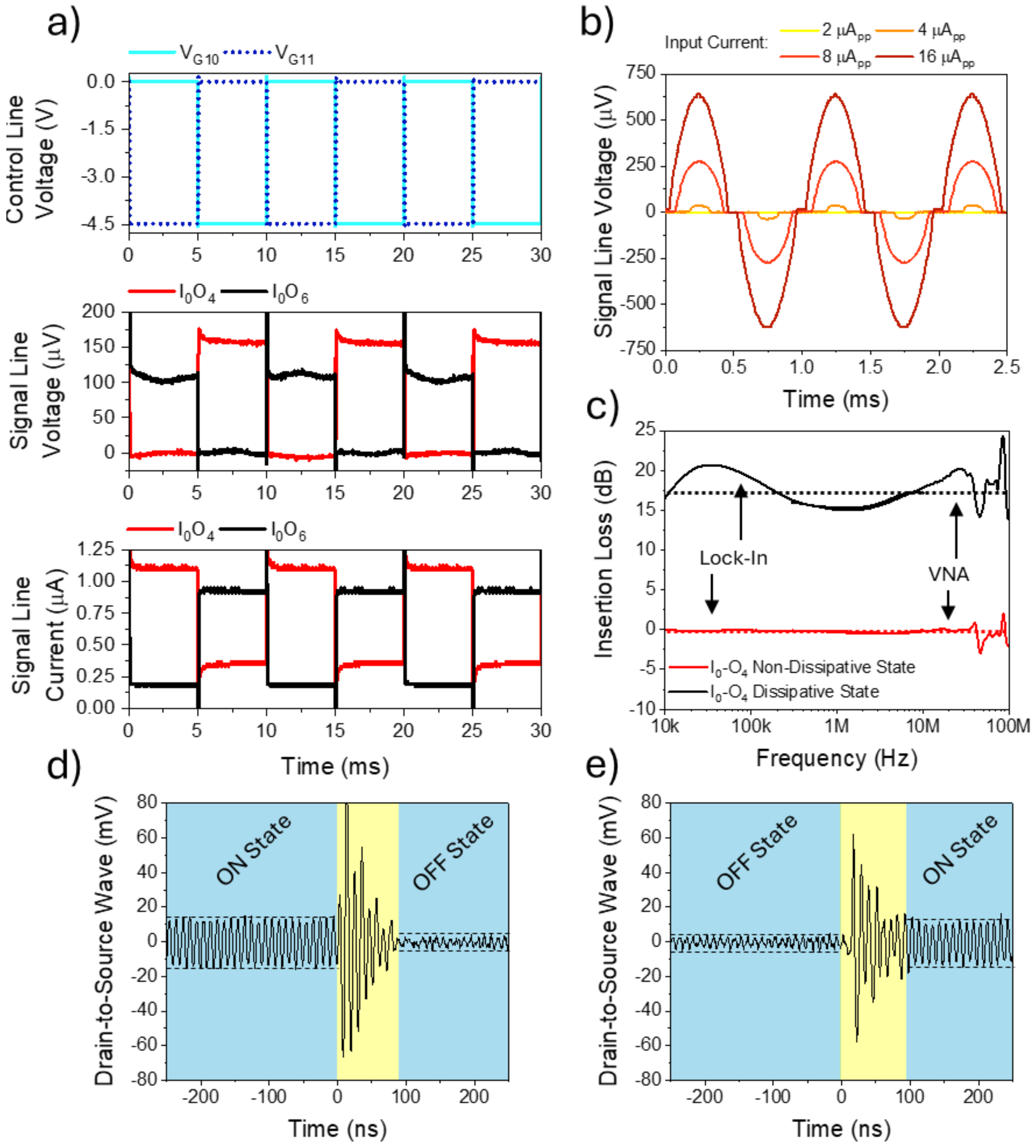

**Fig. 5: Time-resolved and AC electrical characterization of the superconducting 1-input-8-outputs analog demultiplexer.** a) Time-resolved electrical behavior of the superconducting demultiplexer. Time-dependent control line voltages (top), signal line voltages (middle), and signal line currents (bottom). b) Time-resolved $I_0$-$O_4$ signal line voltage with different peak-to-peak input currents. c) Frequency-resolved insertion loss of signal line $I_0$-$O_4$ in the non-dissipative and dissipative states. Measurements are taken with a lock-in amplifier and a VNA for low and high-frequency ranges, respectively. d,e) Drain-to-source wave voltage (f=100 MHz) measured during the ON-to-OFF (d) and OFF-to-ON (e) state transients. The blue areas indicate the steady state behavior while the yellow area indicates the transient.

By applying an AC signal lower than the minimal critical current supported, we then evaluated the electrical behavior of the superconducting 1I8O demultiplexer in the frequency range from 10 kHz to 100 MHz (LF-VHF radio spectrum) in a dilution fridge equipped with a 50Ω-matched cryogenic measurement setup (maximum operation frequency ~ 300 MHz, Fig. S12). A lock-in amplifier and a vector network analyzer (VNA) were used to measure the magnitude of the forward power gain scattering parameter ($|S_{21}|$) in the 10 kHz - 10 MHz and 300 kHz - 100 MHz ranges, respectively. We simultaneously measured the output powers of signal lines $I_0$-$O_4$ and $I_0$-$O_6$, injecting an AC input signal with a -80 dBm power level (10 pW, 447 $nA_{rms}$). Gate settings were chosen to make superconductive only one line at a time. Fig. 5c shows the insertion loss ($IL\ (dB) = -S_{21}(dB)$) of the signal line $I_0$-$O_4$ in the dissipative and non-dissipative states. When the signal line is superconducting, an insertion loss of ~0 dB was measured in the whole frequency range, demonstrating that the input power is efficiently transferred to the output port of the selected signal line without significant dissipation along the path. This is a remarkable improvement compared to the case in which the demultiplexer in ON-state operates above the critical temperature of Al (specifically at 1.65 K) and, therefore, in a fully dissipative configuration. In this case, the insertion loss increases to ~5.5 dB (Fig. S13). The latter result highlights the importance of a fully superconducting path between the input and the output port. Compared with conventional gate-tunable cryogenic non-superconducting solutions, it promises a significant competitive advantage of a switchable superconducting system like the one we proposed. On the other hand, when the line is gated in the dissipative state, the insertion loss increases to ~17.5 dB, which is relatively constant over all the frequency range tested and allow to achieve an $\frac{ON}{OFF}$ ratio of ~ 17.5 dB. Similar results were also obtained for the signal line $I_0$-$O_6$ (Fig. S14).

The turn-OFF and turn-ON time-resolved behaviors of a typical JoFET ($L_{JJ}$ = 500 nm, $W_{JJ}$ = 4 µm, and $L_G$ = 1000 nm) were reported in Fig. 5d,e. The electrical characterization was performed in dilution fridge equipped with a very-high-frequency (VHF) measurement setup (cut-off frequency ~ 100 MHz for the gate control lines, cut-off frequency ~ 500 MHz for the drain-to-source signal lines). An AC sinusoidal wave (-80 dBm power and 100 MHz frequency) was used as the drain-to-source signal; a square wave ($t_r$=$t_f$~130 ns, Fig. S15) was used as the gate control switching signal. Switching from the ON to OFF state (Fig. 5d), the steady-state peak-to-peak voltage value of the drain-to-source wave reduces by ~7.6 times (17.6 dB in power), according to what reported in Fig. 5c. The increased wave amplitude observed during the transient was related to the displacement current generated during the gate capacitor discharge upon application of the square wave control signal. Specifically, this current sums to the drain-to-source signal current increasing the value of the drain-to-source wave voltage measured. A similar trend was observed also for the OFF-to-ON state commutation (Fig. 5e). In this case, $t_r$=$t_f$~100 ns were achieved and traduced in an improved $f_{SW_{MAX}} = 10\ MHz$, as demonstrated by the uncorrelation between $t_r$ and $t_f$ and the gate switching frequency reported in Fig. S16. As previously indicated, $f_{SW_{MAX}}$ is limited by the cut-off frequency of the setup used to drive the gate control lines of the superconducting demultiplexer (Fig. S15). Fig. S17 depicts the AC drain-to-source wave increasing the JoFET gate control switching frequency from 1 kHz to 100 kHz, from which the amplitude of the drain-to-source wave was reliably and reproducibly modulated as a function of the gate voltage value applied in time.

We then theoretically estimated the static ($P_{STAT}$) and dynamic ($P_{DYN}$) power dissipated by the 1I8O superconducting demultiplexer. In the DC-to-$f_{MAX}$ range, $P_{STAT}$ is related to the part of the input current flowing in the dissipative OFF branches (see Section 6 of Supporting Information). $P_{STAT}$ ~ 700 fW was evaluated for the superconducting demultiplexer operating in the measurement setup of Fig. 5c, S12. The dynamic power dissipated by the switching activity was calculated by the well-known formula $P_{DYN} = n \times C_G \times V_{GS}^2 \times f_{SW}$, where $n = 14$ is the number of integrated JoFETs, $C_G = C_{GS} + C_{GD}$ is the gate capacitance, and $f_{SW}$ is the gate control switching frequency (see Section 7 of Supporting Information). Referring to the cooling power of the cryostat we used throughout the experiments at the temperature of 50 mK, namely 16 μW, the 1I8O superconducting demultiplexer can operate with control switching frequencies up to 400 kHz without affecting the temperature of the coldest plate. The calculations were performed in the worst-case scenario of the full $P_{DYN}$ dissipated at mixing chamber stage. The switching frequency can be increased by reducing the size of the JoFETs and the gate voltage required to reach the OFF state.

To envisage TDM of supercurrents in the frequency operational range of transmon qubits, we focused on the development of a μwave superconducting 1-input-to-2-outputs (1I2O) analog demultiplexer operating at 50 mK. The chip was manufactured with the same InAsOI-based technology previously described and features 2 JoFETs ($JoFET_0$ and $JoFET_1$), 1 input line ($I_0$), 2 output lines ($O_0$ and $O_1$), and 2 control lines ($G_0$ and $G_1$). On-chip routing of electromagnetic signals in the μwave regime was performed by using integrated Al superconducting coplanar waveguides (CPWs), to minimize power dissipation along the I/O line in the ON-state due to the comparable values of $\lambda_{GaAs}$ with the dimension of the chip at higher operating frequencies. Fig. 6a shows a false-color optical microscope image of the fabricated die, with signal and control lines highlighted in red and blue.

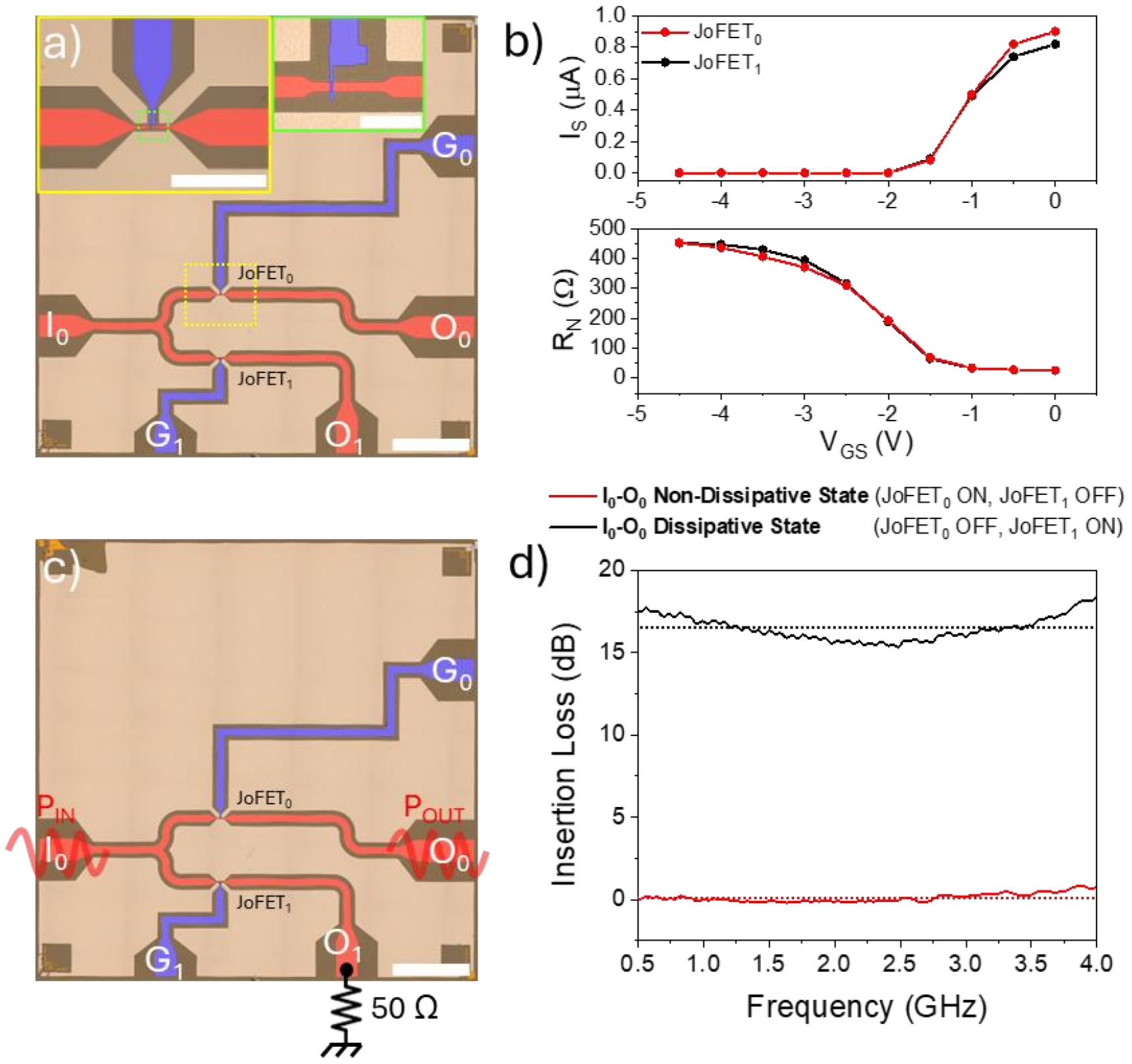

**Fig. 6: Morphological, DC, and RF electrical characterization of the μwave superconducting 1-input-2-outputs analog demultiplexer.** a) False-color optical microscope image of the μwave superconducting 1I2O demultiplexer; the scalebar is 1 mm. Yellow and green insets report false-color optical microscope image of an integrated JoFET; scalebars are 250 μm and 25 μm, respectively. b) Downward gate-dependent switching current and normal-state resistance of the integrated JoFETs featuring $W_{JJ}$ = 6 μm, $L_{JJ}$ = 800 nm, and $L_G$= 1000 nm. c) False-color optical microscope image of the μwave superconducting 1I8O demultiplexer highlighting the electrical configuration used to test chip. d) Frequency-resolved insertion loss of signal line $I_0$-$O_0$ in the non-dissipative and dissipative states.

The chip is 6×6 mm$^2$ and embeds two JoFETs with $W_{JJ}$ = 6 µm, $L_{JJ}$ = 800 nm, and $L_G$= 1000 nm routed with Al CPWs. The gate extension $L_G$ was reduced compared to that of the JoFETs integrated into the 1I8O superconducting demultiplexer to address the gate capacitance limitation on the OFF-state electrical properties at high frequencies, consequently increasing $f_{MAX_{OFF}}$. Both the JoFETs operate as expected, as reported from the $I_S$ and $R_N$ vs. $V_{GS}$ trends depicted in Fig. 6b, from which the full suppression of the switching current as well as ~20 times increase of the normal state resistance was achieved in the gate range 0 to -4.5 V. The behavior of the signal line $I_0$-$O_0$ was collected and considered as representative of all the lines of the chip, with gate settings chosen to make superconductive only one signal line at a time (Fig. 6c). An AC signal with amplitude lower than the minimal critical current supported was used to evaluate the electrical behavior of the µwave 1I2O superconducting demultiplexer in the frequency range from 500 MHz to 4 GHz (VHF-UHF radio spectrum) by involving a dilution fridge equipped with a 50Ω-matched cryogenic measurement setup (maximum operation frequency ~ 8.5 GHz, Fig. S18). An electromagnetic wave with power $P_{IN}$ = -80 dBm (10 pW, 447 $nA_{rms}$) was provided to the input port $I_0$; the power of the electromagnetic wave reaching the output port $O_0$ was collected during while the outport port $O_1$ was connected to a reference load of 50 Ω. Fig. 6d shows IL of line $I_0O_0$ in the non-dissipative and dissipative states. In the non-dissipative state, an IL of 0 dB was observed throughout the frequency range, confirming negligible power consumption along the fully superconducting line in the VHF-UHF radio spectrum range. On the other hand, in the dissipative state, the IL increases to ~16 dB, from which an $\frac{ON}{OFF}$ ratio of 16 dB is calculated. The achieved performance are comparable to what was observed for the 1I8O superconducting demultiplexer chip, which allow us to conclude that the proposed architecture can be employed also for operating TDM of non-dissipative supercurrents in the frequency operational range of transmon qubits. Further considerations on the use of superconducting TDM of non-dissipative signals to transmon qubits for a dramatic reduction of the I/O lines of a cryogenic quantum computer are reported in Section 8 of the Supporting Information.

Our work introduces a significant technological advancement in the form of TDM of non-dissipative signals with a voltage-actuated hybrid superconducting demultiplexer. This innovative device is designed to substantially reduce the number of input signal lines required in a conventional cryostat. The key component of this demultiplexer is an InAsOI-based JoFET, which features Al as a superconductor and $HfO_2$ as a gate insulator. Notably, this JoFET has the capability to completely suppress the switching current and increase the normal-state resistance by 20 times when subjected to a gate voltage of -4.5 V. In this context, the ideal building block should be capable of (i) exhibiting a high switching current in the ON-state, (ii) entirely suppressing the switching current in the OFF-state, and (iii) exhibiting a high normal-state resistance in the OFF-state. We manufactured a superconducting demultiplexer with 1 input and 8 outputs exhibiting a significant non-dissipative dynamic input range of -2÷2 μA and able to operate up to 100 MHz in signal frequency and 100 kHz in switching frequency at 50 mK. The results were remarkable, revealing an insertion loss of approximately 0 dB in the superconducting state and an $\frac{ON}{OFF}$ ratio of approximately 17.5 dB up to 100 MHz in a 50Ω-matched cryogenic measurement setup. These findings vividly illustrate the advantage of superconductivity by establishing a lossless path between the input and output ports. Additionally, the integration of on-chip routing for non-dissipative signals utilizing superconducting coplanar waveguides was employed to expand the operational frequency range of the superconducting demultiplexer. A superconducting demultiplexer with 1 input and 2 outputs was fabricated and validated up to 4 GHz in signal frequency at 50 mK. Also for the latter device, an insertion loss of ~0 dB in the superconducting state and an $\frac{ON}{OFF}$ ratio of ~16 dB were evaluated. These groundbreaking advancements demonstrate the potential for the practical implementation of superconducting TDM, which could substantially reduce in I/O lines, costs, and space occupation within a cryostat.

## Methods

All the methods regarding the sample manufacturing and the cryogenic electric characterization are detailed reported within the Supporting Information file.

### *InAsOI Heterostructure Growth via Molecular Beam Epitaxy*

InAsOI was grown on semi-insulating GaAs (100) substrates using solid-source Molecular Beam Epitaxy. Starting from the GaAs substrate, the sequence of the layer structure includes a 200 nm-thick GaAs layer, a 200 nm-thick GaAs/$Al_{0.16}Ga_{0.84}As$ superlattice, a 200 nm-thick GaAs layer, a 1.250 μm-thick step-graded $In_XAl_{1-X}As$ metamorphic buffer (with X increasing from 0.15 to 0.81), a 400 nm-thick $In_{0.84}Al_{0.16}As$ overshoot layer, and a 100 nm-thick InAs layer. The GaAs layer and GaAs/AlGaAs superlattice are grown at 600 °C ± 5 °C. The metamorphic buffer and the overshoot layers were grown at optimized substrate temperatures of 320 °C ± 5 °C. The InAs epilayer was grown at 480 ± 5 °C.

### *InAsOI Josephson Field Effect Transistor Fabrication*

InAsOI substrates were cut into square samples and cleaned with organic solvents. The air-exposed InAs surface was etched from native InAs oxide ($InAsO_X$) and passivated with S-termination by dipping the InAsOI samples in a $(NH_4)_2S_X$ solution (290 mM $(NH_4)_2S$ and 300 mM S in DIW) at 45°C. The S-terminated InAsOI samples were then rinsed in DIW and loaded into an e-beam evaporator where a 100-nm-thick Al layer was deposited. After Al deposition, the mesa geometry was defined via photolithography and subsequent Al etching in Al Etchant Type D at 40 °C and InAs etching with a $H_3PO_4$:$H_2O_2$ solution (348 mM $H_3PO_4$, 305 mM $H_2O_2$ in DIW). After MESA fabrication, nanometric aligning Ti/Au markers were defined via e-beam lithography, thermal deposition of a 10/50-nm-thick Ti/Au bilayer, and lift-off. Markers were then used to define the Josephson Junction via e-beam lithography and subsequent Al etching in Al Etchant Type D at 40 °C. The 31-nm-thick $HfO_2$ gate insulator layer was deposited via Atomic Layer Deposition at a temperature of 130 °C. Eventually, markers were again used to define the metallic gate via e-beam lithography, e-beam deposition of a 10/100-nm-thick Ti/Al bilayer, and lift-off.

***InAsOI Superconducting 1I8O Demultiplexer Fabrication***

The superconducting 1I8O demultiplexer chip was fabricated according to the step previously reported for the InAsOI Josephson Field Effect Transistor Fabrication. Photolithography was used as the patterning technique to define the metallic gate lines aligned with micrometric fabricated at the mesa stage.

***InAsOI µwave Superconducting 1I2O Demultiplexer Fabrication***

The µwave superconducting 1I2O demultiplexer chip was fabricated according to the step previously reported for the InAsOI Josephson Field Effect Transistor Fabrication.

**Data Availability**

All the data are available upon request from the corresponding author.

**Code availability**

No code is used in this study.

## References

1. Casparis, L. *et al.* Superconducting gatemon qubit based on a proximitized two-dimensional electron gas. *Nat. Nanotechnol.* **13**, 915–919 (2018).
2. Landig, A. J. *et al.* Virtual-photon-mediated spin-qubit–transmon coupling. *Nat. Commun.* **10**, 5037 (2019).
3. Wang, J. I. J. *et al.* Hexagonal boron nitride as a low-loss dielectric for superconducting quantum circuits and qubits. *Nat. Mater.* **21**, 398–403 (2022).
4. Pita-Vidal, M. *et al.* Direct manipulation of a superconducting spin qubit strongly coupled to a transmon qubit. *Nat. Phys.* **19**, 1110–1115 (2023).
5. Pita-Vidal, M. *et al.* Strong tunable coupling between two distant superconducting spin qubits. *Nat. Phys.* **20**, 1158–1163 (2024).
6. Ho Eom, B., Day, P. K., LeDuc, H. G. & Zmuidzinas, J. A wideband, low-noise superconducting amplifier with high dynamic range. *Nat. Phys.* **8**, 623–627 (2012).
7. Sarkar, J. *et al.* Quantum-noise-limited microwave amplification using a graphene Josephson junction. *Nat. Nanotechnol.* **17**, 1147–1152 (2022).
8. Phan, D. *et al.* Gate-Tunable Superconductor-Semiconductor Parametric Amplifier. *Phys. Rev. Appl.* **19**, 064032 (2023).
9. Splitthoff, L. J. *et al.* Gate-tunable kinetic inductance parametric amplifier. *Phys. Rev. Appl.* **21**, 014052 (2024).
10. Ando, F. *et al.* Observation of superconducting diode effect. *Nature* **584**, 373–376 (2020).
11. Baumgartner, C. *et al.* Supercurrent rectification and magnetochiral effects in symmetric Josephson junctions. *Nat. Nanotechnol.* **17**, 39–44 (2022).
12. Margineda, D. *et al.* Back-action supercurrent rectifiers. *Commun. Phys.* **8**, 16 (2025).
13. Marsili, F. *et al.* Efficient Single Photon Detection from 500 nm to 5 μm Wavelength. *Nano Lett.* **12**, 4799–4804 (2012).
14. Korzh, B. *et al.* Demonstration of sub-3 ps temporal resolution with a superconducting nanowire single-photon detector. *Nat. Photonics* **14**, 250–255 (2020).
15. Charaev, I. *et al.* Single-photon detection using high-temperature superconductors. *Nat. Nanotechnol.* **18**, 343–349 (2023).
16. Hart, S. *et al.* Controlled finite momentum pairing and spatially varying order parameter in proximitized HgTe quantum wells. *Nat. Phys.* **13**, 87–93 (2017).
17. Wiedenmann, J. *et al.* 4φ-periodic Josephson supercurrent in HgTe-based topological Josephson junctions. *Nat. Commun.* **7**, 1–7 (2016).
18. Laroche, D. *et al.* Observation of the 4π-periodic Josephson effect in indium arsenide nanowires. *Nat. Commun.* **10**, 245 (2019).
19. Dartiailh, M. C. *et al.* Phase Signature of Topological Transition in Josephson Junctions. *Phys. Rev. Lett.* **126**, 1–6 (2021).
20. Amet, F. *et al.* Supercurrent in the quantum Hall regime. *Science (80-. ).* **352**, 966–969 (2016).
21. Seredinski, A. *et al.* Quantum Hall–based superconducting interference device. *Sci. Adv.* **5**, 2–7 (2019).
22. Vignaud, H. *et al.* Evidence for chiral supercurrent in quantum Hall Josephson junctions. *Nature* **624**, 545–550 (2023).
23. Bolotin, K. I., Ghahari, F., Shulman, M. D., Stormer, H. L. & Kim, P. Observation of the fractional quantum Hall effect in graphene. *Nature* **462**, 196–199 (2009).
24. Kerelsky, A. *et al.* Maximized electron interactions at the magic angle in twisted bilayer graphene. *Nature* **572**, 95–100 (2019).
25. Fu, H. *et al.* 3/2 fractional quantum Hall plateau in confined two-dimensional electron gas. *Nat. Commun.* **10**, 4351 (2019).
26. Arute, F. *et al.* Quantum supremacy using a programmable superconducting processor.

*Nature* **574**, 505–510 (2019).
27. Arute, F. *et al.* Hartree-Fock on a superconducting qubit quantum computer. *Science (80-. ).* **369**, 1084–1089 (2020).
28. Yeh, J.-H. & Anlage, S. M. In situ broadband cryogenic calibration for two-port superconducting microwave resonators. *Rev. Sci. Instrum.* **84**, (2013).
29. Oates, D. E., Slattery, R. L. & Hover, D. J. Cryogenic test fixture for two-port calibration at 4.2 K and above. in *2017 89th ARFTG Microwave Measurement Conference (ARFTG)* 1–4 (IEEE, 2017). doi:10.1109/ARFTG.2017.8000842
30. Shin, S.-H., Stanley, M., Skinner, J., de Graaf, S. E. & Ridler, N. M. Broadband Coaxial S-Parameter Measurements for Cryogenic Quantum Technologies. *IEEE Trans. Microw. Theory Tech.* **72**, 2193–2201 (2024).
31. Paquelet Wuetz, B. *et al.* Multiplexed quantum transport using commercial off-the-shelf CMOS at sub-kelvin temperatures. *npj Quantum Inf.* **6**, 43 (2020).
32. Ruffino, A. *et al.* A cryo-CMOS chip that integrates silicon quantum dots and multiplexed dispersive readout electronics. *Nat. Electron.* **5**, 53–59 (2021).
33. Potočnik, A. *et al.* Millikelvin temperature cryo-CMOS multiplexer for scalable quantum device characterisation. *Quantum Sci. Technol.* **7**, 015004 (2022).
34. Acharya, R. *et al.* Multiplexed superconducting qubit control at millikelvin temperatures with a low-power cryo-CMOS multiplexer. *Nat. Electron.* **6**, 900–909 (2023).
35. Ward, D. R., Savage, D. E., Lagally, M. G., Coppersmith, S. N. & Eriksson, M. A. Integration of on-chip field-effect transistor switches with dopantless Si/SiGe quantum dots for high-throughput testing. *Appl. Phys. Lett.* **102**, (2013).
36. Al-Taie, H. *et al.* Cryogenic on-chip multiplexer for the study of quantum transport in 256 split-gate devices. *Appl. Phys. Lett.* **102**, (2013).
37. Hornibrook, J. M. *et al.* Cryogenic Control Architecture for Large-Scale Quantum Computing. *Phys. Rev. Appl.* **3**, 024010 (2015).
38. Olšteins, D. *et al.* Cryogenic multiplexing using selective area grown nanowires. *Nat. Commun.* **14**, 7738 (2023).
39. Wagner, A., Ranzani, L., Ribeill, G. & Ohki, T. A. Demonstration of a superconducting nanowire microwave switch. *Appl. Phys. Lett.* **115**, (2019).
40. Naaman, O., Abutaleb, M. O., Kirby, C. & Rennie, M. On-chip Josephson junction microwave switch. *Appl. Phys. Lett.* **108**, (2016).
41. Pechal, M. *et al.* Superconducting Switch for Fast On-Chip Routing of Quantum Microwave Fields. *Phys. Rev. Appl.* **6**, 024009 (2016).
42. Akazaki, T., Takayanagi, H., Nitta, J. & Enoki, T. A Josephson field effect transistor using an InAs-inserted-channel In0.52Al0.48As/In0.53Ga0.47As inverted modulation-doped structure. *Appl. Phys. Lett.* **68**, 418–420 (1996).
43. Shabani, J. *et al.* Two-dimensional epitaxial superconductor-semiconductor heterostructures: A platform for topological superconducting networks. *Phys. Rev. B* **93**, 155402 (2016).
44. Doh, Y.-J. *et al.* Tunable Supercurrent Through Semiconductor Nanowires. *Science (80-. ).* **309**, 272–275 (2005).
45. Vigneau, F. *et al.* Germanium Quantum-Well Josephson Field-Effect Transistors and Interferometers. *Nano Lett.* **19**, 1023–1027 (2019).
46. Generalov, A. A. *et al.* Wafer-scale CMOS-compatible graphene Josephson field-effect transistors. *Appl. Phys. Lett.* **125**, (2024).
47. Clark, T. D., Prance, R. J. & Grassie, A. D. C. Feasibility of hybrid Josephson field effect transistors. *J. Appl. Phys.* **51**, 2736–2743 (1980).
48. Amado, M., Fornieri, A., Biasiol, G., Sorba, L. & Giazotto, F. A ballistic two-dimensional-electron-gas Andreev interferometer. *Appl. Phys. Lett.* **104**, 1–5 (2014).
49. Fornieri, A. *et al.* A ballistic quantum ring Josephson interferometer. *Nanotechnology* **24**, 245201 (2013).

50. Amado, M. *et al.* Electrostatic tailoring of magnetic interference in quantum point contact ballistic Josephson junctions. *Phys. Rev. B* **87**, 134506 (2013).
51. Roddaro, S. *et al.* Hot-electron effects in InAs nanowire Josephson junctions. *Nano Res.* **4**, 259–265 (2011).
52. Paghi, A. *et al.* InAs on Insulator: A New Platform for Cryogenic Hybrid Superconducting Electronics. *Adv. Funct. Mater.* **35**, (2025).
53. Battisti, S. *et al.* Extremely weak sub-kelvin electron–phonon coupling in InAs on Insulator. *Appl. Phys. Lett.* **125**, (2024).
54. Paghi, A. *et al.* Josephson Field Effect Transistors with InAs on Insulator and High Permittivity Gate Dielectrics. *ACS Appl. Electron. Mater.* (2025). doi:10.1021/acsaelm.5c00038
55. Affentauschegg, C. & Wieder, H. H. Properties of InAs/InAlAs heterostructures. *Semicond. Sci. Technol.* **16**, 708–714 (2001).
56. Strickland, W. M., Hatefipour, M., Langone, D., Farzaneh, S. M. & Shabani, J. Controlling Fermi level pinning in near-surface InAs quantum wells. *Appl. Phys. Lett.* **121**, 1–6 (2022).
57. Capotondi, F. *et al.* Two-dimensional electron gas formation in undoped In0.75Ga0.25As/In0.75Al0.25As quantum wells. *J. Vac. Sci. Technol. B Microelectron. Nanom. Struct. Process. Meas. Phenom.* **22**, 702–706 (2004).
58. Paghi, A., Battisti, S., Tortorella, S., De Simoni, G. & Giazotto, F. Cryogenic behavior of high-permittivity gate dielectrics: The impact of atomic layer deposition temperature and the lithographic patterning method. *J. Appl. Phys.* **137**, (2025).
59. Sütő, M. *et al.* Near-surface InAs two-dimensional electron gas on a GaAs substrate: Characterization and superconducting proximity effect. *Phys. Rev. B* **106**, 235404 (2022).
60. Kukli, K. *et al.* Effect of selected atomic layer deposition parameters on the structure and dielectric properties of hafnium oxide films. *J. Appl. Phys.* **96**, 5298–5307 (2004).
61. Baik, M. *et al.* Effects of thermal and electrical stress on defect generation in InAs metal–oxide–semiconductor capacitor. *Appl. Surf. Sci.* **467–468**, 1161–1169 (2019).
62. Akazaki, T., Takayanagi, H., Nitta, J. & Enoki, T. A Josephson field effect transistor using an InAs-inserted-channel In0.52Al0.48As/In0.53Ga0.47As inverted modulation-doped structure. *Appl. Phys. Lett.* **68**, 418–420 (1996).
63. Mayer, W. *et al.* Superconducting proximity effect in epitaxial Al-InAs heterostructures. *Appl. Phys. Lett.* **114**, (2019).
64. Aggarwal, K. *et al.* Enhancement of proximity-induced superconductivity in a planar Ge hole gas. *Phys. Rev. Res.* **3**, L022005 (2021).
65. Tosato, A. *et al.* Hard superconducting gap in germanium. *Commun. Mater.* **4**, 23 (2023).
66. De Simoni, G., Paolucci, F., Solinas, P., Strambini, E. & Giazotto, F. Metallic supercurrent field-effect transistor. *Nat. Nanotechnol.* **13**, 802–805 (2018).
67. Yan, S. *et al.* Supercurrent, Multiple Andreev Reflections and Shapiro Steps in InAs Nanosheet Josephson Junctions. *Nano Lett.* **23**, 6497–6503 (2023).
68. Zhu, M. *et al.* Supercurrent and multiple Andreev reflections in micrometer-long ballistic graphene Josephson junctions. *Nanoscale* **10**, 3020–3025 (2018).
69. Courtois, H., Meschke, M., Peltonen, J. T. & Pekola, J. P. Origin of Hysteresis in a Proximity Josephson Junction. *Phys. Rev. Lett.* **101**, 067002 (2008).

**Acknowledgments**

We thank Stefan Heun for helping to carry out room-temperature and cryogenic measurements to estimate InAs mobility and two-dimensional charge density. We thank Andrea Ria for the valuable discussion about the JoFET's lumped element model. This work was partially supported by H2020-EU.1.2. - EXCELLENT SCIENCE - Future and Emerging Technologies (FET) under grant 964398 (SuperGate), by HORIZON.3.1 - The European Innovation Council (EIC) – Transition Open Programme under grant 101057977 (SPECTRUM), and by the Piano Nazionale di Ripresa e Resilienza, Ministero dell'Università e della Ricerca (PNRR MUR) Project under Grant PE0000023-NQSTI.

**Author Contributions**

A.P. and F.G. conceptualized the experiments. A.P. and L.S. fabricated the samples. A.P. and S.T. performed the experiments. A.P., L.B., and S.T. analyzed the data. A.P. performed the electrical simulations. A.P prepared the figures and wrote the original draft. A.P., L.B., S.T., G.D.S., E.S., L.S., and F.G. discussed the results and contributed to the original draft. A.P. led the review process. L.B., E.S., and F.G. contributed to the review process. F.G. supervised the overall activity.

**Ethics Declaration**

*Competing interests*

The authors declare no competing interests.

**Supporting Information**

# Non-Dissipative Supercurrent Time Division Multiplexing with Solid-State Integrated Hybrid Superconducting Electronics

Alessandro Paghi[1*], Simone Tortorella[1], Laura Borgongino[1], Giorgio De Simoni[1], Elia Strambini[1], Lucia Sorba[1], and Francesco Giazotto[1*]

[1]Istituto Nanoscienze-CNR and Scuola Normale Superiore, Piazza San Silvestro 12, 56127 Pisa, Italy.

[2]Dipartimento di Ingegneria Civile e Industriale, Università di Pisa, Largo Lucio Lazzarino, 56122 Pisa, Italy

[*]Corresponding authors: alessandro.paghi@nano.cnr.it, francesco.giazotto@sns.it

**Summary**

## 1. Supporting Tables

**Table S1. State of the art of JoFETs**. For each platform, the most representative JoFETs are reported. In the case of NWs, “W” stays for the NW diameter. “*” means that electrical or morphological properties are extrapolated from the figures of the papers and not directly provided by the authors.

| **InAsOI** | | | | | | | | | | | | | | | |
|---|---|---|---|---|---|---|---|---|---|---|---|---|---|---|---|
| **Specs** | **Superconductor** | **$L_{JJ}$ [nm]** | **$W_{JJ}$ [μm]** | **$I_C$ ($V_{GS}$=0V) [μA]** | **$I_C$/ $W_{JJ}$ ($V_{GS}$=0V) [μA/μm]** | **$I_C$ Suppr. [%]** | **$R_N$ ($V_{GS}$=0V) [Ω]** | **$R_N$ Incr. [times]** | **$\Delta V_{GS}$ [V]** | **Gate Type** | **Gate Insulator** | **Gate Distance [nm]** | **T [mK]** | **Year** | **Ref.** |
| $n_{2D}$=1.35×10$^{12}$ cm$^{-2}$, $\mu_n$=6700 cm$^2$/Vs | Al | 800 | 5 | 1.14 | 0.274 | 100 | 16.8 | 19.6 | -4.5 | Top | $HfO_2$ | 30 | 50 | 2024 | This Work |
| $n_{2D}$=1.35×10$^{12}$ cm$^{-2}$, $\mu_n$=6700 cm$^2$/Vs | Al | 500 | 5 | 4.92 | 0.826 | 99.6 | 28.7 | 12.3 | -4.5 | Top | $HfO_2$ | 30 | 50 | 2024 | This Work |
| **InAs 3D Substrates** | | | | | | | | | | | | | | | |
| **Specs** | **Superconductor** | **$L_{JJ}$ [nm]** | **$W_{JJ}$ [μm]** | **$I_C$ ($V_{GS}$=0V) [μA]** | **$I_C$/ $W_{JJ}$ ($V_{GS}$=0V) [μA/μm]** | **$I_C$ Suppr. [%]** | **$R_N$ ($V_{GS}$=0V) [Ω]** | **$R_N$ Incr. [times]** | **$\Delta V_{GS}$ [V]** | **Gate Type** | **Gate Insulator** | **Gate Distance [nm]** | **T [mK]** | **Year** | **Ref.** |
| *p*-InAs | Nb | 400 | - | 25 | - | 99.9 | 158* | 316* | -20 | Top | SiO/$Nb_XO_Y$ | 100/70 | 20 | 1995 | [1] |
| *p*-InAs | Nb | 200 | 50 | 75* | 1.5* | 100* | 2* | 15* | -20 | Top | $SiO_2$ | - | 1800 | 1999 | [2] |
| **InAs 2D QWs** | | | | | | | | | | | | | | | |
| **Specs** | **Superconductor** | **$L_{JJ}$ [nm]** | **$W_{JJ}$ [μm]** | **$I_C$ ($V_{GS}$=0V) [μA]** | **$I_C$/ $W_{JJ}$ ($V_{GS}$=0V) [μA/μm]** | **$I_C$ Suppr. [%]** | **$R_N$ ($V_{GS}$=0V) [Ω]** | **$R_N$ Incr. [times]** | **$\Delta V_{GS}$ [V]** | **Gate Type** | **Gate Insulator** | **Gate Distance [nm]** | **T [mK]** | **Year** | **Ref.** |
| Buried-QW t=4 nm, $n_{2D}$=1.9×10$^{12}$cm$^{-2}$, $\mu_n$=155000cm$^2$/Vs | Nb | 400 | 80 | 20.7 | 0.259 | 71* | 2.5* | 2* | -15 | Top | $SiO_2$/InAlAs /InGaAs | 100/20 /13.5 | 1000 | 1995 | [3] [4] |
| Buried-QW t=4 nm, $n_{2D}$=2.3×10$^{12}$cm$^{-2}$, $\mu_n$=111000cm$^2$/Vs | Nb | 350 | 40 | 5 | 0.125 | 100 | 16 | 91250 | -1.15 | Top | InAlAs /InGaAs | 20 /13.5 | 1000 | 1995 | [5] [6] [7] |
| Buried-QW $n_{2D}$=1.1×10$^{12}$cm$^{-2}$, $\mu_n$=160000cm$^2$/Vs | Nb | 600 | 20 | 15.6 | 0.780 | 100 | 5.9 | - | -25 | Top | $SiO_2$ | 300 | 1700 | 2002 | [8] [9] |
| Near-Surface-QW t=7 nm, $\mu_n$=17700cm$^2$/Vs | Al | 200 | 3 | 1.45* | 0.483* | 100* | 90* | 10* | -3.5 | Top | $Al_2O_3$ | 40 | 30 | 2016 | [10] |
| Near-Surface-QW t=7 nm, $n_{2D}$=1.3×10$^{12}$cm$^{-2}$ $\mu_n$=15600cm$^2$/Vs | Al | 250 | 3 | 1.85* | 0.616* | 100 | 100* | 17* | -2.5 | Top | $Al_2O_3$ | 40 | 30 | 2017 | [11] |
| Near-Surface-QW t=4 nm, $n_{2D}=7.2\times10^{11}cm^{-2}$, $\mu_n=14400cm^2/Vs$ | Al | 100 | 4 | 4.87* | 1.21* | 100* | 100* | 27* | -10 | Top | $AlO_X$ | 50 | 20 | 2019 | [12] |
| Near-Surface-QW t=7 nm, $n_{2D}=8.7\times10^{11}cm^{-2}$ $\mu_n=52400cm^2/Vs$ | Al | 40 | 1.5 | 1.37* | 0.913* | 100 | 270* | 4* | -2.2 | Top | $HfO_2$ | 30 | 30 | 2019 | [13] |
| Buried-QW t=4 nm, $n_{2D}=6.2\times10^{11}cm^{-2}$, $\mu_n=160000cm^2/Vs$ | Nb | 900 | 0.73 | - | - | 100 | - | - | -3 | Side | - | 1250 | 350 | 2019 | [14] |
| Near-Surface-QW t=4 nm, $n_{2D}=3.7\times10^{11}cm^{-2}$ $\mu_n=200000cm^2/Vs$ | Nb | 900 | 3.6 | 0.095* | 0.026* | 84* | 420* | 3.6* | -25 | Side | - | - | 315 | 2019 | [15] |
| Near-Surface-QW t=4 nm, $n_{2D}=7\times10^{11}cm^{-2}$ | Al | 100 | 4 | 2.25* | 0.563* | 88* | 160* | 4* | -2* | Top | h-BN | 6 | 30 | 2021 | [16] |
| Near-Surface-QW t=7 nm, $n_{2D}=9.6\times10^{11}cm^{-2}$ $\mu_n=16800cm^2/Vs$ | Al | 150 | 4 | 1.8 | 0.45 | 100 | 323 | 6* | -8 | Top | $Al_2O_3$ | 60 | 30 | 2021 | [17] |
| Near-Surface-QW t=4 nm, $n_{2D}=3.7\times10^{12}cm^{-2}$ | Al | 250 | 4 | 1.9* | 0.475* | 100* | 60* | 17.8* | -1.5 | Top | $Al_2O_3$ | 10 | 17 | 2021 | [18] |
| Near-Surface-QW t=4 nm, $n_{2D}=7\times10^{11}cm^{-2}$ | Al | 100 | 4 | 4.85* | 1.213* | 75* | 95* | 2.7* | -7.5* | Top | $AlO_X$ | 50 | 30 | 2021 | [16] |
| NW-QW t=30 nm, $\mu_n=3200cm^2/Vs$ | Al | 120 | 0.185 | 0.048* | 0.259* | 100 | 1930* | 1.9* | -0.3 | Top | $HfO_2$ | 15 | 20 | 2021 | [19] |
| Near-Surface-QW t=4 nm, $n_{2D}=4\text{-}6\times10^{11}cm^{-2}$ $\mu_n=0.9\text{-}1.3\times10^5cm^2/Vs$ | Al | 300 | 9 | 1.1 | 0.122 | 100 | 34* | 48.5* | -3.5 | Top | $Al_2O_3$ | 50 | 20 | 2022 | [20] |
| Near-Surface-QW t=7 nm, $n_{2D}=1.6\times10^{12}cm^{-2}$ $\mu_n=12000cm^2/Vs$ | Al | 150 | 3 | 0.86* | 0.286* | 100* | 10* | 1* | -1 | Top | $HfO_2$ | 15 | 50 | 2023 | [21] |
| Near-Surface-QW t=8 nm, $n_{2D}=8\times10^{11}cm^{-2}$ $\mu_n=18000cm^2/Vs$ | Al | 100 | - | 0.280 | - | 100* | 80* | 9.8* | -0.8 | Top | $Al_2O_3/HfO_2$ | 3/15 | 10 | 2024 | [22] |

**InAs 2D NSs**

| Specs | Super conductor | $L_{JJ}$ [nm] | $W_{JJ}$ [μm] | $I_C$ ($V_{GS}$=0V) [μA] | $I_C/W_{JJ}$ ($V_{GS}$=0V) [μA/μm] | $I_C$ Suppr. [%] | $R_N$ ($V_{GS}$=0V) [Ω] | $R_N$ Incr. [times] | $\Delta V_{GS}$ [V] | Gate Type | Gate Insulator | Gate Distance [nm] | T [mK] | Year | Ref. |
|---|---|---|---|---|---|---|---|---|---|---|---|---|---|---|---|
| t=15-30 nm, $n_{2D}=4\times10^{1}cm^{-2}$ $\mu_n=8300cm^2/Vs$ | Al | 90 | 0.3 | 0.036* | 0.12* | 10* | 1760* | 2.5* | -10 | Back | SiO2 | 300 | 20 | 2023 | [23] |
| t=15-30 nm, $n_{2D}=4\times10^{1}cm^{-2}$ $\mu_n=8300cm^2/Vs$ | Al | 90 | 0.3 | 0.067* | 0.223* | 54* | 402* | 3* | -8 | Top | $Al_2O_3$ | 30 | 20 | 2023 | [23] |

**InAs 1D NWs**

| Specs | Super conductor | $L_{JJ}$ [nm] | $W_{JJ}$ [μm] | $I_C$ ($V_{GS}$=0V) [μA] | $I_C/W_{JJ}$ ($V_{GS}$=0V) [μA/μm] | $I_C$ Suppr. [%] | $R_N$ ($V_{GS}$=0V) [Ω] | $R_N$ Incr. [times] | $\Delta V_{GS}$ [V] | Gate Type | Gate Insulator | Gate Distance [nm] | T [mK] | Year | Ref. |
|---|---|---|---|---|---|---|---|---|---|---|---|---|---|---|---|
| *n*-InAs $n_{3D}$ = 2÷10 ×$10^{18}$ $cm^{-3}$, $\mu_n$ =200÷2000 $cm^2/Vs$ | Al | 100-450 | 0.04 - 0.13 | 0.0012 | - | 100 | 4500 | 15.5 | -71 | Back | $SiO_2$ | 250 | 40 | 2005 | [24] |
| *n*-InAs $n_{3D}$ = 1.0×$10^{18}$ $cm^{-3}$ | Nb | 70 | 0.08 | 0.0028 | 0.035 | 100 | 3800 | 2.01* | -20 | Back | $SiO_2$ | - | 500 | 2012 | [25] |
| *n*-InAs | Nb | 55 | 0.08 | 0.075 | 0.938 | 46.7 | 1523* | 1.56* | -7 | Back | - | - | 15 | 2021 | [26] |
| *n*-InAs | Sn | 100 | - | 0.24* | - | 83.7* | 962* | 5.31* | -0.5 | Top | $HfO_2$ | 10 | 40 | 2023 | [27] |

**Ge 2D QWs**

| Specs | Super conductor | $L_{JJ}$ [nm] | $W_{JJ}$ [μm] | $I_{C\,MAX}$ [μA] | $I_C/W_{JJ}$ [μA/μm] | $I_C$ Suppr. [%] | $R_{N\,min}$ [Ω] | $R_N$ Incr. [times] | $\Delta V_{GS}$ [V] | Gate Type | Gate Insulator | Gate Distance [nm] | T [mK] | Year | Ref. |
|---|---|---|---|---|---|---|---|---|---|---|---|---|---|---|---|
| Near-Surface-QW t=16 nm, $p_{2D}$ = 6×$10^{11}$ $cm^{-2}$ $\mu_h$ =500000 $cm^2/Vs$ | Al | 50 | 1 | 0.043 ($V_{GS}$=-4V) | 0.043 | 100 | 371* ($V_{GS}$=-4V) | 2.70* | 2 | Top | $Al_2O_3$ | - | 10 | 2019 | [28] |
| Buried-QW t= 16 nm $p_{2D}$ = 2÷3 ×$10^{11}$ $cm^{-2}$ $\mu_h$ ~ 2÷3×$10^5$ $cm^2/Vs$ | Al | 1000 | 0.5* | 0.006* ($V_{GS}$=-1.95V) | 0.012* | 100* | 1120* ($V_{GS}$=-1.95V) | 7* | 0.95 | Top | $HfO_2$ | 40 | 15 | 2019 | [29] |
| Buried-QW t=16 nm, $p_{2D}$ = 6×$10^{11}$ $cm^{-2}$ $\mu_h$ = 5×$10^5$ $cm^2/Vs$ | Al/Nb | 150 | - | 1 ($V_{GS}$=-2.5V) | - | 94.3* | 344* ($V_{GS}$=-2.5V) | 2.90* | 1.5 | Top | $Al_2O_3$ | 20 | 20 | 2021 | [30] |
| Buried-QW t=16 nm, $p_{2D}$ = 6×$10^{11}$ $cm^{-2}$ $\mu_h$ ~ 6×$10^5$ $cm^2/Vs$ | PtSiGe | 70 | 0.04 | 0.1* ($V_{GS}$=-2V) | 2.5* | 100* | 510* ($V_{GS}$=-2V) | - | 0.75 | Top | $Al_2O_3$ | 10 | 15 | 2023 | [31] |
| Buried-QW<br>t=20 nm,<br>$p_{2D}$ = $2.7\times10^{11}$ $cm^{-2}$<br>$\mu_h$ = $8.3\times10^{5}$ $cm^2/Vs$ | PtSiGe | 0.3 | 4 | 0.93 ($V_{GS}$=-1.2V) | 0.2325 | 100 | 100 ($V_{GS}$=-1.2V) | >1.2* | 0.5 | Top | $SiO_2$ | 12 | 80 | 2024 | [32] |
| Buried-QW<br>t=16 nm,<br>$p_{2D}$ = $6\times10^{11}$ $cm^{-2}$<br>$\mu_h$ = $5\times10^{5}$ $cm^2/Vs$ | PtSiGe | 0.3* | - | 0.01* ($V_{PG}$=-2.22V, $V_{HS}$=-1V, $V_{LB}$=-0.9V*, $V_{RB}$=-1.5V*) | - | 100* | - | - | - | Top + Side | $Al_2O_3$ | 7 | 9 | 2025 | [33] |
| **Graphene** | | | | | | | | | | | | | | | |
| **Specs** | **Superconductor** | **$L_{JJ}$ [nm]** | **$W_{JJ}$ [μm]** | **$I_C$ [μA]** | **$I_C$/ $W_{JJ}$ [μA/μm]** | **$I_C$ Suppr. [%]** | **$R_N$ [Ω]** | **$R_N$ Incr. [times]** | **$\Delta V_{GS}$ [V]** | **Gate Type** | **Gate Insulator** | **Gate Distance [nm]** | **T [mK]** | **Year** | **Ref.** |
| Monolayer<br>*p*-type regime | Nb | 1500 | 6 | 1.59* ($V_{GS}$=-60 V) | 0.265* | 98.7* ($V_{GS}$=0 V) | 52* ($V_{GS}$=-60 V) | 12.5* | 60 | Back | $SiO_2$ | 300 | 300 | 2018 | [34] |
| *Monolayer*<br>*n*-type regime<br>$\mu_n$ =300000 $cm^2/Vs$ | Nb | 1500 | 6 | 7 ($V_{GS}$=60 V) | 1.167 | 99.7 ($V_{GS}$=0 V) | 19* ($V_{GS}$=60 V) | 30* | -60 | Back | $SiO_2$ | 300 | 300 | 2018 | [34] |
| Monolayer<br>$p_{2D}$=$5.5\times10^{12}$ $cm^{-2}$<br>$\mu_h$ =2100 $cm^2/Vs$ | Nb | 250 | 80 | 1.9 ($V_{GS}$=-50 V) | 0.024 | 89 | 40 ($V_{GS}$=-50 V) | 4 | 113.85 | Back | $SiO_2$ | 300 | 320 | 2019 | [35] |
| Monolayer | Al | 300 | 20 | 0.9* ($V_{GS}$=10 V) | 0.045* | 74* | 24.1* ($V_{GS}$=10 V) | 3.6* | -21.6 | Top | $Al_2O_3$ | 30 | 42 | 2024 | [36] |
| Monolayer | Al | 350 | 20 | 1* ($V_{GS}$=10 V) | 0.05* | 85* | 23.8* ($V_{GS}$=10 V) | 5.6* | -22.2 | Top | $Al_2O_3$ | 30 | 42 | 2024 | [36] |
| Monolayer | Al | 250 | 50 | 2.34* ($V_{GS}$=10 V) | 0.047* | 80* | 14.4* ($V_{GS}$=10 V) | 3* | -22.5 | Top | $Al_2O_3$ | 30 | 42 | 2024 | [36] |
| Monolayer | Al | 300 | 50 | 3.15* ($V_{GS}$=10 V) | 0.063* | 77* | 13.3* ($V_{GS}$=10 V) | 2.1* | -21.1 | Top | $Al_2O_3$ | 30 | 42 | 2024 | [36] |
| **Metal NWs** | | | | | | | | | | | | | | | |
| **Specs** | **Superconductor** | **$L_{JJ}$ [nm]** | **$W_{JJ}$ [μm]** | **$I_C$ ($V_{GS}$=0V) [μA]** | **$I_C$/ $W_{JJ}$ ($V_{GS}$=0V) [μA/μm]** | **$I_C$ Suppr. [%]** | **$R_N$ ($V_{GS}$=0V) [Ω]** | **$R_N$ Incr. [times]** | **$\Delta V_{GS}$ [V]** | **Gate Type** | **Gate Insulator** | **Gate Distance [nm]** | **T [mK]** | **Year** | **Ref.** |
| NW<br>t =30 nm | Ti | 900 | 0.2 | 11 | 55 | 100 | 45 | 1 | 40 | Back<br>+<br>Side | - | 112* | 5 | 2018 | [37] |
| NW<br>t = 11 nm | Al | 800 | 0.03 | 12.3 | 410 | 35 | 320 | 1 | 70 | Back | - | 112* | 5 | 2018 | [37] |
| DB | Nb | 100 | 0.9 | 0.03 | 0.033 | 90 | 30 | 1 | 40 | Side | - | 70 | 30 | 2020 | [38] |
| DB<br>t = 60 nm | V | 160 | 0.09 | 1420 | 15778 | 100 | 110 | 1 | 15 | Side | - | 70 | 2 | 2020 | [39] |

**Table S2. Theoretical frequency-dependent electrical behavior of the lumped element model of the JoFET fabricated in the superconducting 1I8O demultiplexer.**

| $f$ [Hz] | $L_K$ [pH] | $\lvert Z_{L_K} \rvert$ [Ω] *ON-state* | $R_N$ [Ω] *OFF-state* | $C_{GS} = C_{GD}$ [fF] | $\lvert Z_{C_{GS}} \rvert = \lvert Z_{C_{GD}} \rvert$ [Ω] | $\lambda_{GaAs}$ [mm] |
|---|---|---|---|---|---|---|
| 10M | 163 | 0.01 | 564 | 90 | 176839 | 2500 |
| 100M | 163 | 0.10 | 564 | 90 | 17683.9 | 250 |
| 1G | 163 | 1.02 | 564 | 90 | 1768.39 | 25 |
| 10G | 163 | 10.2 | 564 | 90 | 176.839 | 2.5 |

**Table S3. Superconducting 1I8O demultiplexer truth table.** The highlighted rows are the gate settings reported in the paper.

| **Gate Control Lines** | | | | | | **Line Resistance (Referred To $I_O$)** | | | | | | | |
|---|---|---|---|---|---|---|---|---|---|---|---|---|---|
| $G_{00}$ | $G_{01}$ | $G_{10}$ | $G_{11}$ | $G_{20}$ | $G_{21}$ | $O_0$ | $O_1$ | $O_2$ | $O_3$ | $O_4$ | $O_5$ | $O_6$ | $O_7$ |
| -4.5 V | 0 | -4.5 V | 0 | -4.5 V | 0 | 0 | $1R_N$ | $1R_N$ | $2R_N$ | $1R_N$ | $2R_N$ | $2R_N$ | $3R_N$ |
| -4.5 V | 0 | -4.5 V | 0 | 0 | -4.5 V | $1R_N$ | 0 | $2R_N$ | $1R_N$ | $2R_N$ | $1R_N$ | $3R_N$ | $2R_N$ |
| -4.5 V | 0 | 0 | -4.5 V | -4.5 V | 0 | $1R_N$ | $2R_N$ | 0 | $1R_N$ | $2R_N$ | $3R_N$ | $1R_N$ | $2R_N$ |
| -4.5 V | 0 | 0 | -4.5 V | 0 | -4.5 V | $2R_N$ | $1R_N$ | $1R_N$ | 0 | $3R_N$ | $2R_N$ | $2R_N$ | $1R_N$ |
| **0** | **-4.5 V** | **-4.5 V** | **0** | **-4.5 V** | **0** | **$1R_N$** | **$2R_N$** | **$2R_N$** | **$3R_N$** | **0** | **$1R_N$** | **$1R_N$** | **$2R_N$** |
| 0 | -4.5 V | -4.5 V | 0 | 0 | -4.5 V | $2R_N$ | $1R_N$ | $3R_N$ | $2R_N$ | $1R_N$ | 0 | $2R_N$ | $1R_N$ |
| **0** | **-4.5 V** | **0** | **-4.5 V** | **-4.5 V** | **0** | **$2R_N$** | **$3R_N$** | **$1R_N$** | **$2R_N$** | **$1R_N$** | **$2R_N$** | **0** | **$1R_N$** |
| 0 | -4.5 V | 0 | -4.5 V | 0 | -4.5 V | $3R_N$ | $2R_N$ | $2R_N$ | $1R_N$ | $2R_N$ | $1R_N$ | $1R_N$ | 0 |

## 2. Supporting Figures

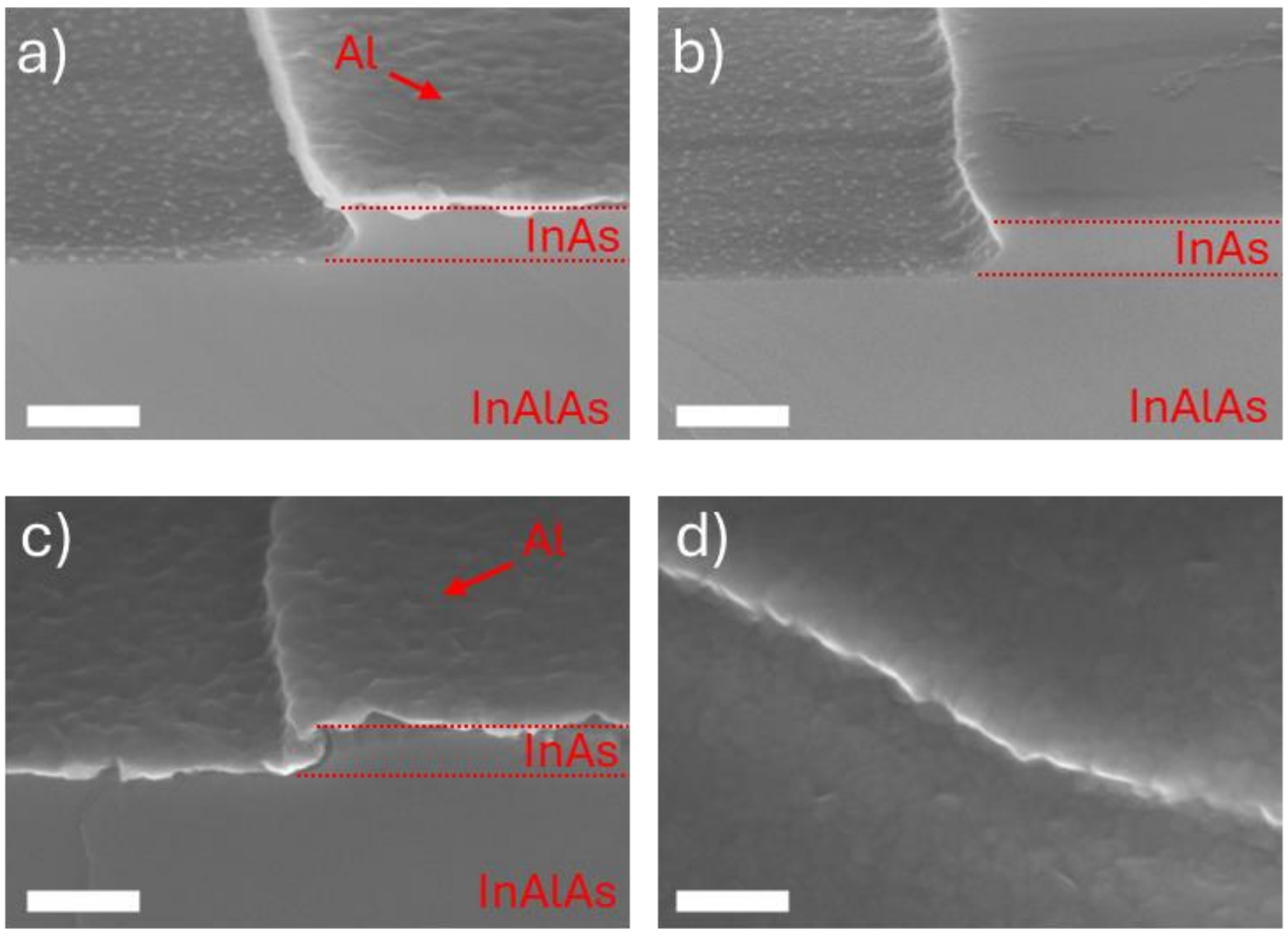

**Figure S1: Tilted scanning electron microscopy analysis (100k×) of InAs MESA edge coating by metal gate.** (a) InAs profile after MESA definition via wet etching. (b) InAs profile after Al removal from MESA. (c,d) InAs profile coverage via tilted Ti/Al deposition. The scalebar is 200 nm.

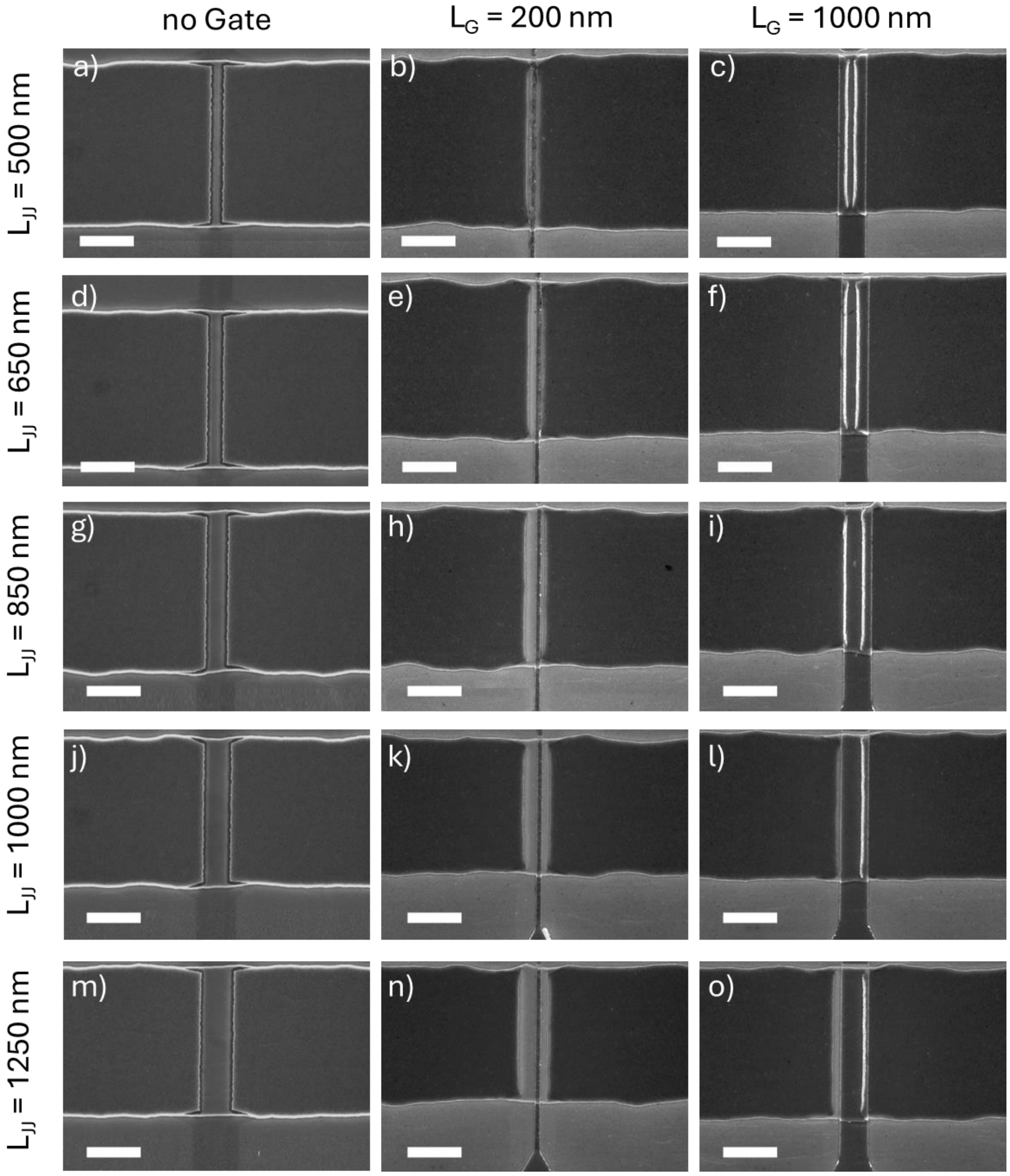

**Figure S2: Top-view scanning electron microscopy analysis (10k×) of JoFETs.** a,d,g,j,m) SEM images of JoFETs before gate architecture manufacturing, i.e., Josephson Junctions, with different interelectrode separations. b,e,h,k,n) SEM images of JoFETs featuring $L_G$ = 200 nm with different interelectrode separations. c,f,i,l,o) SEM images of JoFETs featuring $L_G$ = 1000 nm with different interelectrode separations. The scalebar is 2 μm.

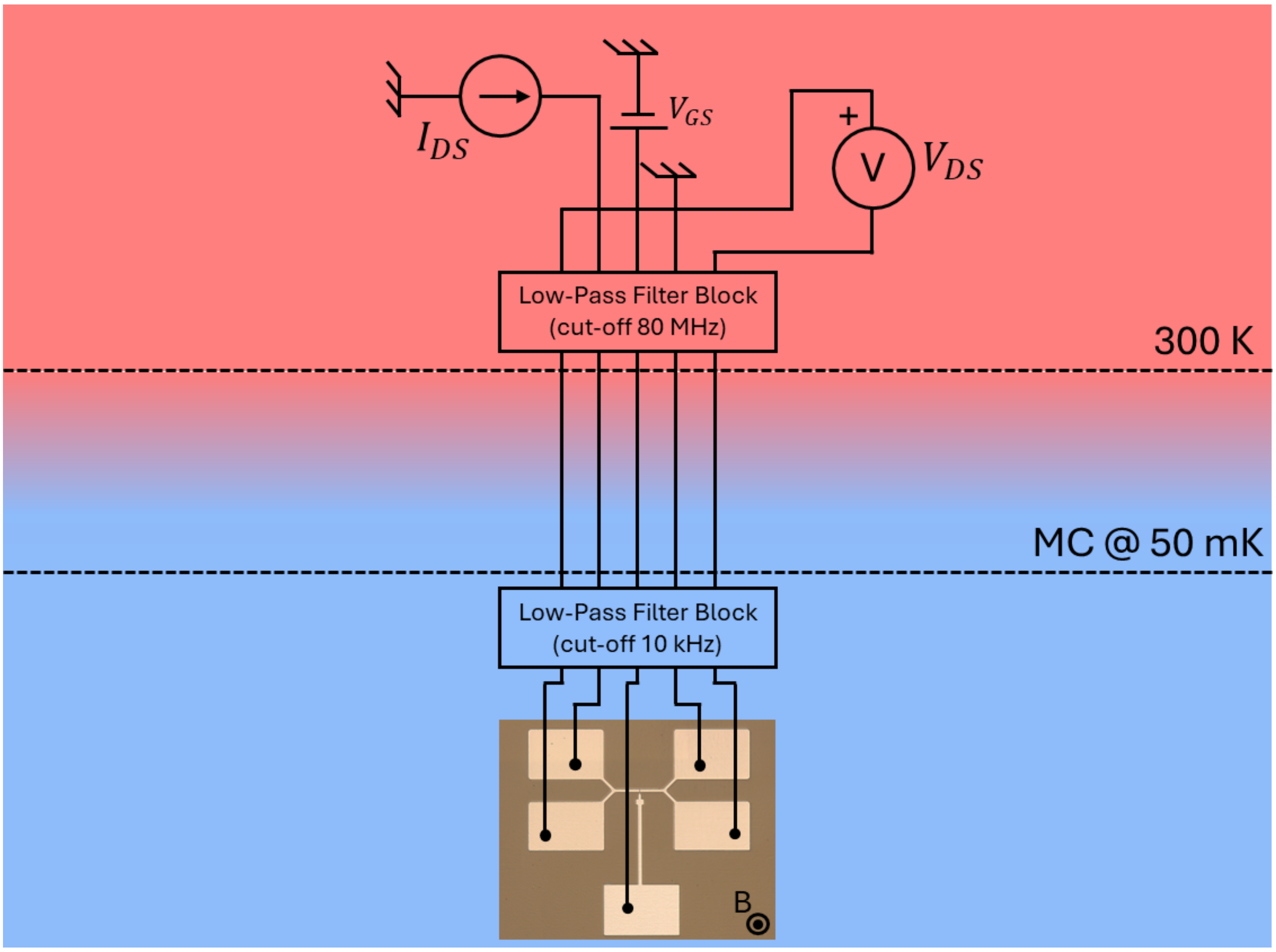

**Figure S3: DC measurement setup used to characterize the JoFETs in a dilution fridge equipped with a z-axis superconducting magnet.**

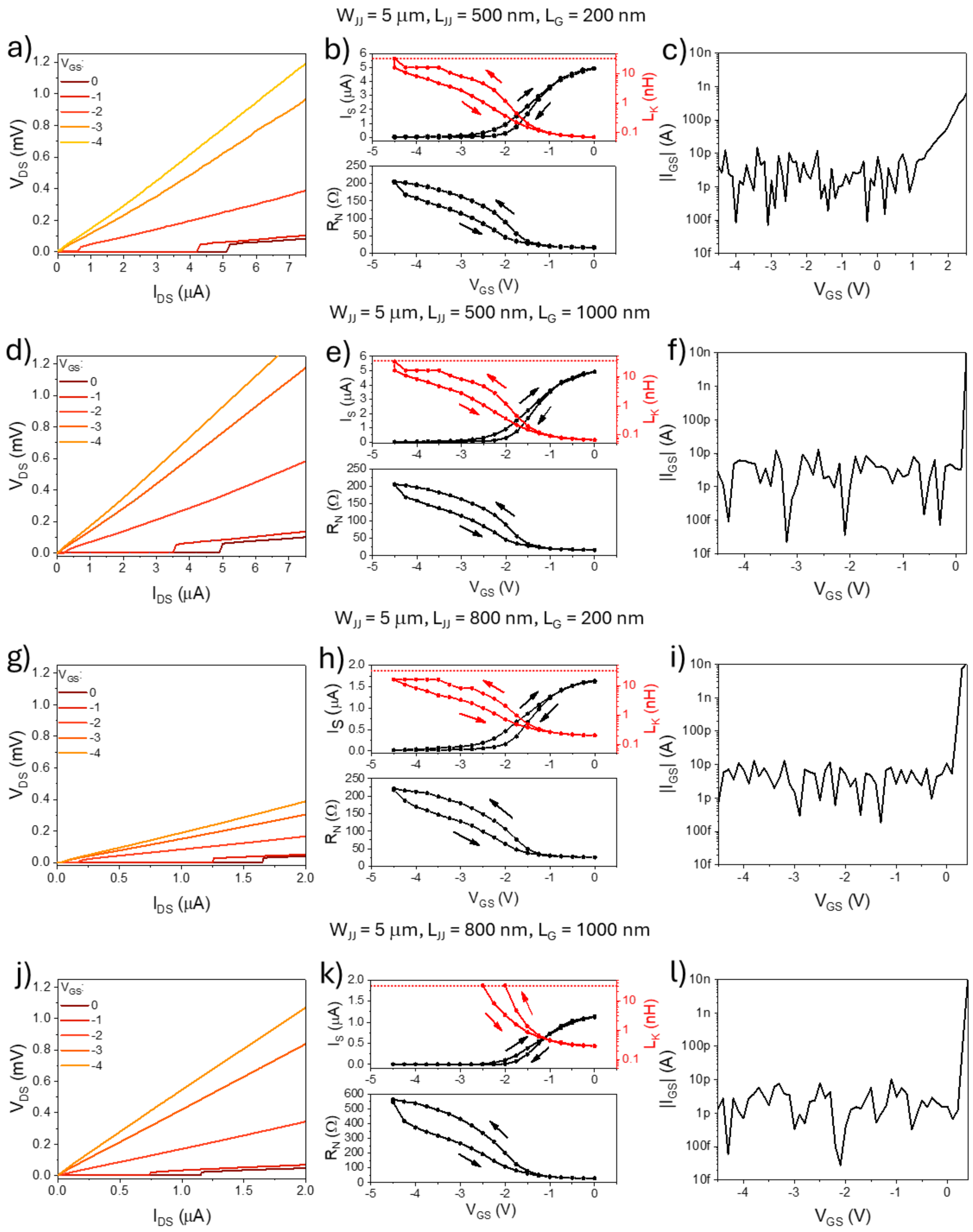

**Figure S4: InAsOI-based Josephson Field Effect Transistors electrical characterization.** a,d,g,j) Gate-dependent voltage vs. current characteristic of JoFETs. b,e,h,k) Upward and downward gate-dependent switching current, kinetic inductance, and normal state resistance of JoFETs. c,f,i,l,) Gate leakage current vs. gate voltage of JoFETs. Measurements were performed at 50 mK.

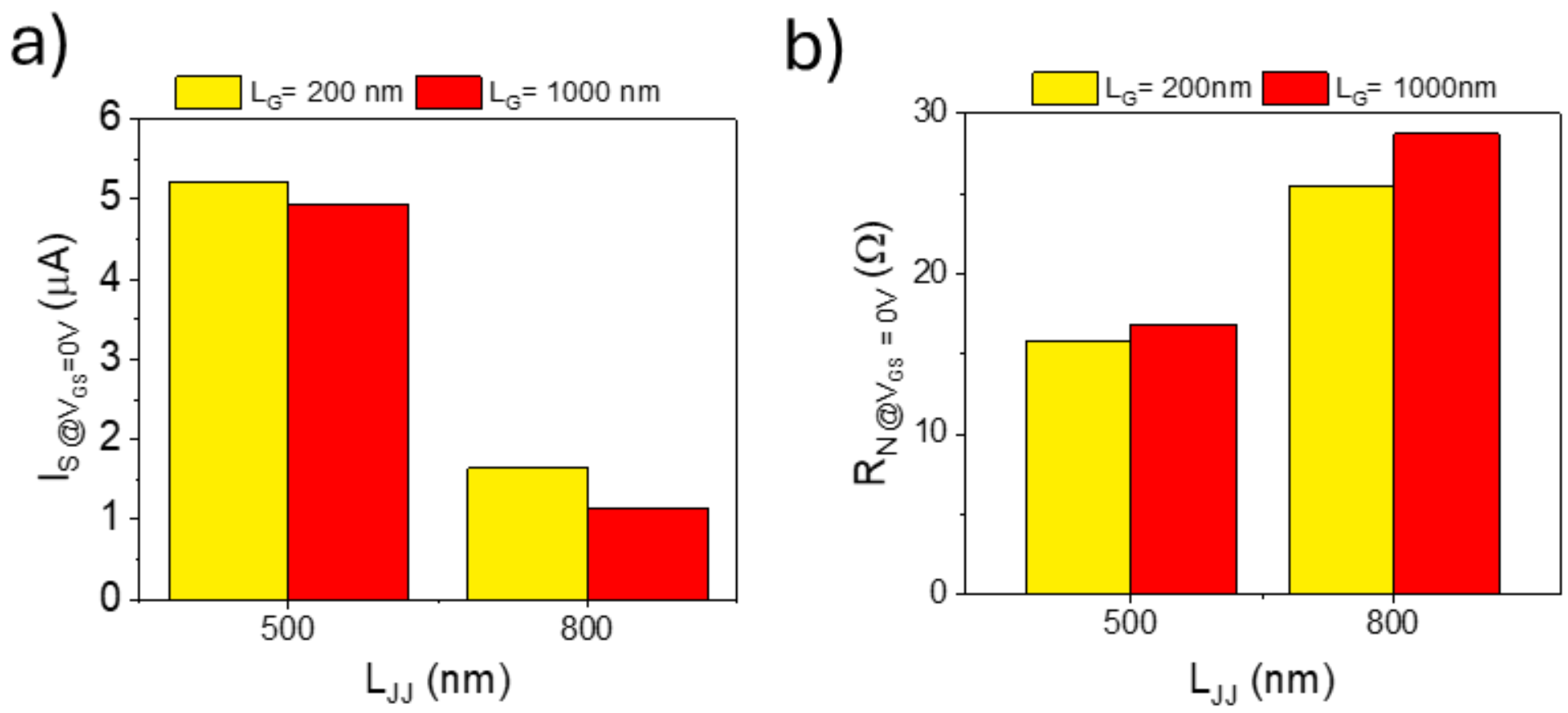

**Figure S5: Zero-gate-voltage electrical properties of JoFETs.** a,b) Switching current (a) and normal state resistance (b) of JoFETs with different gate lengths and interelectrode separations.

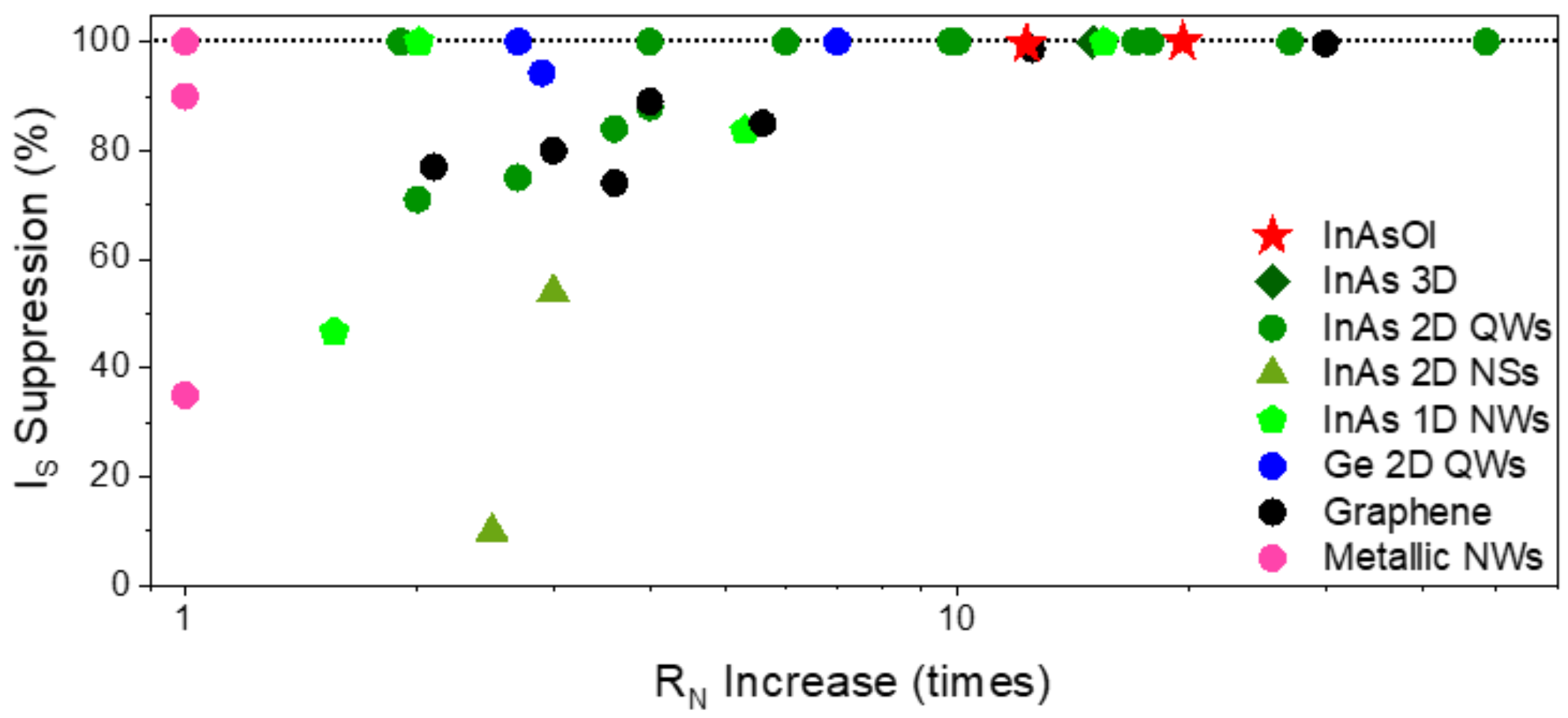

**Figure S6: State of the art of JoFETs**. *$I_S$ suppression* and *$R_N$ increase* factors achieved for InAsOI-based JoFETs are perfectly in agreement with the top values obtain in the research field.

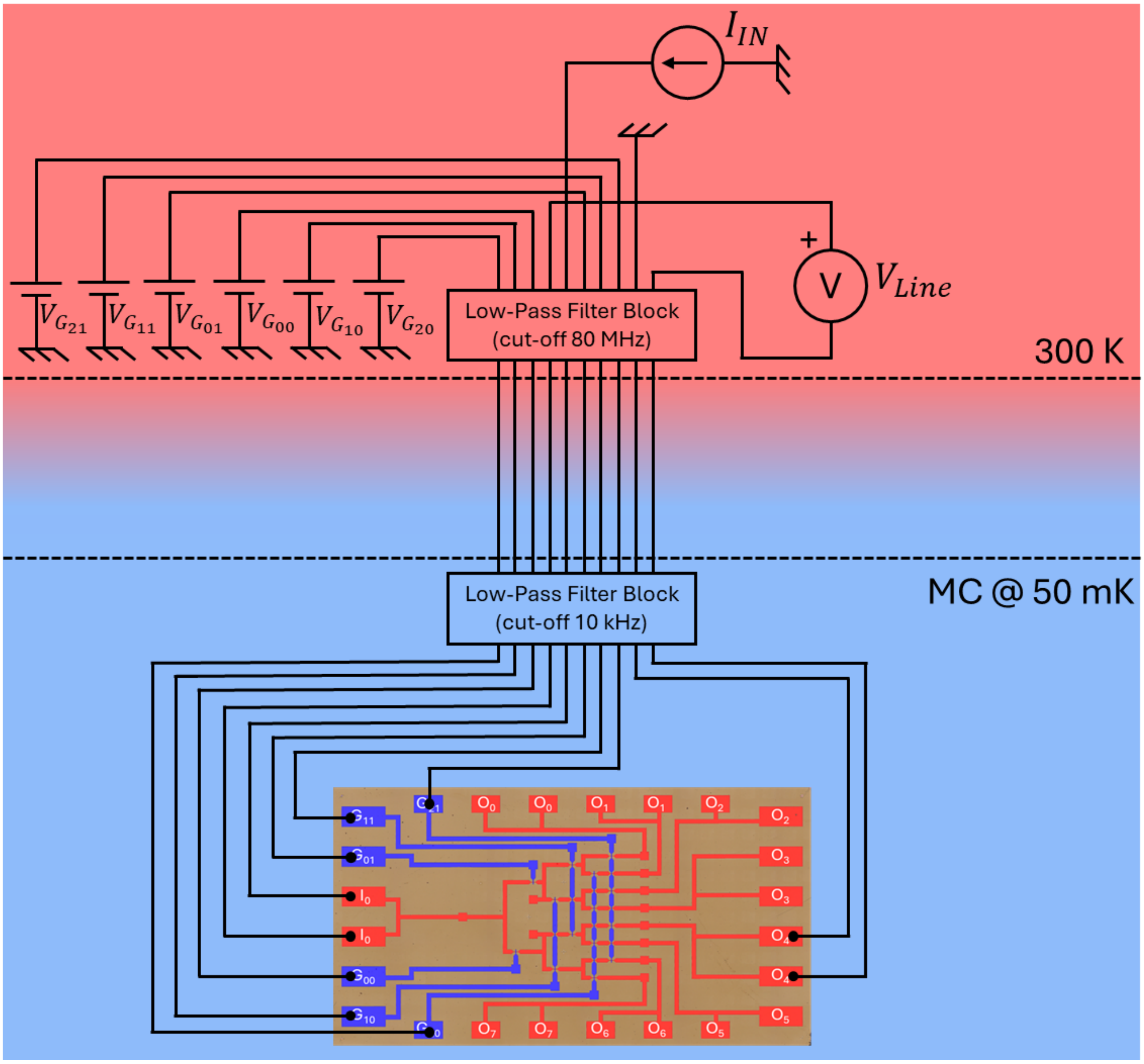

**Figure S7: DC measurement setup used to characterize the DC electrical behavior of the superconducting 1I8O analog demultiplexer.** The input current ($I_{IN}$) was provided to the demultiplexer input while the signal line voltage drops ($V_{Line}$) are measured changing the gate settings. The figure shows only one line connected.

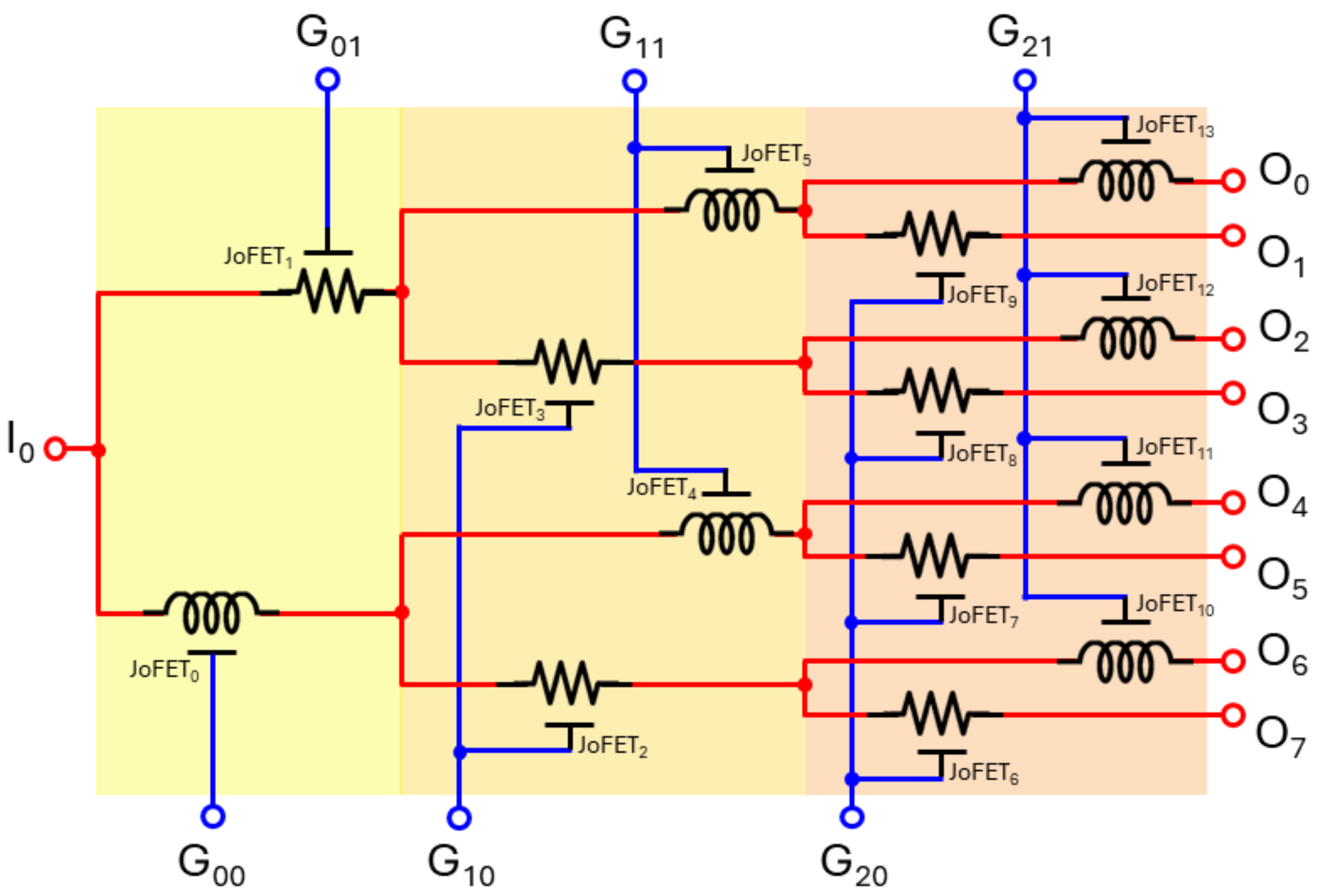

**Figure S8: Schematic lumped elements diagram of the superconducting 1I8O demultiplexer with the signal line $I_0$-$O_4$ in superconducting state.** The gate settings are $V_{G00}$,$V_{G11}$,$V_{G21}$ = 0 V and $V_{G01}$,$V_{G10}$,$V_{G20}$ = -4.5 V.

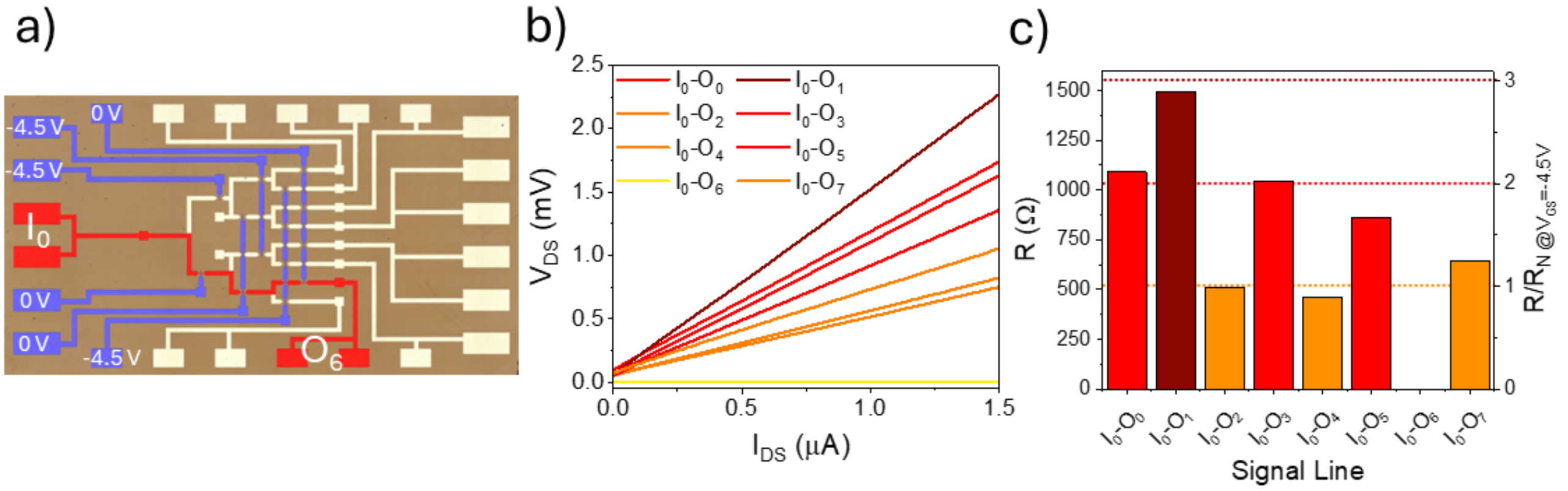

**Figure S9: DC electrical characterization of the superconducting 1I8O demultiplexer with the signal line $I_0$-$O_6$ in superconducting state.** a) False-colors optical microscope image of the superconducting 1I8O demultiplexer highlighting the chosen superconductive path and the gate control configuration. b) V-I characteristics of all the 1I8O switch signal lines with the gate configuration shown in (a). Line resistances (left) and normalized line resistances (right) of all the 1I8O demultiplexer signal lines with the gate configuration shown in (a).

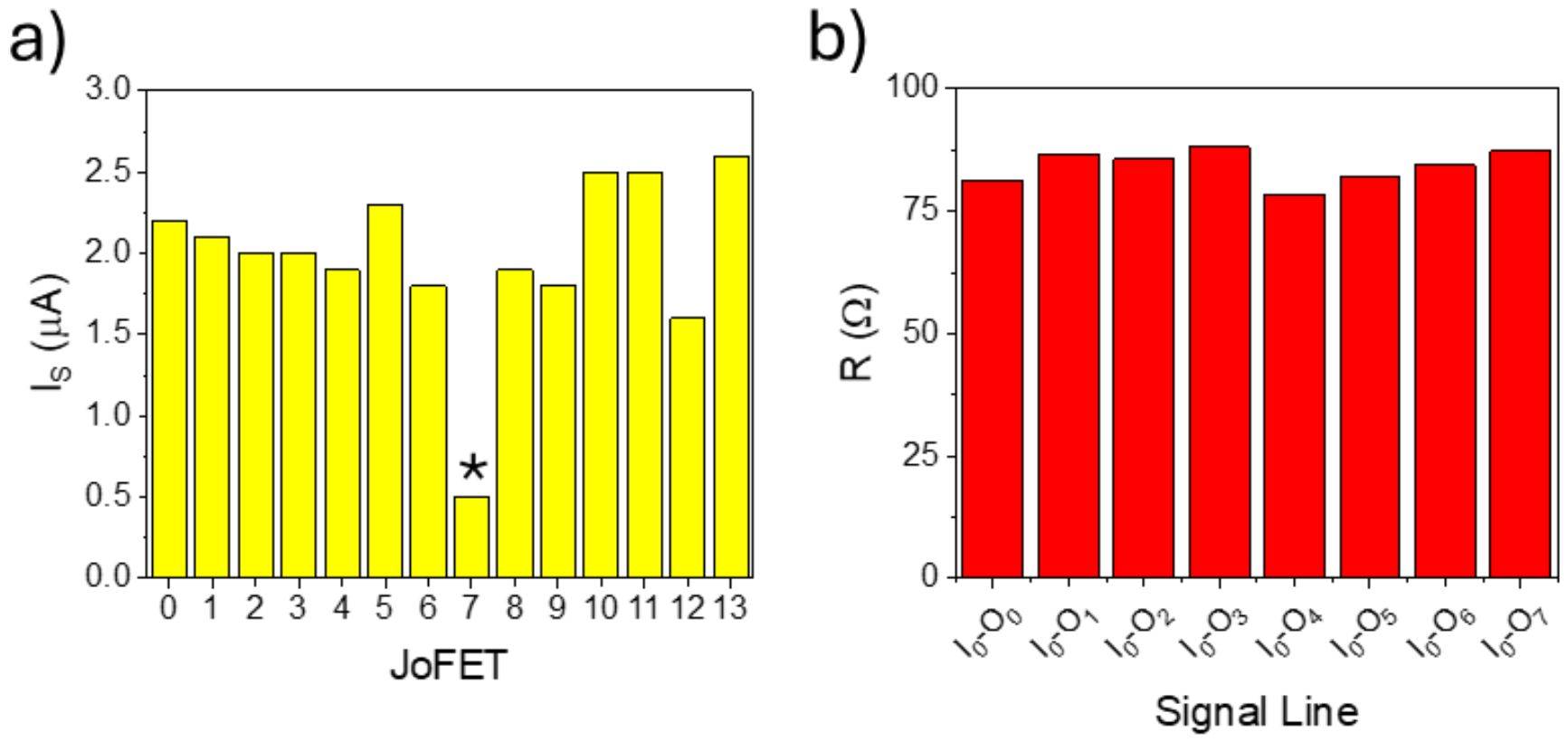

**Figure S10: Zero-gate-voltage electrical properties of JoFETs and signal lines of the superconducting 1I8O analog demultiplexer.** a) Zero-gate-voltage switching currents of the JoFETs; only the JoFET$_7$ exhibits a switching current far from the average. b) Zero-gate-voltage signal line resistances for input currents higher than the switching current.

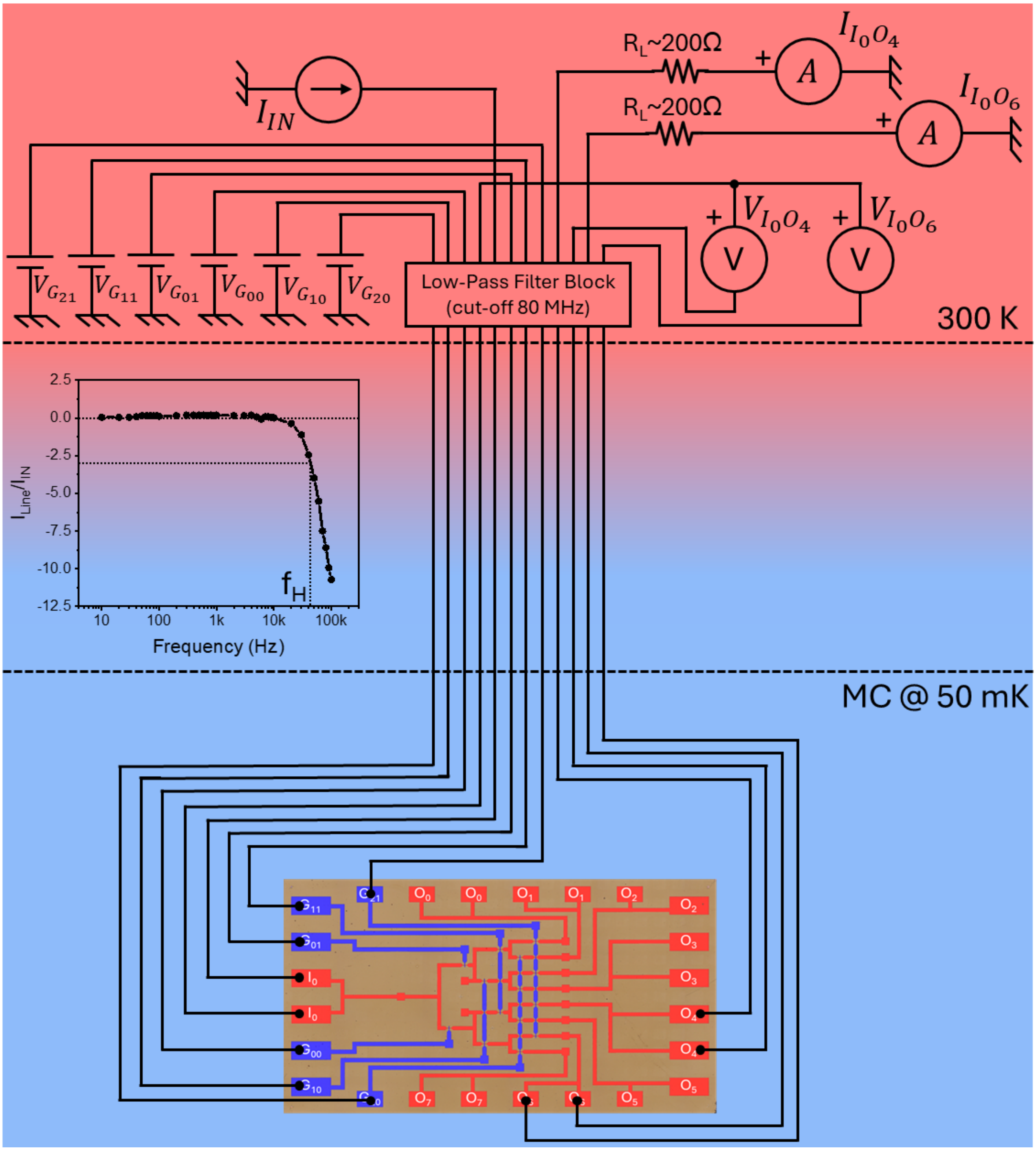

**Figure S11: DC-VLF measurement setup used to characterize the real-time electrical behavior of the superconducting 1I8O analog demultiplexer.** The input current ($I_{IN}$) was provided to the demultiplexer input while the signal line output currents ($I_{I_0O_4}$ and $I_{I_0O_6}$) and the signal line voltage drops ($V_{I_0O_4}$ and $V_{I_0O_6}$) are measured changing the gate settings. Cryostat's wires account as a load resistance of 200 Ω. The system exhibits a cut-off frequency of ~45 kHz at 300 K.

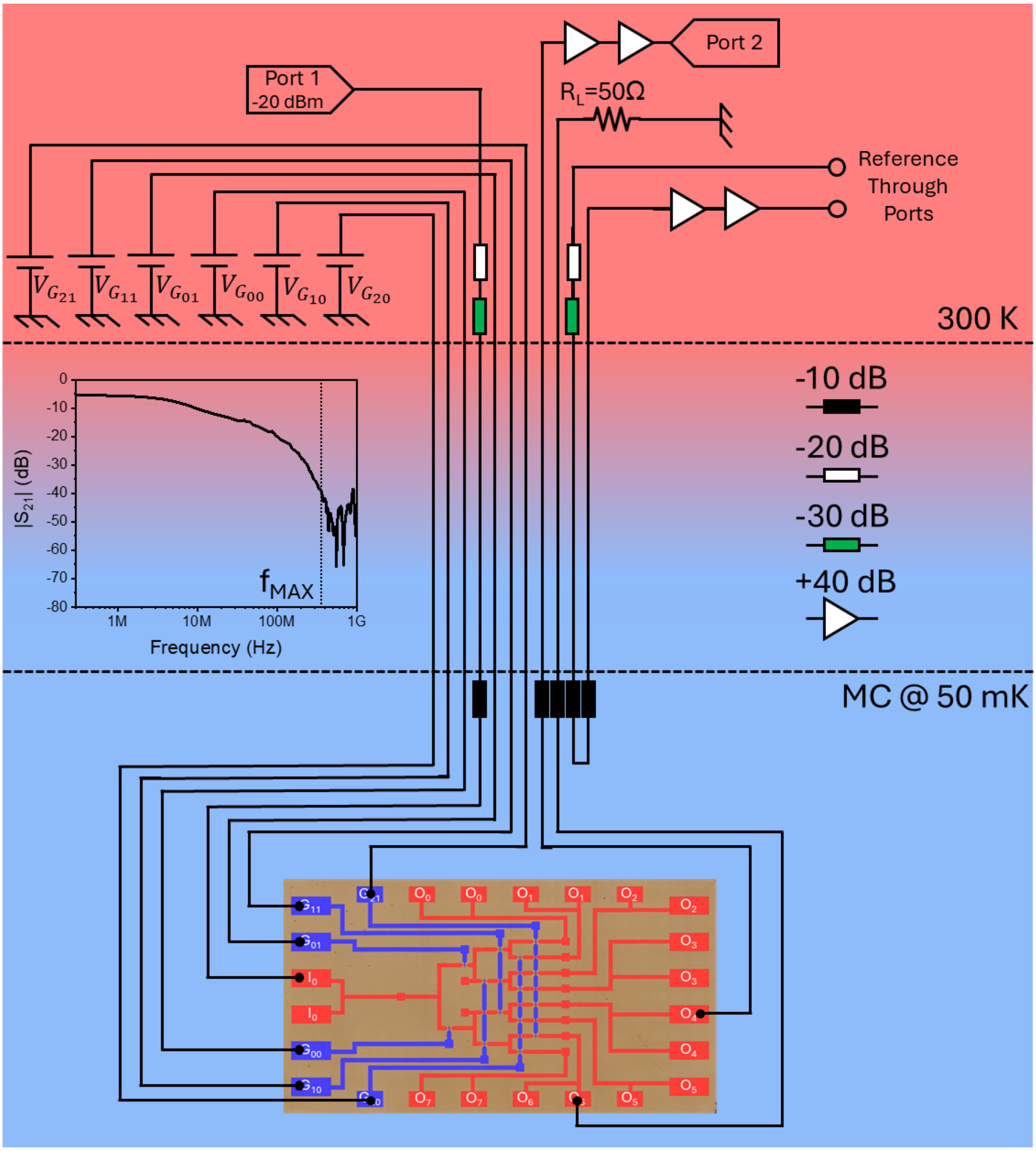

**Figure S12: LF-VHF measurement setup used to characterize the frequency-resolved electrical behavior of the superconducting 1I8O analog demultiplexer.** The input power ($P_{IN} = -20\ dBm$) supplied by the Port$_1$ was attenuated both at 300 K (-50 dB) and 50 mK (-10 dB) then provided to the demultiplexer input. The line $O_4$ output power was attenuated at 50 mK (-10 dB) and amplified at 300 K (+80 dB) then provided to the Port$_2$, while line $O_6$ was terminated with 50 Ω. The output power was measured by changing the gate settings. The same was performed to check the line $O_6$ output power. Reference through ports were used to collect the trough calibration data. The system exhibits a maximum operation frequency of ~300 MHz at 300 K.

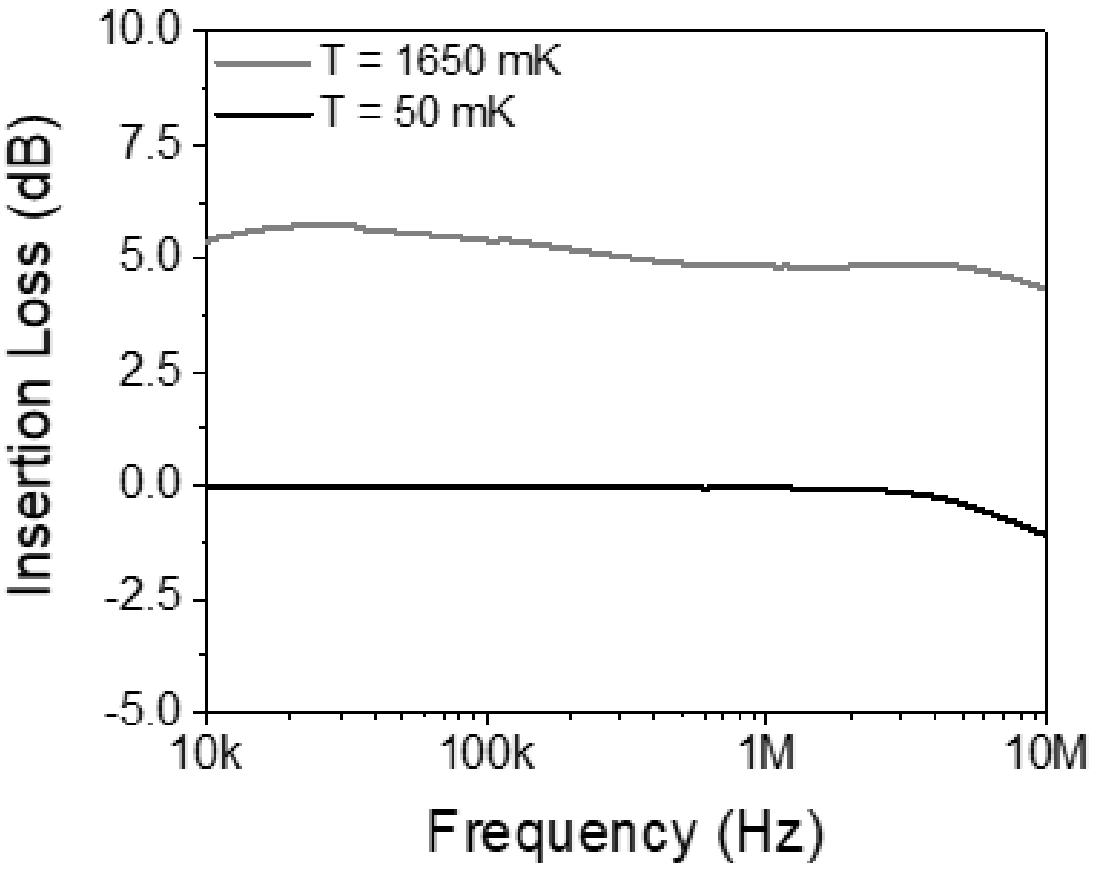

**Figure S13: Frequency-resolved insertion loss of signal line $I_0$-$O_4$ gated in the non-dissipative state measured at 50 mK and 1650 mK.** At 1650 mK, the 1I8O demultiplexer is in the normal state which increases the insertion loss of the signal line.

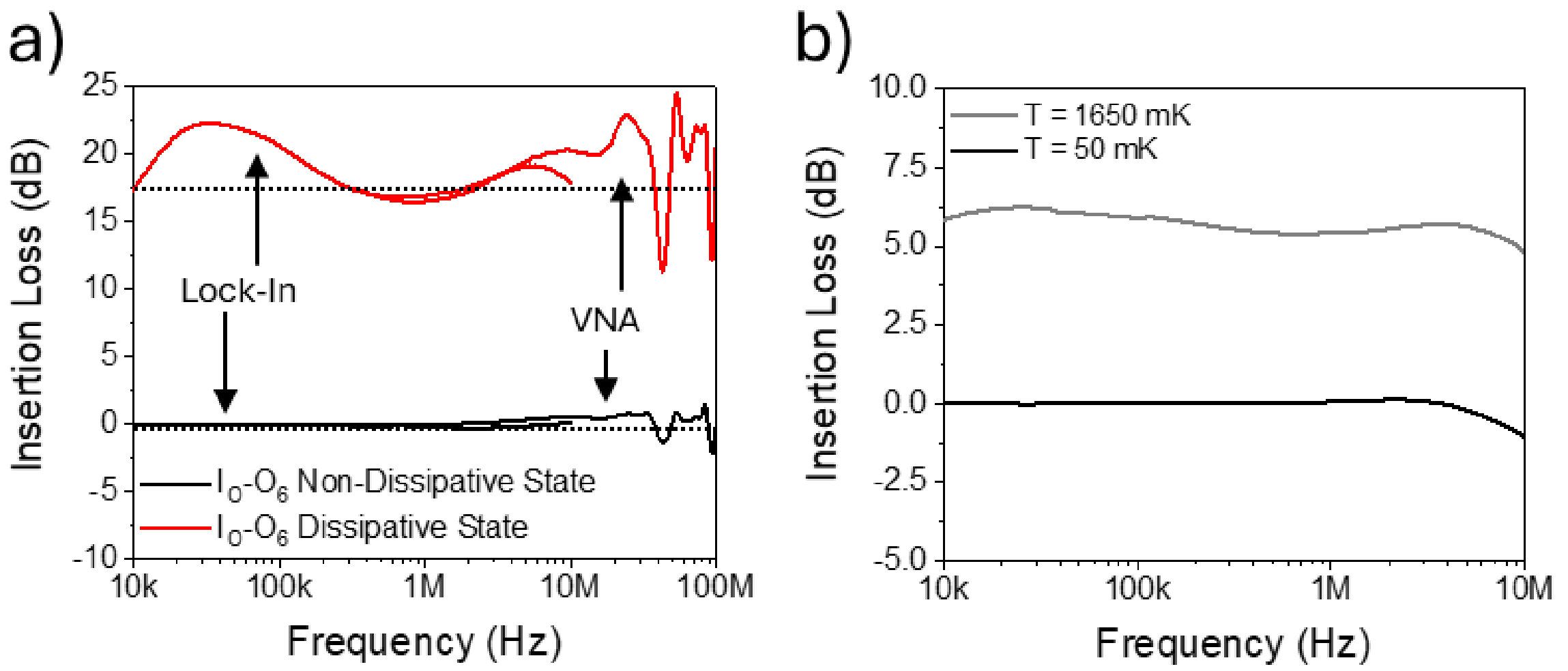

**Figure S14: Frequency-resolved insertion loss of signal line $I_0$-$O_4$.** a) Frequency-resolved insertion loss of signal line $I_0$-$O_6$ in the non-dissipative and dissipative states at 50 mK. b) Frequency-resolved insertion loss of signal line $I_0$-$O_6$ gated in the non-dissipative state measured at 50 mK and 1650 mK. At 1650 mK, the 1I8O demultiplexer is in the normal state which increases the insertion loss of the signal line.

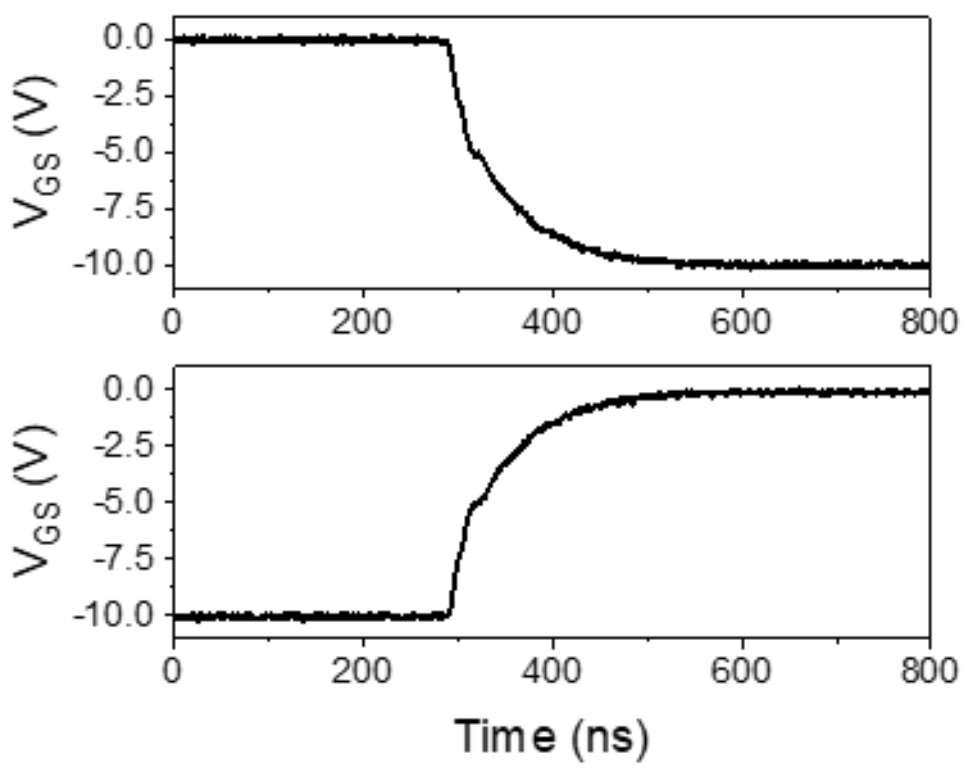

**Figure S15: Gate control switching voltage during the ON-to-OFF (up) and OFF-to-ON (bottom) state transients.** The square wave gate control signal travels into the cryostat and comes out for measurement. The returned room temperature control signal exhibited $t_r$ and $t_f$ of ~130 ns.

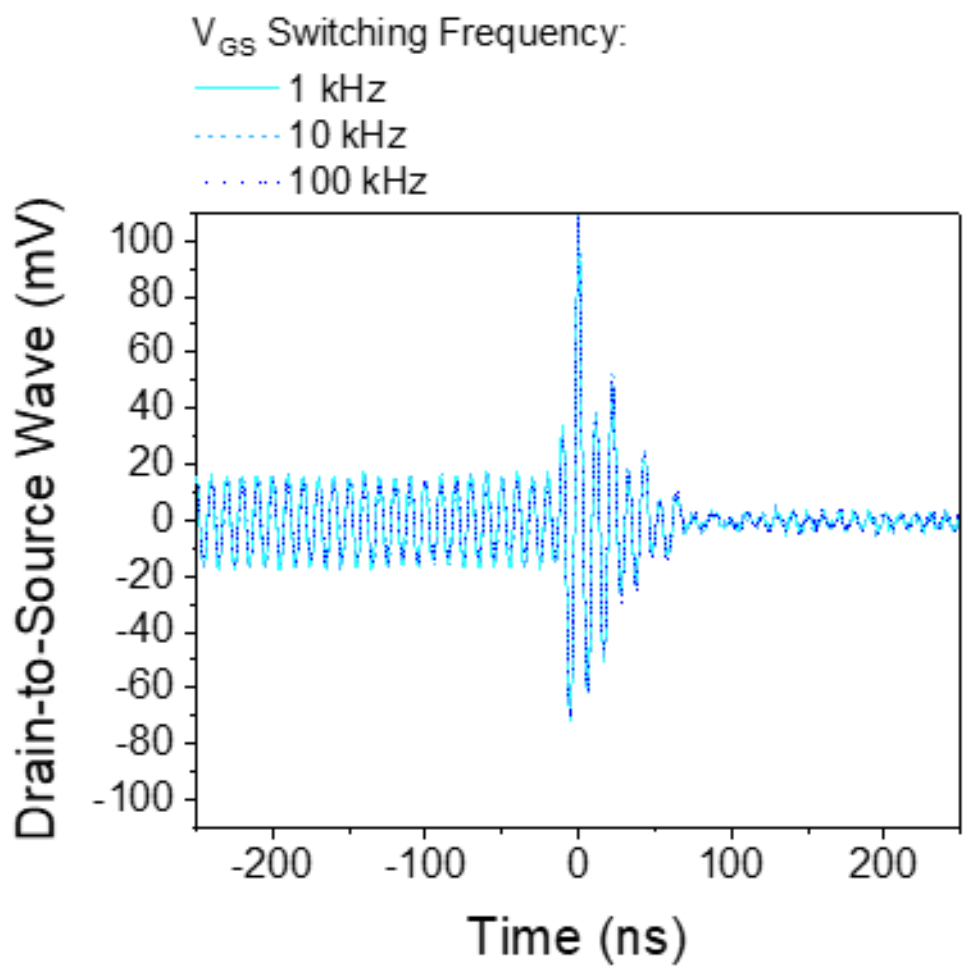

**Figure S16: Drain-to-source wave voltage (f=100 MHz) measured during the ON-to-OFF state transient with different JoFET gate control switching frequencies.**

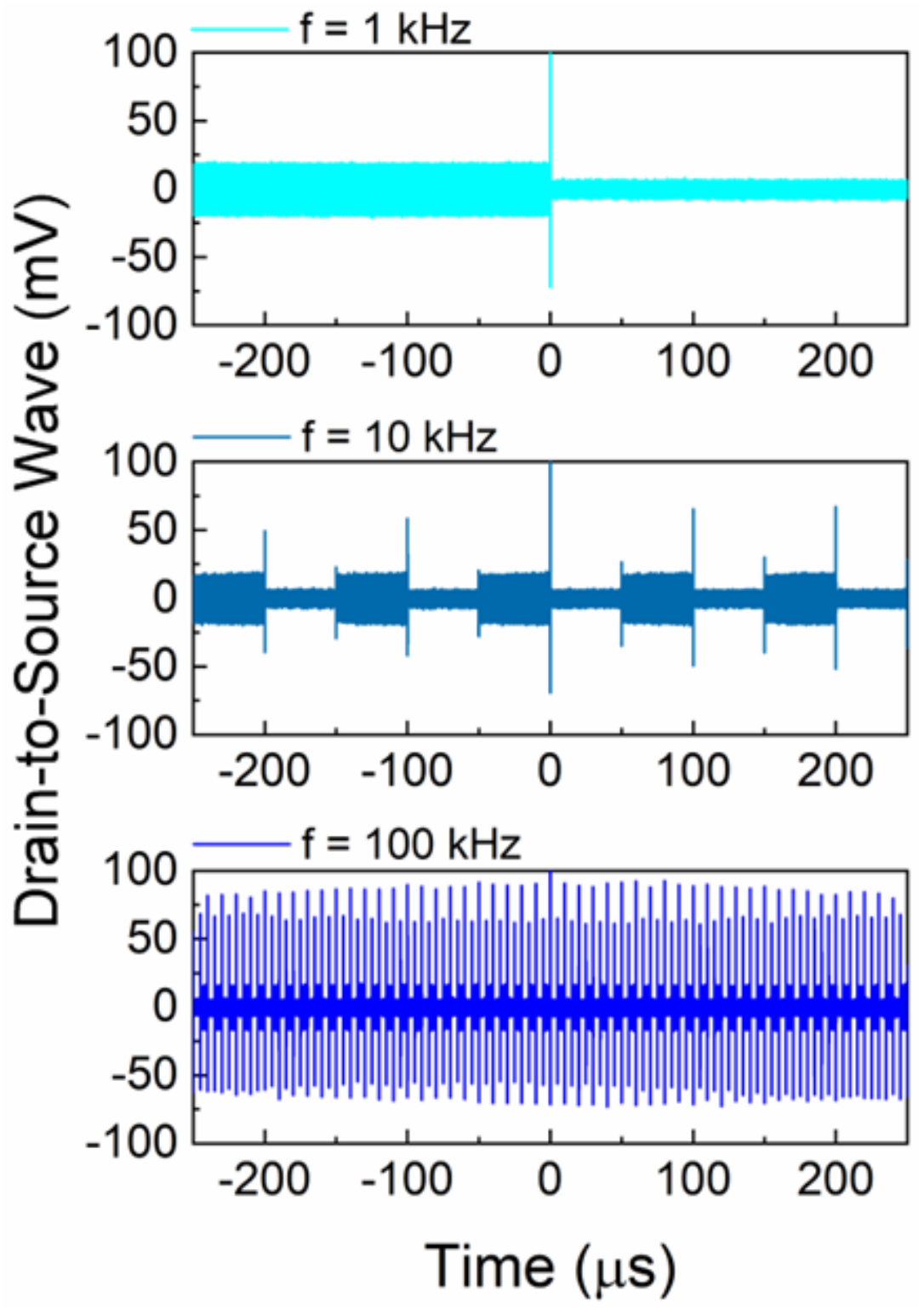

**Figure S17: Drain-to-source wave voltage (f=100 MHz) measured with different JoFET gate control switching frequencies.**

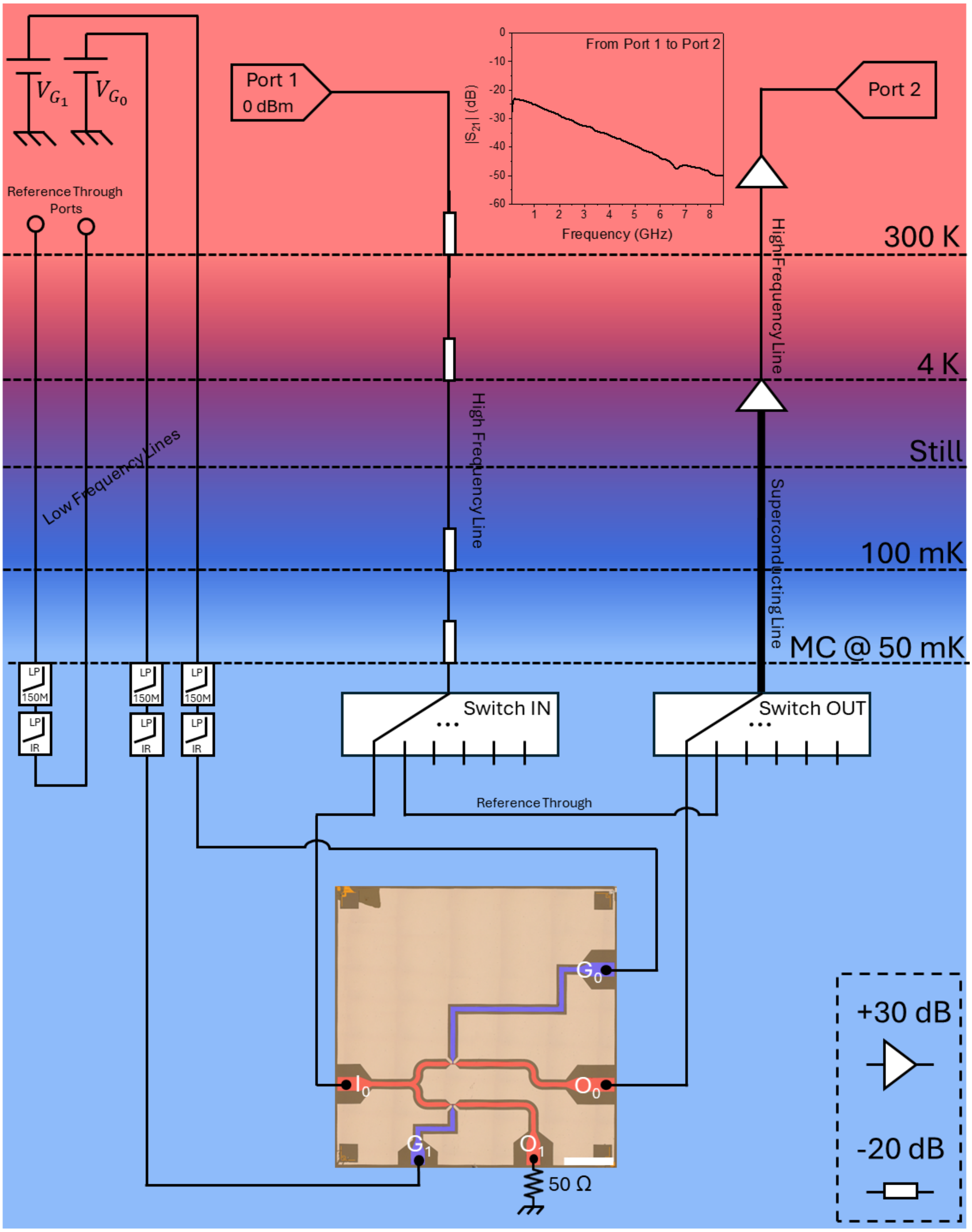

**Figure S18: VHF-UHF measurement setup used to characterize the frequency-resolved electrical behavior of the µwave superconducting 1I2O analog demultiplexer.** The input power ($P_{IN} = 0\ dBm$) supplied by the $Port_1$ was attenuated by -80 dB then provided to the demultiplexer input $I_0$. The line $O_0$ output power was amplified at 4 K (+30 dB) and 300 K (+30 dB) then provided to $Port_2$, while line $O_1$ was terminated with 50 Ω. The output power was measured changing the gate settings. Reference through ports were used to collect the trough calibration data. The system was operated up to a maximum operation frequency of 8.5 GHz at 50 mK.

## 3. Materials and Manufacturing Methods

### 3.1 Materials and Chemicals

GaAs wafers (2'' diameter, (100) orientation, $\rho=5.9\times10^7$ Ω×cm) were purchased from Wafer Technology LTD. Materials used for the Molecular Beam Epitaxy growth (Gallium 7N5, Aluminum 6N5, Indium 7N5, and Arsenic 7N5) were purchased from Azelis S.A.. Acetone (ACE, ULSI electric grade, MicroChemicals), 2-propanol (IPA, ULSI electric grade, MicroChemicals), S1805 G2 Positive Photoresist (S1805, Microposit, positive photoresist), AR-P 679.04 (AllResist, positive e-beam resist), MF319 Developer (MF319, Microposit), AR 600-56 Developer (AR 600-56, AllResist), AR600-71 (AllResist, remover for photo- and e-beam resist), Aluminum Etchant Type D (Transene), Phosphoric acid ($H_3PO_4$, Sigma Aldrich, semiconductor grade ≥85% in water), Hydrogen peroxide ($H_2O_2$, Carlo Erba Reagents, RSE-For electronic use-Stabilized, 30% in water), Nitrogen ($N_2$, 5.0, Nippon Gases) was provided by the Clean Room Facility of the National Enterprise for nanoScience and nanotechnology (NEST, Pisa, Italy). Diammonium sulfide ($(NH_4)_2S$, Carlo Erba Reagents, 20% in water) was provided by the Chemical Lab Facility of the NEST. Sulfur pieces (S, Alfa Aesar, 99.999% pure) was purchased from Carlo Erba Reagents S.r.l. Aluminum pellets (99.999% pure) were purchased from Kurt J. Lesker Company. Aqueous solutions were prepared using deionized water (DIW, 15.0 MΩ×cm) filtered by Elix® (Merck Millipore) provided by the Clean Room Facility of the NEST.

### 3.2 InAsOI Heterostructure Growth via Molecular Beam Epitaxy

InAsOI was grown on semi-insulating GaAs (100) substrates using solid-source Molecular Beam Epitaxy (MBE, Compact 21 DZ, Riber). Starting from the GaAs substrate, the sequence of the layer structure includes a 200 nm-thick GaAs layer, a 200 nm-thick GaAs/$Al_{0.16}Ga_{0.84}As$ superlattice, a 200 nm-thick GaAs layer, a 1.250 μm-thick step-graded $In_XAl_{1-X}As$ metamorphic buffer (with X increasing from 0.15 to 0.81), a 400 nm-thick $In_{0.84}Al_{0.16}As$ overshoot layer, and a 100 nm-thick InAs layer. The GaAs layer and the GaAs/$Al_{0.16}Ga_{0.84}As$ superlattice below the $In_XAl_{1-X}As$ buffer layer are grown to planarize the starting GaAs surface and to reduce surface roughness caused by the oxide desorption process. Both are grown with a group V/III beam flux ratio of 6.

The metamorphic buffer consists of two regions with different misfit gradients. The first $In_XAl_{1-X}As$ region is composed of twelve 50 nm-thick layers with X ramping from 0.15 to 0.58. The second $In_XAl_{1-X}As$ region is composed of twelve 50 nm-thick layers with X ramping from 0.58 to 0.81. The Al flux was kept constant during the buffer layer growth, while the In flux was increased at each step without growth interruptions. At the end of the buffer, the overshoot layer was grown to increase the strain relaxation of the $In_XAl_{1-x}As$ metamorphic layer [40]. The As flux was adjusted during the growth of the metamorphic buffer and the overshoot layer to keep a constant group V/III beam flux ratio of 8. The InAs layer was growth with a group V/III beam flux ratio of 8 and a growth rate of 0.96 μm/h. The GaAs layer and GaAs/AlGaAs superlattice are grown at 600 °C ± 5 °C. The metamorphic buffer and the overshoot layers were grown at optimized substrate temperatures of 320 °C ± 5 °C. The InAs epilayer was grown at 480 ± 5 °C.

In the following, all wetting steps were performed in cleaned glass Beckers using stainless steel tweezers provided with carbon tips. Teflon-coated tweezers were used for all the steps requiring acid or base solutions. The Julabo TW2 was used to heat the solution to a specific temperature.

### 3.3 InAsOI Josephson Field Effect Transistor Fabrication

*Superconductor Deposition on InAsOI*

InAsOI substrates were cut into square samples (7×7 mm×mm) and sonicated in ACE and IPA for 5 min to remove GaAs dusts. The air-exposed InAs surface was etched from native InAs oxide ($InAsO_X$) and passivated with S-termination by dipping the InAsOI samples in a $(NH_4)_2S_X$ solution (290 mM $(NH_4)_2S$ and 300 mM S in DIW) at 45°C for 90 s. The S-terminated InAsOI samples were then rinsed twice in DIW for 30 s and immediately loaded (~ 90 s exposure time in the air) into the load-lock vacuum chamber of an e-beam evaporator (acceleration voltage 7 kV). Samples were transferred into the deposition chamber, where a 100-nm-thick Al layer was deposited at a rate of 2 A/s at a residual chamber pressure of 1E-8 ÷ 5E-9 Torr.

*InAsOI MESA Fabrication*

After Al deposition, a layer of S1805 positive photoresist was spin-coated at 5000 RPM for 60 s (spin coating acceleration of 5000 RPM/s) and soft-baked at 115 °C for 60 s. The resist was then exposed via direct writing UV lithography (UVL, DMO, ML3 laser writer, λ=385 nm) with a dose of 60 $mJcm^{-2}$, resolution of 0.6 μm, high exposure quality, and laser-assisted real-time focus correction to define the MESA geometry. Unless otherwise stated, all the rinsing steps were performed at room temperature (RT, 21 °C). The UV-exposed samples were developed in MF319 for 45 s with soft agitation to remove exposed photoresist, then rinsed in DIW for 30 s to stop the development and dried with $N_2$. The exposed Al layer was removed by dipping the sample in Al Etchant Type D at 40 °C for 65 seconds with soft agitation, then rinsed in DIW for 30 seconds to stop the etching and dried with $N_2$. The exposed InAs epilayer was etched by dipping the samples in a $H_3PO_4$:$H_2O_2$ solution (348 mM $H_3PO_4$, 305 mM $H_2O_2$ in DIW) for 60 s with soft agitation, then rinsed in DIW for 30 s to stop the etching and dried with $N_2$. Eventually, the photoresist was removed by rinsing the InAsOI samples in ACE at 60 °C for 5 minutes and IPA for 60 s, then dried with $N_2$. At the end of this step, the width (W) of the JoFET was set to 5 μm.

*Markers Deposition for Aligned Steps*

After MESA fabrication, a layer of AR-P 679.04 positive e-beam resist was spin-coated at 4000 RPM for 60 s (spin coating acceleration of 10000 RPM/s) and soft-baked at 160 °C for 60 s. The resist was then exposed via UVL-marker-aligned e-beam lithography (EBL, ZEISS, Ultra Plus) with a dose of 350 $\mu Ccm^{-2}$, voltage acceleration of 30 kV, aperture of 7.5 or 120 μm, line step size of 1 nm or 200 nm, to define EBL-markers for the next alignment steps. The electron-exposed samples were developed in AR 600-56 for 90 s with soft agitation to remove the exposed e-beam resist, then rinsed in IPA for 30 s to stop the development and dried with $N_2$. The samples were loaded into a thermal evaporator (Sistec prototype) where a 10/50-nm-thick Ti/Au bilayer was deposited at a rate of 1 A/s at a residual chamber pressure of 2E-6 mbar. The deposited film was lifted-off in ACE at 70 °C for 5 min with strong agitation, then rinsed in IPA for 60 s, and dried with $N_2$.

*Aligned Josephson Junction Fabrication*

After EBL marker definition, a layer of AR-P 679.04 positive e-beam resist was spin-coated at 4000 RPM for 60 s (spin coating acceleration of 10000 RPM/s) and soft-baked at 160 °C for 60 s. The resist was then exposed via marker-aligned EBL (ZEISS, Ultra Plus) with a dose of 350 $\mu Ccm^{-2}$, voltage acceleration of 30 kV, aperture of 7.5 μm, and line step size of 1 nm to define the Josephson junction length ($L_{JJ}$). The electron-exposed samples were developed in AR 600-56 for 90 s with soft agitation to remove the exposed e-beam resist, then rinsed in IPA for 30 s to stop the development and dried with $N_2$. Subsequently, the exposed Al layer was removed by dipping the sample in Al Etchant Type D at 40 °C for 65 seconds with soft agitation, then rinsed in DIW for 30 seconds to stop the etching and dried with $N_2$. Eventually, the e-beam resist was removed by rinsing the InAsOI samples in ACE at 70 °C for 5 min, IPA for 60 s, and dried with $N_2$. At the end of this step, we achieved $L_{JJ}$ ranging from 500 to 1250 nm.

*Gate Insulator Deposition*

Samples were loaded into the vacuum chamber of an Atomic Layer Deposition system (ALD, Oxford Instruments, OpAL) where the gate insulator was uniformly deposited at a temperature of 130 °C. Tetrakis(ethylmethylamino)hafnium (TEMAH) and $H_2O$ were used as $HfO_2$ precursors, while Ar was used as carrier gas. Ar bubbling was also involved to increase the volatility of TEMAH. After reaching a base pressure of ~3-4 mTorr, the chamber pressure was increased to ~350 mTorr injecting Ar. The deposition process follows 4 steps: (i) TEMAH dose, (ii) TEMAH purge, (iii) $H_2O$ dose, and (iv) $H_2O$ purge:

(i) TEMAH dose: TEMAH valve on; Ar bubbler: 250 sccm; Ar purge: 10 sccm; step duration: 0.9 s.

(ii) TEMAH: purge: TEMAH valve off; Ar purge: 250 sccm; step duration: 110 s.

(iii) $H_2O$ dose: $H_2O$ valve on; Ar purge: 10 sccm; step duration: 0.03 s.

(iv) $H_2O$ purge: $H_2O$ valve off; Ar purge: 250 sccm; step duration: 90 s.

We performed 250 ALD cycles to achieve a total insulator thickness of ~31 nm [41]. Eventually, the chamber was pumped to reach a base pressure of ~3-4 mTorr and then vented using $N_2$. The entire process takes ~16 h.

*Aligned Metal Gate Deposition*

After gate insulator deposition, a layer of AR-P 679.04 positive e-beam resist was spin-coated at 4000 RPM for 60 s (spin coating acceleration of 10000 RPM/s) and soft-baked at 160 °C for 60 s. The resist was then exposed via marker-aligned EBL (ZEISS, Ultra Plus) with a dose of 350 $\mu Ccm^{-2}$, voltage acceleration of 30 kV, aperture of 7.5 or 120 μm, line step size of 1 or 200 nm, to define the JoFET gate length ($L_G$). The electron-exposed samples were developed in AR 600-56 for 90 s with soft agitation to remove the exposed e-beam resist, then rinsed in IPA for 30 s to stop the development and dried with $N_2$. Samples were mounted into an e-beam evaporator (acceleration voltage 7 kV), where a 10/100-nm-thick Ti/Al bilayer was deposited at a rate of 0.5/2 A/s with a tilt angle of 53° at a residual chamber pressure of 1E-8 ÷ 1E-9 Torr. The deposited film was lifted-off in AR600-71 at 80 °C for 5 min with strong agitation, then rinsed in IPA for 30 s, and dried with $N_2$.

### 3.4 InAsOI Superconducting 1I8O Demultiplexer Fabrication

*Superconductor Deposition on InAsOI*

This step was performed in accordance with what is reported in Section 3.3 "InAsOI Josephson Field Effect Transistor Fabrication: Superconductor Deposition on InAsOI".

*InAsOI MESA Fabrication*

This step was performed in accordance with what is reported in Section 3.3 "InAsOI Josephson Field Effect Transistor Fabrication: InAsOI MESA Fabrication".

*Aligned Josephson Junction Fabrication*

This step was performed in accordance with what is reported in Section 3.3 "InAsOI Josephson Field Effect Transistor Fabrication: Aligned Josephson Junction Fabrication" using UVL markers defined in the previous step instead of EBL markers.

*Gate Insulator Deposition*

This step was performed in accordance with what is reported in Section 3.3 "InAsOI Josephson Field Effect Transistor Fabrication: Gate Insulator Deposition".

*Aligned Metal Gate Deposition*

After the insulator deposition, samples were removed from the ALD system and heated up to 170 °C for 180 s to desorb adsorbed water molecules from air. After that, a LOR3A/S1805 bilayer was deposited. LOR3A was spin-coated at 4000 RPM for 60 s (spin coating acceleration of 4000 RPM/s) and soft-baked at 170 °C for 180 s. The LOR3A spin coating and soft bake procedure was performed 2 times to increase the LOR3A thickness. Then, S1805 positive photoresist was spin-coated at 5000 RPM for 60 s (spin coating acceleration of 5000 RPM/s) and soft-baked at 115 °C for 60 s. The resist was exposed via direct writing UV lithography (UVL, DMO, ML3 laser writer, $\lambda$=385 nm) with a dose of 60 $mJcm^{-2}$, resolution of 0.6 $\mu$m, high exposure quality, and no real-time focus correction to define gate pattern. The UV-exposed samples were developed in MF319 for 120 s with soft agitation to remove exposed S1805 and the underlying layer of LOR3A, then rinsed in DIW for 30 s to stop the development and dried with $N_2$.

Samples were mounted into an e-beam evaporator (acceleration voltage 7 kV), where a 10/250-nm-thick Ti/Al bilayer was deposited at a rate of 0.5/2 A/s at a residual chamber pressure of 1E-8 ÷ 1E-9 Torr. The evaporation was performed with the following tilt angles: (i) 0° for 10 nm Ti and 100 nm Al; (ii) -45° for 75 nm Al; (iii) 45° for 75 nm Al. The deposited film was lifted-off in AR600-71 at 80 °C for 5 min with strong agitation, then rinsed in IPA for 30 s, and dried with $N_2$.

Compared to the electron beam-aligned lithographic process involved in fabrication of the aligned metallic gates in InAsOI-based JoFETs, the use of an UV-aligned lithographic step to fabricate aligned metallic gates and traces in the superconducting demultiplexer allows to reduce the time required for the manufacturing step.

### 3.5 InAsOI μwave Superconducting 1I2O Demultiplexer Fabrication

*Superconductor Deposition on InAsOI*

This step was performed in accordance with what is reported in Section 3.3 "InAsOI Josephson Field Effect Transistor Fabrication: Superconductor Deposition on InAsOI".

*InAsOI MESA Fabrication*

This step was performed in accordance with what is reported in Section 3.3 "InAsOI Josephson Field Effect Transistor Fabrication: InAsOI MESA Fabrication".

*Aligned Josephson Junction Fabrication*

This step was performed in accordance with what is reported in Section 3.3 "InAsOI Josephson Field Effect Transistor Fabrication: Aligned Josephson Junction Fabrication" using UVL markers defined in the previous step instead of EBL markers.

*Gate Insulator Deposition*

This step was performed in accordance with what is reported in Section 3.3 "InAsOI Josephson Field Effect Transistor Fabrication: Gate Insulator Deposition".

*Aligned Metal Gate Deposition*

After gate insulator deposition, a layer of AR-P 679.04 positive e-beam resist was spin-coated at 4000 RPM for 60 s (spin coating acceleration of 10000 RPM/s) and soft-baked at 160 °C for 60 s. The resist was then exposed via marker-aligned EBL (ZEISS, Ultra Plus) with a dose of 350 $\mu Ccm^{-2}$, voltage acceleration of 30 kV, aperture of 7.5 or 120 μm, line step size of 1 or 200 nm, to define the JoFET gate length ($L_G$). The electron-exposed samples were developed in AR 600-56 for 90 s with soft agitation to remove the exposed e-beam resist, then rinsed in IPA for 30 s to stop the development and dried with $N_2$. Samples were mounted into an e-beam evaporator (acceleration voltage 7 kV), where a 10/250-nm-thick Ti/Al bilayer was deposited at a rate of 0.5/2 A/s at a residual chamber pressure of 1E-8 ÷ 1E-9 Torr. The evaporation was performed with the following tilt angles: (i) 0° for 10 nm Ti and 100 nm Al; (ii) -45° for 75 nm Al; (iii) 45° for 75 nm Al. The deposited film was lifted-off in AR600-71 at 80 °C for 5 min with strong agitation, then rinsed in IPA for 30 s, and dried with $N_2$.

## 3.6 InAsOI Hall Bars Fabrication

*Metal Deposition on InAsOI*

This step was performed in accordance with what is reported in Section 3.3 "InAsOI Josephson Field Effect Transistor Fabrication: Superconductor Deposition on InAsOI".

*Metal Pads Fabrication*

After Al deposition, a layer of S1805 positive photoresist was spin-coated at 5000 RPM for 60 s (spin coating acceleration of 5000 RPM/s) and soft-baked at 115 °C for 60 s. Then, the photoresist was exposed via direct writing UV lithography (DMO ML3 laser writer, λ=385 nm) with a dose of 60 $mJcm^{-2}$, resolution of 0.6 μm, high exposure quality, and laser-assisted real-time focus correction to define the Al geometry for fabrication of 6-terminals Hall Bars. Unless stated otherwise, all the rinsing steps were performed at room temperature (RT, 21 °C). The UV-exposed samples were developed in MF319 for 45 s with soft agitation to remove exposed photoresist, then rinsed in DIW for 30 s to stop the development and dried with $N_2$. The exposed Al layer was removed by dipping the sample in Al Etchant Type D at 40 °C for 65 seconds with soft agitation, then rinsed in DIW for 30 seconds to stop the etching and dried with $N_2$. Eventually, the photoresist was removed by rinsing the InAsOI samples in ACE at 60 °C for 5 min and IPA for 60 s, which was then dried with $N_2$.

*InAsOI MESA Fabrication*

A layer of S1805 positive photoresist was spin-coated at 5000 RPM for 60 s (spin coating acceleration of 5000 RPM/s) and soft-baked at 115 °C for 60 s. Then, the photoresist was exposed via a second marker-aligned direct writing UV lithography (DMO ML3 laser writer, λ=385 nm) with a dose of 60 $mJcm^{-2}$, resolution of 0.6 μm, high exposure quality, and no real-time focus correction, to define the InAs/InAlAs geometry for fabrication of Hall Bars. The UV-exposed samples were developed in MF319 for 45 s with soft agitation to remove exposed photoresist, then rinsed in DIW for 30 s to stop the development and dried with $N_2$. Then, the exposed InAs epilayer was etched by dipping the samples in a $H_3PO_4$:$H_2O_2$ solution (348 mM $H_3PO_4$, 305 mM $H_2O_2$ in DIW) for 60 s with soft agitation, then rinsed in DIW for 30 s to stop the etching and dried with $N_2$. Eventually, the photoresist was removed by rinsing the InAsOI samples in ACE at 60 °C for 5 min, IPA for 60 s, and drying with $N_2$. We fabricated 6-contact Hall bars with a width of 100 μm, source-to-drain length of 500 μm, and probe-to-probe length of 250 μm.

## 3.7 Sample Bonding via Wire Wedge Bonding

All the fabricated samples were provided with bonding pads ranging from 150 × 150 to 200 × 200 μm×μm and then used to connect the device with the chip carrier. Samples were glued using a small drop of AR-P 679.04, then left dry at RT for 1 hour on a 24-pin dual-in-line (DIL) chip carrier. Samples were bonded via wire wedge bonding (MP iBond5000 Wedge) using an Al/Si wire (1%, 25 μm wire diameter), leaving the user-bonder and the DIL chip carrier electrically connected to the ground.

# 4. Characterization Methods

## 4.1 Morphological Characterization

***via Scanning Electron Microscopy***

Top view morphological characterization of JJs was carried out via scanning electron microscopy (SEM, ZEISS Merlin) with 5 kV acceleration voltage, 178 pA filament current, back scattered electron relevator, at different magnifications (2.5k and 50k).

***via Optical Microscopy***

Optical microscopy (Leica, DM8000 M, provided with LEICA MC190 HD camera) was used to verify all the steps without photoresist. An optical microscope (Nikon, Eclipse ME600, provided with Nikon TV Lens C-0.6× and a UV filter) was used to evaluate all the steps involving the photoresist.

***via Atomic Force Microscopy***

Atomic force microscopy (AFM, Bruker, DIMENSION edge with ScanAsyst provided with an ASYLEC-01-R2 tip - silicon tip Ti/Ir coated, $f_0$=75 kHz, k=2.8 N/m - in tapping mode) was used to evaluate the profile of the fabricated JoFETs. All the AFM photos were processed using Gwyddion.

## 4.2 Electrical Characterization

### *4.2.1 Cryogenic Electrical Characterization of InAsOI Josephson Field Effect Transistors*

*DC Electrical Characterization*

Electrical characterization of JoFETs was carried out by measuring 4-wires V-I curves at 50 mK. Out-of-plane magnetic field ($B_\perp$) was used to maximize the switching current at 0 gate voltage, then maintained for all the characterization. The sample was mounted in contact with the mixing chamber (MC) plate of the Leiden CF-CS81-1400 cryostat. Electrical configuration of the measurement setup is shown in Figure S3. Source-drain current ($I_{DS}$) was injected applying an increasing DC voltage (Voltage Source, YOKOGAWA GS200) over an input series resistor (R= 1 MΩ) at least 100 times larger than the total resistance of the remaining measurement setup. The voltage drop across the probe contacts ($V_{DS}$) was amplified (Voltage Amplifier, DL Instruments 1201, Gain = 10k, High pass filter = DC, Low pass filter = 100 Hz) and read (Multimeter, Agilent, 34410A, NPLC = 2). The gate-source voltage ($V_{GS}$) was changed between 0 and -4.5 V (SMU, Keithley, 2400).

The switching current ($I_S$) was estimated as the last applied current in the V-I curve before reading a voltage drop different from the noise floor, while the normal state resistance ($R_N$) was evaluated as the angular coefficient of the V-I linear best fitting curve for $I>I_S$.

Gate current leakage was evaluated measuring the gate I-V curve (with source terminal grounded and the other terminals left open) upon application of an increasing voltage (SMU, Keithley, 2400, absolute voltage step = 100 mV, step delay = 1 s) and collecting the flowing current (TIA, FEMTO, DDPCA-300, gain = $10^{10}$ V/A, rise time = fast; Multimeter, Agilent, 34410A, NPLC = 2).

*Real-Time AC VHF Electrical Characterization*

Real-time VHF AC electrical characterization of JoFETs was carried out using a lock-in amplifier (Lock-in, Zurich Instruments, UHF Lock-in Amplifier 600 MHz, 1.8 GSa/s) at 100 mK. The output port of the lock-in (Port1, power 0 dBm, frequency 100 MHz) was attenuated by -20 dB at 300 K, -20 dB at the 4K plate, -20 dB at the 100 mK plate, and -20 dB at the mixing chamber (MC) plate, then low-pass filtered (cut-off frequency of 12 GHz) and IR-filtered with an Ecosorb filter, and provided as input power of the JoFET to the drain terminal. Then, the signal line connected to the source terminal was attenuated amplified by +30 dB at 4 K plate (Low Noise Factory, LNF-LNC03_14A), +30 dB at 300 K (MITEQ, LNA-30-00101200-42-20P), and provided to the input port (Port2, averages = 50) of the lock-in. A wave function generator (Agilent 33220A) was used to provide the control switching signal to the gate terminal of the JoFET. The output port of the signal generator ($V_{DC}$ = -2.25 V, $V_{PP}$ = 4.5 V, frequency from 1 to 100 kHz) was then low-pass filtered (cut-off frequency of 150 MHz) and IR-filtered with an Ecosorb filter, then provided as control signal of the JoFET to the gate terminal. We collected the voltage measured by the lock-in amplifier in Scope mode at different switching control frequencies.

#### *4.2.2 Cryogenic Electrical Characterization of the Superconducting 1I8O Analog Demultiplexer*

*DC Electrical Characterization*

DC electrical characterization of the signal lines was carried out by measuring 4-wires V-I curves at 50 mK. The sample was mounted in contact with the mixing chamber (MC) plate of the Oxford Triton 200. Electrical configuration of the measurement setup is shown in Figure S7. Signal line input current ($I_{IN}$) was injected into the $I_0$ terminal applying an increasing DC voltage (Voltage Source, YOKOGAWA GS200) over an input series resistor (R= 1 MΩ) at least 100 times larger than the total resistance of the remaining measurement setup. The voltage drop across the probe contacts of the signal line ($V_{Line}$) was amplified (Voltage Amplifier, DL Instruments 1201, Gain = 10k, High pass filter = DC, Low pass filter = 100 Hz) and read (Multimeter, Agilent, 34410A, NPLC = 1). Gate control voltages ($V_{Gij}$) were changed between 0 and -4.5 V (SMU, Keithley, 2400). Terminals of other lines were left open during the measurement.

Each signal line features 3 different $I_S$ of the 3 JoFETs along the path. $I_S$ were estimated as the last applied current in the V-I curve before reading a step in the voltage drop, while the signal line resistance (R) was evaluated as the angular coefficient of the V-I linear best fitting curve for input currents larger than the third switching current detected.

Gate current leakage was evaluated measuring the gate I-V curve (with $I_0$ terminal grounded and the other terminals left open) upon application of an increasing voltage (SMU, Keithley, 2400, absolute voltage step = 100 mV, step delay = 1 s) and collecting the flowing current (TIA, DL, 1211, gain=$10^{10}$ V/A, rise time = 0.3 s; Multimeter, Agilent, 34410A, NPLC = 2).

*Real-Time Electrical Characterization*

Real-time electrical characterization of the signal lines $I_0O_4$ and $I_0O_6$ was carried out by measuring 4-wires V-I curves at 50 mK. Electrical configuration of the measurement setup is shown in Figure S10. Signal line input current ($I_{IN}$) of ~1.3 µA was injected into the $I_0$ terminal applying an DC voltage of 1.3 V (Voltage Source, YOKOGAWA GS200) over an input series resistor (R= 1 MΩ) at least 100 times larger than the total resistance of the remaining measurement setup. Voltage drops across the probe contacts of each signal line ($V_{Line}$) was amplified (Voltage Amplifier, DL Instruments 1201, Gain = 500, High pass filter = DC, Low pass filter = MAX) and real-time collected (Oscilloscope, Tektronix, TDS2024B, averages = 128). The current flowing in each signal line ($I_{Line}$) was also measured (TIA, DL, 1211, gain=$10^5$ V/A, rise time = min; Oscilloscope, Tektronix, TDS2024B, averages = 128). Gate control voltages ($V_{Gij}$) were changed between 0 and -4.5 V (SMU, Keithley, 2400). Terminals of other lines were left open during the measurement. The cryostat's wires account as a load resistance of ~200 Ω.

Real-time VLF AC electrical characterization of the signal line $I_0O_4$ was carried out by measuring 4-wires V-I curves at 50 mK. A signal line input current ($I_{IN}$) with 1 kHz frequency and peak-to-peak values from 2 to 16 μA was injected into the $I_0$ terminal applying an AC voltage (Wavefunction Generator, Agilent, 33220A) over an input series resistor (R= 1 MΩ) at least 100 times larger than the total resistance of the remaining measurement setup. Voltage drops across the probe contacts of the signal line $I_0O_4$ was amplified (Voltage Amplifier, DL Instruments 1201, Gain = 500, High pass filter = DC, Low pass filter = MAX) and real-time collected (Oscilloscope, Tektronix, TDS2024B, averages = 128). Gate control terminals and other signal line terminals were left open during the measurement.

*VLF-VHF AC Electrical Characterization*

VLF-HF AC electrical characterization of signal lines $I_0O_4$ and $I_0O_6$ was carried using a lock-in amplifier (Lock-in, Zurich Instruments, UHF Lock-in Amplifier 600 MHz, 1.8 GSa/s) at 50 and 1650 mK. Electrical configuration of the measurement setup is shown in Figure S11. The output port of the lock-in (Port1, power -20 dBm, frequency 10 kHz - 10 MHz) was attenuated by -50 dB at 300 K and -10 dB on the mixing chamber (MC), then provided as input power of the demultiplexer to the $I_0$ terminal. Then, the signal line output power was attenuated by -10 dB on the MC, amplified by +80 dB at 300 K (FEMTO, HSA-X-I-2-40), and provided to input port (Port2, averages = 25) of the lock-in. When measuring the signal line $I_0O_4$, the $I_0O_6$ was terminated to 50 Ω, and vice versa. To retrieve the "through" calibration file of the measurement system, two cryostat ports were connected with a feed-through connector on the MC and provided with the same attenuation and amplification of the measurement lines. For both the signal line ($r_{line+sys}$) and calibration ($r_{sys}$) measurements, we collected the magnitude of the voltage measured by the lock-in (r). The forward power gain scattering parameter, i.e., the $|S_{21}|$ (dB), of the signal line was obtained as $|S_{21}| = 20\log(r_{line+sys}/r_{sys})$. Gate control voltages ($V_{Gij}$) were changed between 0 and -4.5 V (SMU, Keithley, 2400).

MF-VHF AC electrical characterization of signal lines $I_0O_4$ and $I_0O_6$ was carried using a vector network analyzer (VNA, PicoTech, PicoVNA 108) at 50 mK. Electrical configuration of the measurement setup is shown in Figure S11. The Port1 of the VNA (power -20 dBm, frequency 300 kHz - 100 MHz) was attenuated by -50 dB at 300 K and -10 dB on the mixing chamber (MC), then provided as input power of the demultiplexer to the $I_0$ terminal. Then, the signal line output power was attenuated by -10 dB on the MC, amplified by +80 dB at 300 K (FEMTO, HSA-X-I-2-40), and provided to the Port2 (bandwidth = 1 kHz) of the VNA. When measuring the signal line $I_0O_4$, the $I_0O_6$ was terminated to 50 Ω, and vice versa. To retrieve the "through" calibration file of the measurement system, two cryostat ports were connected with a feed-through connector on the MC and provided with the same attenuation and amplification of the measurement lines. For both the signal line ($|S_{21}|_{line+sys}$) and calibration ($|S_{21}|_{sys}$) measurements, we collected the $|S_{21}|$ parameter. The $|S_{21}|$ (dB) of the signal line was obtained as $|S_{21}|_{line} = |S_{21}|_{line+sys} - |S_{21}|_{sys}$. Gate control voltages ($V_{Gij}$) were changed between 0 and -4.5 V (SMU, Keithley, 2400).

#### *4.2.3 Cryogenic Electrical Characterization of the μwave Superconducting 1I2O Analog Demultiplexer*

*DC Electrical Characterization*

DC electrical characterization of the signal lines was carried out by measuring 4-wires V-I curves at 50 mK. The sample was mounted in contact with the mixing chamber (MC) plate of the Oxford Triton 200. Electrical configuration of the measurement setup is similar to that shown in Figure S7. Signal line input current ($I_{IN}$) was injected into the $I_0$ terminal applying an increasing DC voltage (Voltage Source, YOKOGAWA GS200) over an input series resistor (R= 1 MΩ) at least 100 times larger than the total resistance of the remaining measurement setup. The voltage drop across the probe contacts of the signal line ($V_{Line}$) was amplified (Voltage Amplifier, DL Instruments 1201, Gain = 10k, High pass filter = DC, Low pass filter = 100 Hz) and read (Multimeter, Agilent, 34410A, NPLC = 1). Gate control voltages ($V_{Gij}$) were changed between 0 and -4.5 V (SMU, Keithley, 2400). Terminals of other lines were left open during the measurement.

The switching current ($I_S$) was estimated as the last applied current in the V-I curve before reading a voltage drop different from the noise floor, while the normal state resistance ($R_N$) was evaluated as the angular coefficient of the V-I linear best fitting curve for $I>I_S$.

Gate current leakage was evaluated measuring the gate I-V curve (with $I_0$ terminal grounded and the other terminals left open) upon application of an increasing voltage (SMU, Keithley, 2400, absolute voltage step = 100 mV, step delay = 1 s) and collecting the flowing current (TIA, DL, 1211, gain=$10^{10}$ V/A, rise time = 0.3 s; Multimeter, Agilent, 34410A, NPLC = 2).

*VHF-UHF AC Electrical Characterization*

VHF-UHF AC electrical characterization of signal line $I_0O_0$ was carried using a vector network analyzer (VNA, PicoTech, PicoVNA 108) at 50 mK. Electrical configuration of the measurement setup is shown in Figure S18. The Port1 of the VNA (power 0 dBm, frequency 300 kHz - 8.5 GHz) was attenuated by -20 dB at 300 K, -20 dB at 4K, -20 dB at 100 mK, and -20 dB on the mixing chamber (MC), then provided as input power of the demultiplexer to the $I_0$ terminal. The output line $O_0$ power was amplified by +30 dB at 4 K (Low Noise Factory, LNF-LNC0.3_14A), +30 dB at 300 K (MITEQ, LNA-30-00101200-42-20P) and provided to the Port2 (bandwidth = 1 kHz, averages = 10) of the VNA. $O_6$ was terminated to 50 Ω at the MC stage. To retrieve the "through" calibration file of the measurement system, a reference "gold standard" through was mounted on the MC and provided with the same attenuation and amplification of the measurement lines thanks to the input and output cryogenic switches mounted at the MC stage. For both the signal line ($|S_{21}|_{line+sys}$) and calibration ($|S_{21}|_{sys}$) measurements, we collected the $|S_{21}|$ parameter. The $|S_{21}|$ of the signal line was obtained as $|S_{21}|_{line} = |S_{21}|_{line+sys} - |S_{21}|_{sys}$. Gate control voltages ($V_{Gi}$) were changed between 0 and -4.5 V (SMU, Keithley, 2400).

#### *4.2.4 Cryogenic and Room Temperature Electrical Characterization of InAsOI*

Hall measurements were performed on Hall Bars (HBs) with a standard lock-in-amplifier-based technique. The first lock-in amplifier oscillator voltage ($V_{OSC,RMS}$ = 1 V, f = 13.321 Hz) was applied across a series resistor (R=10 MΩ) at least 100 times larger than the total resistance of the remaining measurement setup to use the AC current ($I_{OSC, RMS}$). The AC current is injected into the sample's source contact. In contrast, the flowing current is measured with the drain contact, and the second lock-in amplifier measures the voltage drop's magnitude and phase across two other contacts. The contacts are manually switched to inject the current and measure the voltage. The Hall measurements were performed in three steps. In the first step, the resistivity ($\rho$) is calculated at zero magnetic fields using the formula $\rho = R_{xx} \times \frac{W \times t}{l}$, where $R_{xx}$ is the resistance measured between longitudinal (same-side) contacts of the HB for a current passing between source and drain, $W$ is the HB width, $t$ is the InAs thickness, and $l$ is the distance between probe contacts. In the second and third steps, the Hall voltage ($V_H$) is measured by applying positive and negative out-of-plane magnetic fields (|B| = 250 mT), respectively. $V_H$ is measured between opposite probe contacts of the HB for a current passing through the source and drain. The sheet electron density ($n_{2D}$) is calculated using the formula $n_{2D} = \frac{n_{2D}^1 + n_{2D}^2}{2}$, with $n_{2D}^i = \frac{I_{OSC,RMS} \times B_i}{q \times V_H}$, where $q$ is the fundamental charge, and $B_i$ is the magnitude of the applied out-of-plane magnetic field. The electron mobility ($\mu_n$) is calculated using the formula $\mu_n = \frac{\mu_n^1 + \mu_n^2}{2}$, with $\mu_n^i = \frac{1}{n_{2D}^i \times q \times R_{xx}}$.

## 5. Estimation of the JoFET Maximum Operational Frequency

The JoFET acts as an inductor with value $L_K$ and as a resistor with value $R_N$ in the ON and OFF state, respectively (see Figure 4b). The capacitors between gate and drain ($C_{GD}$) and gate and source ($C_{GS}$) terminals represent the major parasitic elements for both the operating states. Figure S19a reports the usage of the JoFET in the ON state where three simulation ports $P_1$, $P_2$, and $P_3$ were involved as the $V_{DS}$ source, $R_L$ load, and $V_{GS}$ source, respectively. Figure S19b shows the frequency-resolved simulations of the transmission scattering parameter magnitudes $|S_{21}|$ and $|S_{31}|$.

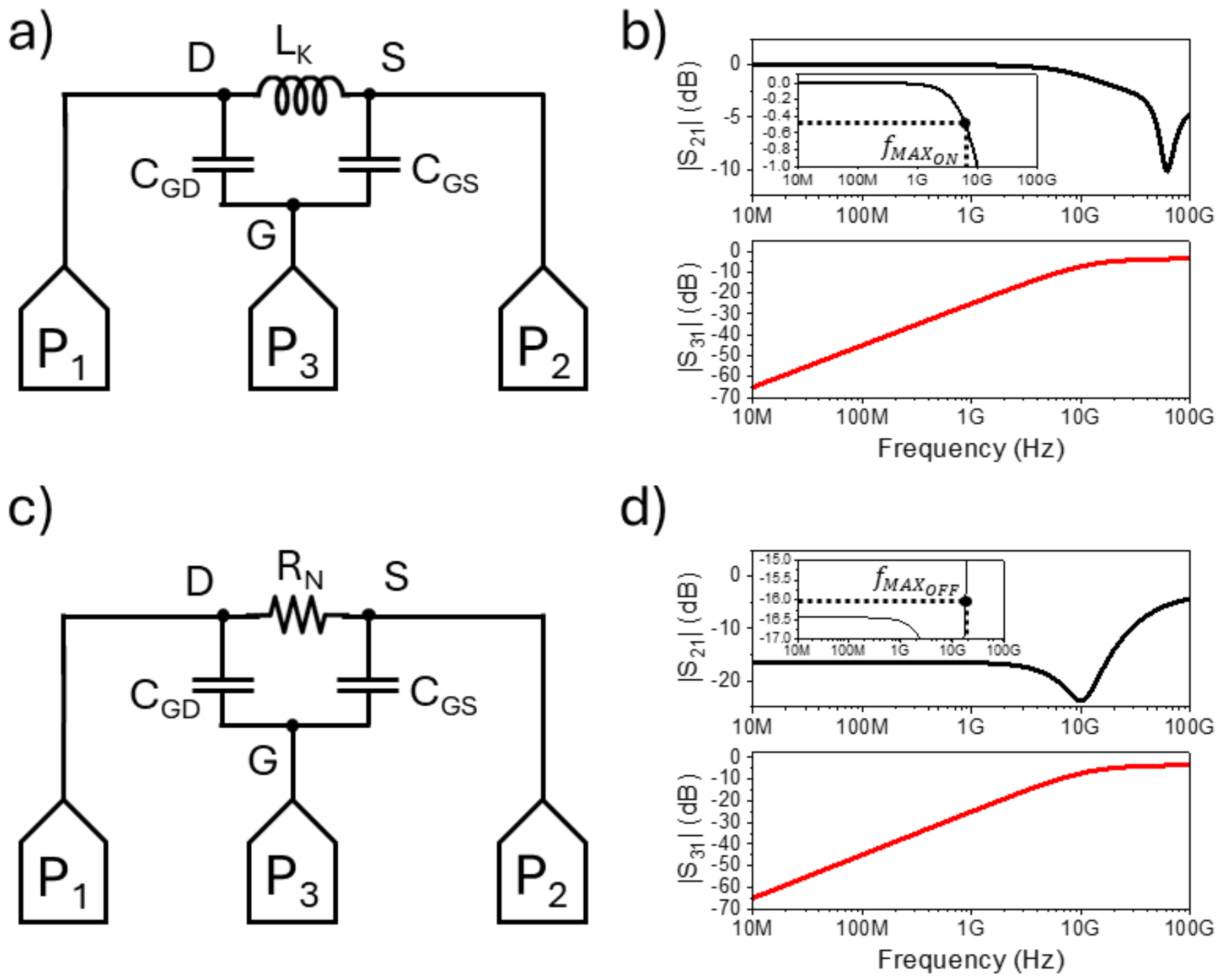

**Figure S19: Theoretical frequency-dependent operating behavior of the JoFET.** a) Usage of the JoFET in the ON state with three simulation ports involved. b) Frequency-resolved simulations of the transmission scattering parameters of the JoFET in the ON state. c) Usage of the JoFET in the OFF state with three simulation ports involved. d) Frequency-resolved simulations of the transmission scattering parameters of the JoFET in the OFF state. a,c) The internal series resistance of the three ports is 50 Ω. Circuital simulations were computed with LTSpice XVII.

In the DC ON state, the parasitic capacitors act as open circuits so that the power provided by the input port $P_1$ is totally provided to the port $P_2$, thanks to the negligible impedance of the kinetic inductance represented by the JoFET. In this case, an $|S_{21}| \sim 0$ dB (IL ~ 0 dB) is achieved. With the increase of the operating frequency, the gate capacitors' impedance decreases, linking port $P_1$ with the AC-grounded gate terminal represented by $P_3$. As observed from Figure S19b, $|S_{31}|$ increases with the frequency with the concomitant reduction of $|S_{21}|$. We defined the JoFET ON-state maximum operational frequency ($f_{MAX_{ON}}$) as the frequency at which $|S_{21}|$ decrease of 10% (-0.46 dB) compared to its DC value. This is a crucial point from which the gate capacitance significantly sinks the input power to the AC-grounded gate terminal, compared to that achieved at zero-frequency. Based on the values reported in Table S2, $f_{MAX_{ON}} = 6\ GHz$ was retrieved from the simulations.

Figure S19c reports the usage of the JoFET in the OFF state where three simulation ports $P_1$, $P_2$, and $P_3$ were involved as the $V_{DS}$ source, $R_L$ load, and $V_{GS}$ source, respectively. Figure S19d shows the frequency-resolved simulations of the transmission scattering parameter magnitudes $|S_{21}|$ and $|S_{31}|$. Also in this case, at low-frequency the parasitic capacitors act as open circuits so that the power provided by the input port $P_1$ is in large part reflected to the port itself due to the impedance mismatch between $R_N$ and the internal series resistance of $P_1$, namely 50 Ω. The remaining part flows into the resistive branch reaching the load port $P_2$, consequently achieving $|S_{21}| \sim -16.5$ dB (IL ~ 16.5 dB). Increasing the operating frequency, the gate capacitors start to shunt $R_N$ to both the AC-grounded gate terminal represented by $P_3$ and the load resistance represented by $P_2$. As observed from Figure S19d, $|S_{31}|$ increases with frequency because of the impedance reduction of the gate capacitors, finally shunting the resistive behavior of the JoFET. At the same time, $|S_{21}|$ first decreases due to the sink from the gate terminal, then increases from the $R_N$ shunt reducing the input-to-output isolation of the JoFET. We defined the JoFET OFF-state maximum operational frequency ($f_{MAX_{OFF}}$) as the frequency at which $|S_{21}|$ increase of 10% (+0.41 dB) compared to the DC value. This is a crucial point from which the gate capacitance significantly shunts the resistive behavior of the JoFET, compared to that achieved at zero-frequency. Based on the values reported in Table S2, $f_{MAX_{OFF}} = 17\ GHz$ was retrieved from the simulations. The overall maximum operating frequency ($f_{MAX}$) of the JoFET can be expressed as min $(f_{MAX_{ON}}, f_{MAX_{OFF}})$, resulting in $f_{MAX} = f_{MAX_{ON}} = 6\ GHz$.

## 6. Estimation of the Static Power Dissipation

The static power dissipation ($P_{STAT}$) of a JoFET, which is the power dissipated with no switching activities, is expressed as:

$$(1)\ P_{STAT} = Re\{Z_{DS}(f)\} \times I_{DS_{RMS}}^2$$

where $Re\{Z_{DS}(f)\}$ is the real part of the drain-to-source impedance at the frequency $f$, $I_{DS_{RMS}} = \sqrt{\frac{1}{T}\int_0^T |i_{DS}(t)|dt}$ is the RMS value of $I_{DS}$, and $T = \frac{1}{f}$ is the signal period. In the DC case, equation 1 can be rewritten as the most popular formula:

$$(2)\ P_{STAT} = R \times I^2$$

where $R$ is the DC drain-to-source resistance, and $I$ is the DC current.

$P_{STAT}$ is equal to 0 W when the JoFET is in the non-dissipative ON state with $I_{DS} < I_S$ (Figure 4b top), while $P_{STAT} = Re\{Z_{DS}(f)\} \times I_{DS_{RMS}}^2 = R_N \times I_{DS_{RMS}}^2$ when the JoFET is in the dissipative OFF state (Figure 4b bottom). The reported OFF-state dissipation is valid until the JoFET maximum operational frequency ($f_{MAX}$ ~ 700 MHz), calculated as reported in *Section 5: Estimation of the JoFET Maximum Operational Frequency*. Specifically, for $f > f_{MAX}$ the gate capacitance $Z_{C_G}$ theoretically starts to shunt the resistive behavior of the JoFET in the OFF state, consequently reducing the static power dissipated by the JoFET. In conclusion, $P_{STAT}$ remains unchanged from DC to $f_{MAX}$.

Figure S20a reports the representative case of study of a 1I2O superconducting demultiplexer (1I2O DeMUX) featuring 2 JoFETs, the easiest case of study for proposed architecture.

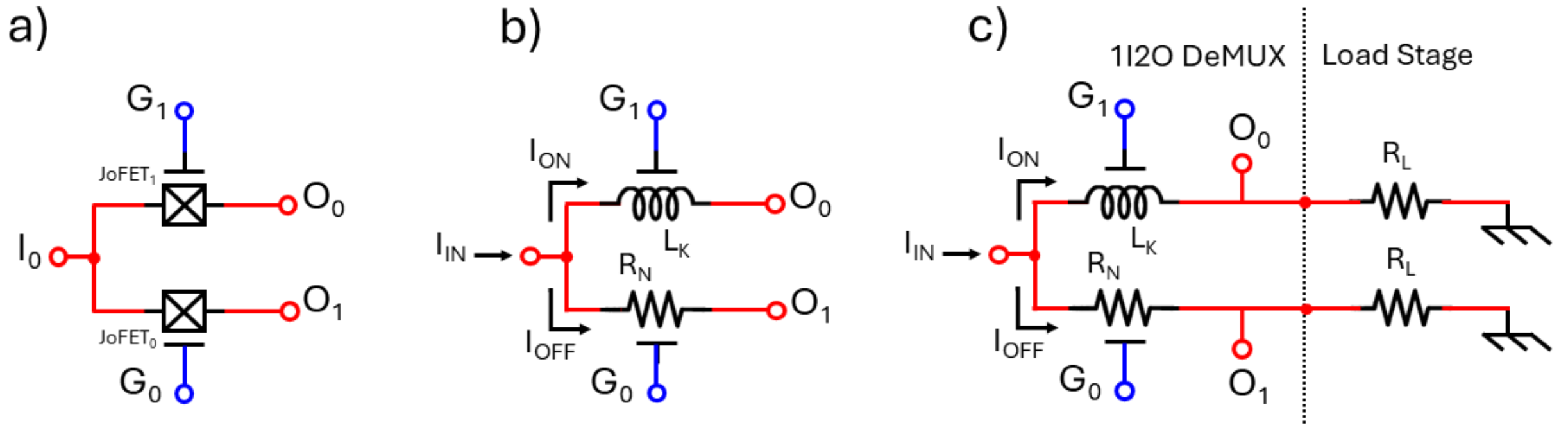

**Figure S20: 1I2O superconducting demultiplexer architecture.** a) Generic architecture of a 1I2O superconducting demultiplexer employing 2 JoFETs. b) Specific architecture of a 1I2O superconducting demultiplexer with the $JoFET_0$ in the OFF state and $JoFET_1$ in the ON state. c) Specific architecture of a 1I2O superconducting demultiplexer with a load stage.

Figure S20b depicts a specific operation of the proposed circuit with the $JoFET_0$ and $JoFET_1$ in OFF and ON states, respectively. The reported lumped element equivalent circuit is valid for $f<f_{MAX}$ ~ 700 MHz, which is in the range of measurements of this work. Based on this assumption, $P_{STAT}$ remains unchanged from DC to $f_{MAX}$. Figure S20c also reports the addition of the load stage with the load resistances $R_L$ to the 1I2O superconducting demultiplexer. Under the previous conditions, the static power dissipated by the 1I2O DeMUX is related to the part of the input current ($I_{IN}$) flowing in the dissipative OFF branch ($I_{OFF}$), namely:

$$(3)\ I_{OFF_{RMS}} = I_{IN_{RMS}} \times \frac{R_L}{R_N+2R_L}.$$

Consequently:

$$(4)\ P_{STAT} = R_N \times I_{OFF_{RMS}}^2 = R_N I_{IN_{RMS}}^2 \times \left(\frac{R_L}{R_N+2R_L}\right)^2.$$

We defined $P_{STAT_{MAX}} = R_N I_{IN_{RMS}}^2$ as the maximum static power which can be dissipated in the resistive OFF branch. By defining a new variable $\alpha = \frac{R_N}{R_L}$, equation 4 becomes:

$$(5)\ \frac{P_{STAT}}{P_{STAT_{MAX}}} = \left(\frac{1}{\alpha+2}\right)^2$$

Figure S21 reports the $I_{OFF}/I_{IN}$ ratio as well as the static power dissipated normalized to $P_{STAT_{MAX}}$ as a function of $\alpha$.

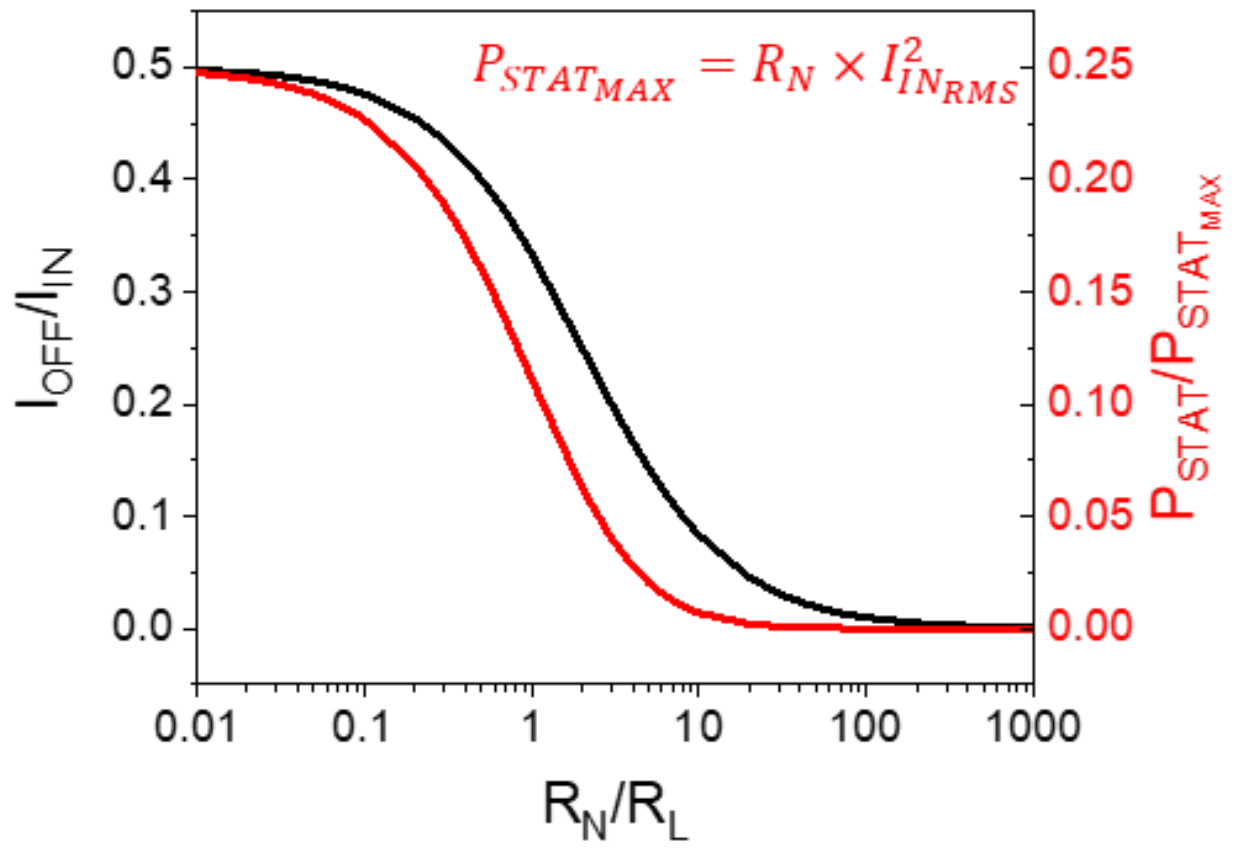

**Figure S21.** $\boldsymbol{I_{OFF}/I_{IN}}$ **and** $\boldsymbol{P_{STAT}/P_{STAT_{MAX}}}$ **as a function of** $\boldsymbol{R_N/R_L}$**.**

The higher the $R_N/R_L$ ratio, the lower the current flowing through the resistive OFF branch, and consequently the lower the static power dissipated. When $R_N<<R_L$, there is no impact of the JoFET's state on the supercurrent division so that the input current is perfectly divided into the two branches with a $P_{STAT}/P_{STAT_{MAX}} = 0.25$. In a different way, when $R_N=R_L$, the input supercurrent flowing in the dissipative branch is $I_0/3$, with $P_{STAT}/P_{STAT_{MAX}} = 0.1$. Eventually, when $R_N>>R_L$, the input supercurrent flows only in the non-dissipative branch with $P_{STAT}/P_{STAT_{MAX}} \sim 0$.

Referring to the case of study of:

- Figure 5a, where $R_L$ = 200 Ω, $R_N$ ~ 500 Ω, and $I_{IN}$ = 1.3 µA, we achieved $P_{STAT}/P_{STAT_{MAX}} \sim 0.05$ with $P_{STAT} \sim 40\ pW$ in the range 0-to-700 MHz;

- Figure 5c, where $R_L = 50\ \Omega$, $R_N \sim 500\ \Omega$, and $I_{IN} = 400\ nA_{rms}$, we achieved $P_{STAT}/P_{STAT_{MAX}} \sim 0.007$ with $P_{STAT} \sim 700\ fW$ in the range 0-to-700 MHz.

**7. Estimation of the Dynamic Power Dissipation**

The dynamic power dissipation ($P_{DYN}$) of a FET is expressed as:

$$(6)\ P_{DYN} = C_{eff} \times V_{DD}^2 \times f_{SW}$$

where $C_{eff}$ is the sum of the load and parasitic capacitances of the FET, $V_{DD}$ is the supply voltage, and $f_{SW}$ is the switching frequency.

For a JoFET, we consider the gate capacitance $C_G = C_{GS} + C_{GD}$ (Figure 4b) as the main capacitive element in the evaluation of $C_{eff}$, as well as $V_{DD}$ equals to the absolute maximum gate voltage applied ($V_{GS}$). Based on this hypothesis, equation 6 can be written as:

$$(7)\ P_{DYN} = C_G \times V_{GS}^2 \times f_{SW}$$

In this context, since both conventional FETs and JoFETs feature charge/discharge of the gate metal-oxide-semiconductor capacitor as the core of the switching activity, no significant improvements are related to the use of superconducting leads instead of conventional metals in reduction of the overall dynamic power dissipation. The improvement of the JoFET compared to the FET can be achieved in "where" the dynamic power is dissipated. Under certain operating conditions, the JoFET and the full chip can be in a non-dissipative state so that the dissipation of the dynamic power happens out from the die. This is a possible gain in using JoFET compared to FET to achieve negligible on-chip dynamic power dissipation and to move the dissipation of power away from the coldest plate.

When *n* JoFETs are taken into account, $P_{DYN}$ can be calculated as:

$$(8)\ P_{DYN} = n \times C_G \times V_{GS}^2 \times f_{SW}$$

Figure S22 reports $P_{DYN}$ estimated for the fabricated JoFETs and the 1I8O superconducting demultiplexer.

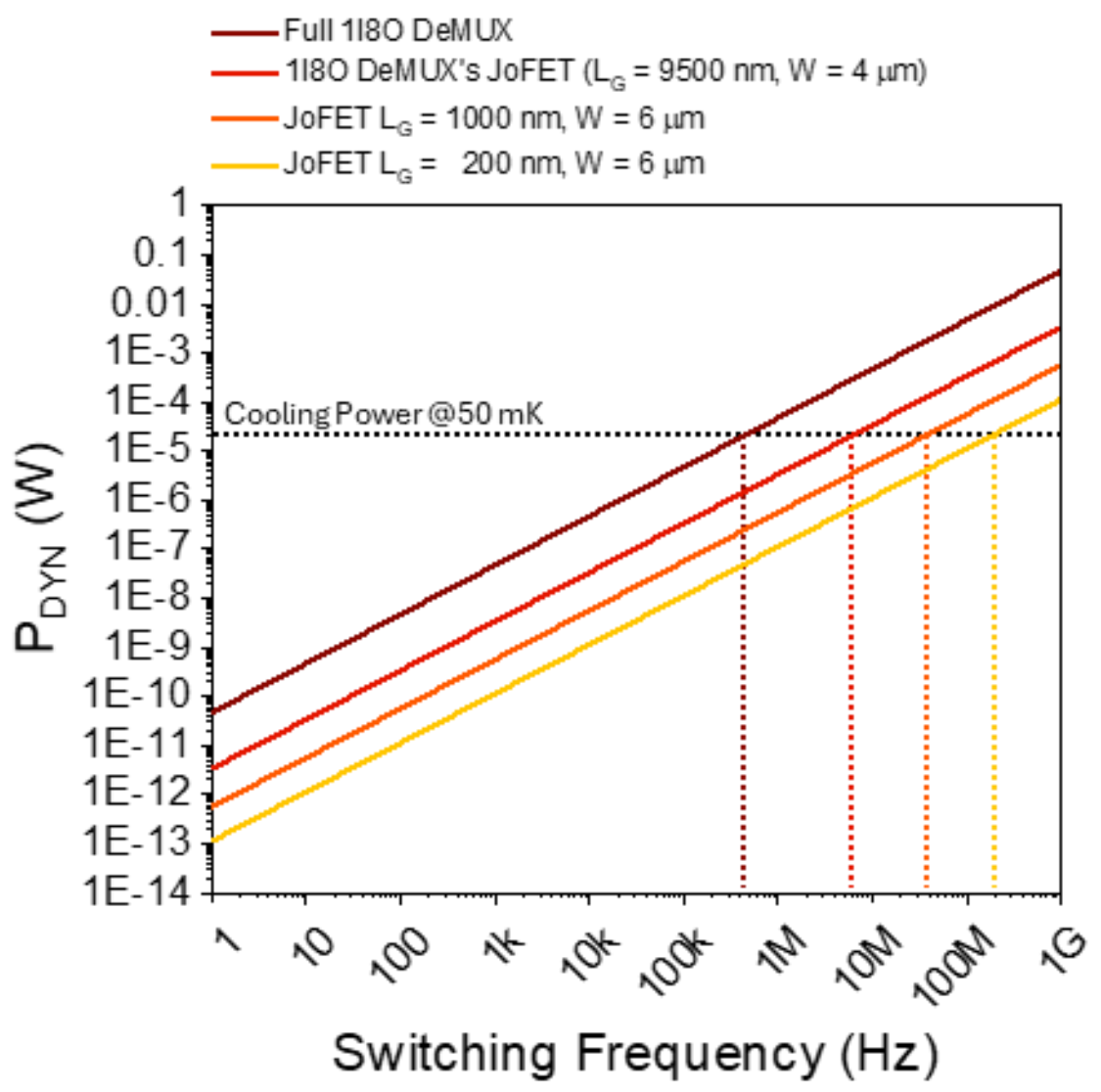

**Figure S22: Dynamic power dissipation vs. frequency of the fabricated JoFETs and the 1I8O superconducting demultiplexer.**

The black dashed line represents the cooling power of the cryostat we used throughout the experiments at a temperature of 50 mK, namely 16 mW. In the following, we have considered the full dissipation of $P_{DYN}$ at the coldest plate as worst-case scenario. The red solid line is the dynamic power dissipated by the typical JoFET integrated in the superconducting 1I8O demultiplexer ($L_G$ = 9500 nm, $W_{JJ}$ = 4 mm), while the brown solid line is the dynamic power dissipation of the entire chip (*n* = 14 JoFETs). The full demultiplexer die can operate until 400 kHz without affecting the temperature of the coldest plate, while the single JoFET can operate until 6 MHz. From equation 7, it is immediately apparent that the dynamic power dissipation of the JoFET can be reduced by reducing both the gate capacitance and the gate voltage. By reducing the size of the JoFET, consequently reducing the $W_{JJ}\times L_{JJ}$ area and the gate capacitance, the dynamic power dissipation decreases as reported for the orange and yellow solid lines, where JoFETs with $L_G$ = 1000 nm / $W_{JJ}$ = 6 mm and $L_G$ = 200 nm / $W_{JJ}$ = 6 mm were reported, respectively. Along the same path, the reduction of the gate voltage necessary to switch OFF the JoFET can quadratically reduce the dynamic power requirements.

## 8. Envisioning of Superconducting Time Division Multiplexing of Transmon Qubits

Superconducting Time Division Multiplexing (TDM) can be performed to read and control a specific transmon qubit of the Quantum Processing Unit (QPU) to reduce the number of cables required for the Quantum Computer (QC). In a transmon, microwave pulses are applied to the qubit drive line to modify the state of the transmon itself. A flux bias line is used to dynamically tune the operating frequency of the transmon. The recognition of the transmon's state, e.g., ground or excited state, can be performed via microwave transmission measurements made between the two readout ports. In this context, TDM can be applied to:

1. Perform microwave transmission measurements to a specific read-out line.
2. Route the non-dissipative supercurrent to a specific qubit flux bias line.
3. Route the microwave control pulses to a specific qubit drive line.

Usually, Frequency Division Multiplexing (FDM) of the operating frequencies of a series of transmons present on the same superconducting read-out line was applied to reduce the number of read-out cables. Each transmon operates at a specific frequency so that a selective read-out can be performed by FDM. This technique is the standard way used to increase the transmon integration in a QC. Superconducting TDM of several superconducting read-out lines embedding building blocks with arrays of qubits operating at the same frequencies represents a promising way to scale-up superconducting QCs.

At the same time, each transmon is provided with a superconducting flux bias line used to tune its operating frequency. Superconducting TDM promises to reduce the number of flux bias lines linked to the 300 K-located electronics.

Eventually, the superconducting TDM of control pulses to a specific qubit drive line can drastically reduce the number of control cables linked to the 300 K-located electronics. Unlike the previous applications, where a superconducting non-dissipative line is driven by the superconducting demultiplexer, in this case TDM signals need to be routed to high-impedance capacitive loads represented by the control ports of the transmons, from which a desired ON/OFF ratio is required. In conclusion, superconducting TDM of transmons requires a superconducting demultiplexer able to:

- Operates at a temperature of 50 mK.
- Operates with a signal frequency in the 4÷8 GHz range.
- Operates with control switching signals in the qubit coherence time range to run quantum algorithms (from a few ns to a few ms).
- Exhibits an ON-state insertion loss of 0 dB to preserve static power dissipation.
- Exhibits an ON/OFF ratio of 30÷40 dBs to route the control signal to the desired qubit.

To date, the proposed superconducting demultiplexers can operate at the mixing chamber base temperature of 50 mK with a suitable frequency operating range, but an increased port-to-port isolation and a reduced switching time need to be reached to perform TDM with transmon qubits. The reduced ON/OFF ratio of the fabricated chip is related to the ability of the InAsOI-based JoFET to reach a high normal state resistance value when gated in the OFF state. The reduced normal state resistance increase is related to the poor ability in depleting the InAs epilayer from charge carriers applying negative gate voltages. To overpass this limit, engineering the InAsOI platform to improve the gate tunable electrical properties is mandatory. As a walk-around of the InAsOI-related poor gate-tunable performance, JoFETs embedding $n$ ($n$>1) Josephson Junctions (JJs) can be manufactured to reach a higher JoFET's ON/OFF ratio. This allows to reach a normal state resistance in the OFF state $n$ times higher than that of the JoFET with a single JJ.
Table S4 reports the specs required for TDM of superconducting transmon qubits [2] and the performance achieved by the proposed superconducting demultiplexer and achievable in the near future by improving the JoFET and the entire chip.

**Table S4. Specs required for TDM of superconducting transmon qubits [2] and performance achieved by the proposed superconducting demultiplexer and achievable in the near future by improving the JoFET and the entire chip.**

| Specs | Required Value [2] | Achieved With This Work | Achievable With a Proper Design |
|---|---|---|---|
| *Operating Temperature* | 50 mK and lower | 50 mK and lower | 50 mK and lower |
| *Maximum Full Power Consumption* | <20 mW for the entire chip | Static Power Consumption for the single JoFET: 700 fW<br><br>Dynamic Power Consumption for the single JoFET: 20 mW @ 5 MHz | Static Power Consumption for the single JoFET: 100 fW<br><br>Dynamic Power Consumption for the single JoFET: 20 mW @ 200 MHz |
| *Frequency Operating Range* | 4÷8 GHz | 4 GHz | 4÷8 GHz |
| *Input Signal Dynamic Range* | -60÷-120 dBm | < -70 dBm | < -60 dBm |
| *Switching Time* | 1 ns | 100 ns | 1 ns |
| *Port-to-Port Isolation* | >30 dB | 17.5 dB | 30 dB |

**References**

[1] H. Takayanagi, J.B. Hansen, J. Nitta, Localization Effects on the Critical Current of a Superconductor—Normal-Metal—Superconductor Junction, Phys. Rev. Lett. 74 (1995) 162–165. https://doi.org/10.1103/PhysRevLett.74.162.

[2] A. Chrestin, R. Kürsten, K. Biedermann, T. Matsuyama, U. Merkt, Superconductor/semiconductor nanostructures on p-type InAs, Superlattices Microstruct. 25 (1999) 711–720. https://doi.org/10.1006/spmi.1999.0719.

[3] T. Akazaki, J. Nitta, H. Takayanagi, K. Arai, Superconducting junctions using a 2DEG in a strained InAs quantum well inserted into an InAlAs/InGaAs MD structure, IEEE Trans. Appiled Supercond. 5 (1995) 2887–2891. https://doi.org/10.1109/77.403195.

[4] H. Takayanagi, T. Akazaki, J. Nitta, T. Enoki, Superconducting Three-Terminal Devices Using an InAs-Based Two-Dimensional Electron Gas, Jpn. J. Appl. Phys. 34 (1995) 1391. https://doi.org/10.1143/JJAP.34.1391.

[5] H. Takayanagi, T.A. Tatsushi Akazaki, Submicron Gate-Fitted Superconducting Junction Using a Two-Dimensional Electron Gas, Jpn. J. Appl. Phys. 34 (1995) 6977. https://doi.org/10.1143/JJAP.34.6977.

[6] T. Akazaki, H. Takayanagi, J. Nitta, T. Enoki, A Josephson field effect transistor using an InAs-inserted-channel In0.52Al0.48As/In0.53Ga0.47As inverted modulation-doped structure, Appl. Phys. Lett. 68 (1996) 418–420. https://doi.org/10.1063/1.116704.

[7] T. Akazaki, J. Nitta, H. Takayanagi, T. Enoki, Superconducting transistors using InAs-inserted-channel InAlAs/InGaAs inverted HEMTs, Supercond. Sci. Technol. 9 (1996) A83–A86. https://doi.org/10.1088/0953-2048/9/4A/022.

[8] A. Richter, M. Koch, T. Matsuyama, U. Merkt, Transport properties of Nb/InAs(2DEG)/Nb Josephson field-effect transistors, Supercond. Sci. Technol. 12 (1999) 874–876. https://doi.org/10.1088/0953-2048/12/11/354.

[9] A. Richter, P. Baars, U. Merkt, Supercurrents in two-dimensional electron systems, Phys. E Low-Dimensional Syst. Nanostructures. 12 (2002) 911–917. https://doi.org/10.1016/S1386-9477(01)00408-8.

[10] J. Shabani, M. Kjaergaard, H.J. Suominen, Y. Kim, F. Nichele, K. Pakrouski, T. Stankevic, R.M. Lutchyn, P. Krogstrup, R. Feidenhans'l, S. Kraemer, C. Nayak, M. Troyer, C.M. Marcus, C.J. Palmstrøm, Two-dimensional epitaxial superconductor-semiconductor heterostructures: A platform for topological superconducting networks, Phys. Rev. B. 93 (2016) 155402. https://doi.org/10.1103/PhysRevB.93.155402.

[11] M. Kjaergaard, H.J. Suominen, M.P. Nowak, A.R. Akhmerov, J. Shabani, C.J. Palmstrøm, F. Nichele, C.M. Marcus, Transparent Semiconductor-Superconductor Interface and Induced Gap in an Epitaxial Heterostructure Josephson Junction, Phys. Rev. Appl. 7 (2017) 034029. https://doi.org/10.1103/PhysRevApplied.7.034029.

[12] W. Mayer, J. Yuan, K.S. Wickramasinghe, T. Nguyen, M.C. Dartiailh, J. Shabani, Superconducting proximity effect in epitaxial Al-InAs heterostructures, Appl. Phys. Lett. 114 (2019). https://doi.org/10.1063/1.5067363.

[13] J.S. Lee, B. Shojaei, M. Pendharkar, A.P. McFadden, Y. Kim, H.J. Suominen, M. Kjaergaard, F. Nichele, H. Zhang, C.M. Marcus, C.J. Palmstrøm, Transport Studies of Epi-Al/InAs Two-Dimensional Electron Gas Systems for Required Building-Blocks in Topological Superconductor Networks, Nano Lett. 19 (2019) 3083–3090. https://doi.org/10.1021/acs.nanolett.9b00494.

[14] S. Guiducci, M. Carrega, G. Biasiol, L. Sorba, F. Beltram, S. Heun, Toward Quantum Hall Effect in a Josephson Junction, Phys. Status Solidi – Rapid Res. Lett. 13 (2019) 1–5. https://doi.org/10.1002/pssr.201800222.

[15] S. Guiducci, M. Carrega, F. Taddei, G. Biasiol, H. Courtois, F. Beltram, S. Heun, Full electrostatic control of quantum interference in an extended trenched Josephson junction, Phys. Rev. B. 99 (2019) 235419. https://doi.org/10.1103/PhysRevB.99.235419.

[16] F. Barati, J.P. Thompson, M.C. Dartiailh, K. Sardashti, W. Mayer, J. Yuan, K. Wickramasinghe, K. Watanabe, T. Taniguchi, H. Churchill, J. Shabani, Tuning Supercurrent in Josephson Field-Effect Transistors Using h-BN Dielectric, Nano Lett. 21 (2021) 1915–1920. https://doi.org/10.1021/acs.nanolett.0c03183.

[17] B.H. Elfeky, N. Lotfizadeh, W.F. Schiela, W.M. Strickland, M. Dartiailh, K. Sardashti, M. Hatefipour, P. Yu, N. Pankratova, H. Lee, V.E. Manucharyan, J. Shabani, Local Control of Supercurrent Density in Epitaxial Planar Josephson Junctions, Nano Lett. 21 (2021) 8274–8280. https://doi.org/10.1021/acs.nanolett.1c02771.

[18] F. Wen, J. Yuan, K.S. Wickramasinghe, W. Mayer, J. Shabani, E. Tutuc, Epitaxial Al-InAs Heterostructures as Platform for Josephson Junction Field-Effect Transistor Logic Devices, IEEE Trans. Electron Devices. 68 (2021) 1524–1529. https://doi.org/10.1109/TED.2021.3057790.

[19] A. Hertel, L.O. Andersen, D.M.T. van Zanten, M. Eichinger, P. Scarlino, S. Yadav, J. Karthik, S. Gronin, G.C. Gardner, M.J. Manfra, C.M. Marcus, K.D. Petersson, Electrical Properties of Selective-Area-Grown Superconductor-Semiconductor Hybrid Structures on Silicon, Phys. Rev. Appl. 16 (2021) 044015. https://doi.org/10.1103/PhysRevApplied.16.044015.

[20] M. Sütő, T. Prok, P. Makk, M. Kirti, G. Biasiol, S. Csonka, E. Tóvári, Near-surface InAs

two-dimensional electron gas on a GaAs substrate: Characterization and superconducting proximity effect, Phys. Rev. B. 106 (2022) 235404. https://doi.org/10.1103/PhysRevB.106.235404.

[21] C. Ciaccia, R. Haller, A.C.C. Drachmann, T. Lindemann, M.J. Manfra, C. Schrade, C. Schönenberger, Gate-tunable Josephson diode in proximitized InAs supercurrent interferometers, Phys. Rev. Res. 5 (2023) 033131. https://doi.org/10.1103/PhysRevResearch.5.033131.

[22] M. Coraiola, A.E. Svetogorov, D.Z. Haxell, D. Sabonis, M. Hinderling, S.C. ten Kate, E. Cheah, F. Krizek, R. Schott, W. Wegscheider, J.C. Cuevas, W. Belzig, F. Nichele, Flux-Tunable Josephson Diode Effect in a Hybrid Four-Terminal Josephson Junction, ACS Nano. 18 (2024) 9221–9231. https://doi.org/10.1021/acsnano.4c01642.

[23] S. Yan, H. Su, D. Pan, W. Li, Z. Lyu, M. Chen, X. Wu, L. Lu, J. Zhao, J.-Y. Wang, H. Xu, Supercurrent, Multiple Andreev Reflections and Shapiro Steps in InAs Nanosheet Josephson Junctions, Nano Lett. 23 (2023) 6497–6503. https://doi.org/10.1021/acs.nanolett.3c01450.

[24] Y.-J. Doh, J.A. van Dam, A.L. Roest, E.P.A.M. Bakkers, L.P. Kouwenhoven, S. De Franceschi, Tunable Supercurrent Through Semiconductor Nanowires, Science (80-. ). 309 (2005) 272–275. https://doi.org/10.1126/science.1113523.

[25] H.Y. Günel, I.E. Batov, H. Hardtdegen, K. Sladek, A. Winden, K. Weis, G. Panaitov, D. Grützmacher, T. Schäpers, Supercurrent in Nb/InAs-nanowire/Nb Josephson junctions, J. Appl. Phys. 112 (2012). https://doi.org/10.1063/1.4745024.

[26] P. Perla, H.A. Fonseka, P. Zellekens, R. Deacon, Y. Han, J. Kölzer, T. Mörstedt, B. Bennemann, A. Espiari, K. Ishibashi, D. Grützmacher, A.M. Sanchez, M.I. Lepsa, T. Schäpers, Fully in situ Nb/InAs-nanowire Josephson junctions by selective-area growth and shadow evaporation, Nanoscale Adv. 3 (2021) 1413–1421. https://doi.org/10.1039/D0NA00999G.

[27] A. Goswami, S.R. Mudi, C. Dempsey, P. Zhang, H. Wu, B. Zhang, W.J. Mitchell, J.S. Lee, S.M. Frolov, C.J. Palmstrøm, Sn/InAs Josephson Junctions on Selective Area Grown Nanowires with in Situ Shadowed Superconductor Evaporation, Nano Lett. 23 (2023) 7311–7318. https://doi.org/10.1021/acs.nanolett.3c01320.

[28] N.W. Hendrickx, M.L. V. Tagliaferri, M. Kouwenhoven, R. Li, D.P. Franke, A. Sammak, A. Brinkman, G. Scappucci, M. Veldhorst, Ballistic supercurrent discretization and micrometer-long Josephson coupling in germanium, Phys. Rev. B. 99 (2019) 075435. https://doi.org/10.1103/PhysRevB.99.075435.

[29] F. Vigneau, R. Mizokuchi, D.C. Zanuz, X. Huang, S. Tan, R. Maurand, S. Frolov, A. Sammak, G. Scappucci, F. Lefloch, S. De Franceschi, Germanium Quantum-Well Josephson Field-Effect Transistors and Interferometers, Nano Lett. 19 (2019) 1023–1027. https://doi.org/10.1021/acs.nanolett.8b04275.

[30] K. Aggarwal, A. Hofmann, D. Jirovec, I. Prieto, A. Sammak, M. Botifoll, S. Martí-Sánchez, M. Veldhorst, J. Arbiol, G. Scappucci, J. Danon, G. Katsaros, Enhancement of proximity-induced superconductivity in a planar Ge hole gas, Phys. Rev. Res. 3 (2021) L022005. https://doi.org/10.1103/PhysRevResearch.3.L022005.

[31] A. Tosato, V. Levajac, J.-Y. Wang, C.J. Boor, F. Borsoi, M. Botifoll, C.N. Borja, S. Martí-Sánchez, J. Arbiol, A. Sammak, M. Veldhorst, G. Scappucci, Hard superconducting gap in germanium, Commun. Mater. 4 (2023) 23. https://doi.org/10.1038/s43246-023-00351-w.

[32] M. Hinderling, S.C. ten Kate, M. Coraiola, D.Z. Haxell, M. Stiefel, M. Mergenthaler, S. Paredes, S.W. Bedell, D. Sabonis, F. Nichele, Direct Microwave Spectroscopy of Andreev Bound States in Planar Ge Josephson Junctions, PRX Quantum. 5 (2024) 030357. https://doi.org/10.1103/PRXQuantum.5.030357.

[33] L. Lakic, W.I.L. Lawrie, D. van Driel, L.E.A. Stehouwer, Y. Su, M. Veldhorst, G. Scappucci, F. Kuemmeth, A. Chatterjee, A quantum dot in germanium proximitized by a superconductor, Nat. Mater. 24 (2025) 552–558. https://doi.org/10.1038/s41563-024-02095-5.

[34] M. Zhu, M. Ben Shalom, A. Mishchsenko, V. Fal'ko, K. Novoselov, A. Geim, Supercurrent and multiple Andreev reflections in micrometer-long ballistic graphene Josephson junctions, Nanoscale. 10 (2018) 3020–3025. https://doi.org/10.1039/C7NR05904C.

[35] T. Li, J.C. Gallop, L. Hao, E.J. Romans, Scalable, Tunable Josephson Junctions and DC SQUIDs Based on CVD Graphene, IEEE Trans. Appl. Supercond. 29 (2019) 1–4. https://doi.org/10.1109/TASC.2019.2897999.

[36] A.A. Generalov, K.L. Viisanen, J. Senior, B.R. Ferreira, J. Ma, M. Möttönen, M. Prunnila, H. Bohuslavskyi, Wafer-scale CMOS-compatible graphene Josephson field-effect transistors, Appl. Phys. Lett. 125 (2024). https://doi.org/10.1063/5.0203515.

[37] G. De Simoni, F. Paolucci, P. Solinas, E. Strambini, F. Giazotto, Metallic supercurrent field-effect transistor, Nat. Nanotechnol. 13 (2018) 802–805. https://doi.org/10.1038/s41565-018-0190-3.

[38] G. De Simoni, C. Puglia, F. Giazotto, Niobium Dayem nano-bridge Josephson gate-controlled transistors, Appl. Phys. Lett. 116 (2020). https://doi.org/10.1063/5.0011304.

[39] C. Puglia, G. De Simoni, N. Ligato, F. Giazotto, Vanadium gate-controlled Josephson half-wave nanorectifier, Appl. Phys. Lett. 116 (2020). https://doi.org/10.1063/5.0013512.

[40] O. Arif, L. Canal, E. Ferrari, C. Ferrari, L. Lazzarini, L. Nasi, A. Paghi, S. Heun, L. Sorba, Influence of an Overshoot Layer on the Morphological, Structural, Strain, and Transport Properties of InAs Quantum Wells, Nanomaterials. 14 (2024) 592. https://doi.org/10.3390/nano14070592.

[41] A. Paghi, S. Battisti, S. Tortorella, G. De Simoni, F. Giazotto, Cryogenic Behavior of High-Permittivity Gate Dielectrics: The Impact of the Atomic Layer Deposition Temperature and the Lithography Pattering Method, Prepr. Http//Arxiv.Org/Abs/2407.04501. (2024). http://arxiv.org/abs/2407.04501.

[42] R. Acharya, S. Brebels, A. Grill, J. Verjauw, T. Ivanov, D.P. Lozano, D. Wan, J. Van Damme, A.M. Vadiraj, M. Mongillo, B. Govoreanu, J. Craninckx, I.P. Radu, K. De Greve, G. Gielen, F. Catthoor, A. Potočnik, Multiplexed superconducting qubit control at millikelvin temperatures with a low-power cryo-CMOS multiplexer, Nat. Electron. 6 (2023) 900–909. https://doi.org/10.1038/s41928-023-01033-8.